\documentclass[10pt,a4paper,oneside,headinclude,footinclude]{scrartcl}
\usepackage{geometry}
\usepackage{authblk}

\usepackage{nicefrac}
\usepackage{ifthen}
\usepackage[nottoc]{tocbibind} 
\usepackage{caption}

\usepackage[singlespacing]{setspace}
\usepackage{scrextend}
\usepackage[linktocpage]{hyperref}
\hypersetup{
    pdfstartview={Fit},
    pdftitle={A Quantum von Neumann Architecture for Large-Scale Quantum Computing},
    pdfauthor={Matthias Brandl},
    pdfcreator={Matthias Brandl},
    pdfproducer={Matthias Brandl},
    pdfsubject={Quantum computer architecture},
    pdfkeywords={physics, computer science, computer architecture, quantum optics, quantum computing, trapped ions},
    colorlinks=true,
    linkcolor=blue,
    citecolor=blue,
    filecolor=blue,
    urlcolor=blue
} 
\usepackage{graphicx}
\usepackage{epstopdf}
\title{A Quantum von Neumann Architecture for Large-Scale Quantum Computing}
\author[]{Matthias F. Brandl\thanks{Matthias.Brandl (at) sqip-project.org}}
\affil[]{\normalsize Institut f\"ur Experimentalphysik, Universit\"at Innsbruck, Technikerstra{\ss}e 25, A-6020 Innsbruck, Austria}

\begin{document}
	\maketitle

	\begin{abstract}
		As the size of quantum systems becomes bigger, more 
		complicated hardware is required to control these systems.  
		In order to reduce the complexity, I discuss the amount of 
		parallelism required for a fault-tolerant quantum computer and what 
		computation speed can be achieved in different architectures.
		To build a large-scale quantum computer, one can use architectural
		principles, from classical computer architecture, like multiplexing 
		or pipelining.  In this document, a Quantum von Neumann architecture
		is introduced which uses specialized hardware for the different 
		tasks of a quantum computer, like computation or storage.  
		Furthermore, it requires long qubit coherence and the
		capability to move quantum information between the different parts
		of the quantum computer. As an example, a Quantum von Neumann 
		architecture for trapped ions is presented which incorporates 
		multiplexing in the memory region for large-scale quantum computation.		
		To illustrate the capability of this architecture, a model trapped ion 
		quantum computer based on Quantum von Neumann architecture, the 
		Quantum~4004, is introduced. Its hardware is optimized for simplicity and
		uses the classical Intel~4004 CPU from 1971 as a blueprint. 
		The Quantum~4004 has only a single processing 
		zone and is structured in 4 qubit packages.  Its quantum memory can 
		store up to 32768 qubit ions and its computation speed is 10~$\mu$s 
		for single qubit operations and 20~$\mu$s for two-qubit operations.
	\end{abstract}
	
	\addcontentsline{toc}{section}{Table of Contents}
	\tableofcontents 

	\section{Introduction}
	\label{sec: intro}

	Since the 1960s, the number of components, e.g. transistors, in integrated 
	circuits (ICs) has doubled approximately every two years.  This exponential growth is 
	described by Moore's law \cite{MooresLaw,MooresLaw2} and it enabled exponential 
	increase in calculation power.  Before the early 2000s, the clock speeds of ICs 
	grew exponentially as well \cite{EndofMooresLaw}. But in the early 2000s, 
	the clock speeds of ICs reached levels for which cooling the ICs limited the 
	clock speed.  In order to maintain an exponential increase in calculation power,
	the size of transistors was decreased and multiple cores in ICs were incorporated.  
	But eventually, the size of transistors will reach the ultimate size limit, the 
	atomic level.  Then, one will have to find new ways of computation or settle 
	with a lower increase in computation power over time.

	One way to speed up computation for specialized applications is quantum
	computation (QC). In QC, the information is stored in quantum mechanical
	systems.  Thus, a quantum computer can make use of quantum mechanical
	properties, such as for example superposition or entanglement, to speed 
	up computation and the execution of quantum algorithms.  
	One of the best known quantum algorithms is Shor's algorithm \cite{ShorAlg}.  
	It allows factoring of large numbers exponentially faster than with 
	classical algorithms and undermines the security of public-key cryptosystems.
	Other quantum algorithms allow the implementation of an oracle \cite{DeutschJozsa}
	or fast searches in databases \cite{GroverAlg}.  Another application of 
	quantum computers is quantum simulation \cite{FeynmanQS,QSoverview}.  
	There, the idea is to use a well-controlled
	quantum mechanical system to simulate the behavior of another quantum
	mechanical system and gain insight into the behavior.  Possible applications
	are in condensed-matter physics, atomic physics, or quantum chemistry 
	\cite{QSoverview,QSoverview2}.

	As the quantum mechanical systems for information storage are subject to
	decoherence, quantum error correction (QEC) has to be performed for
	fault-tolerant QC \cite{KnillFaultTolerance,MikeIke}.  Experimentally, 
	fault-tolerant QC has not yet been performed on any quantum mechanical system.  
	Hence, building a quantum computer still remains a challenge.  
	There are many different systems under investigation as possible candidates 
	for a fault-tolerant quantum computer.  Two promising systems are 
	superconducting circuit quantum electro dynamics (QED) systems 
	\cite{TransmonQubit,MartinisShor} and trapped ions \cite{QIPwithTIWineland,MonzShor}.
	But as the interactions in these systems vary vastly, it proved difficult
	to find a common architecture for a quantum computer \cite{VanMeterQC}, similar to 
	von Neumann architecture for classical computers.

	In a lot of modern quantum computer architectures, quantum information storage 
	and actual computation regions are built up from the same hardware on the same 
	site \cite{DiVincenzoFTA4SQ,HugeIonTrap,QC_Monroe_Kim}.  This "maximum parallelism"
	approach led to "sea of qubits" architectures \cite{Parallelism_QC}
	which enable the execution of QEC on all qubits 
	simultaneously and shift the error threshold of gate operations for fault-tolerant 
	QC up \cite{KnillFaultTolerance,KnillFaultTolerantLimit}.
	For superconducting circuit-QED systems, two-dimensional lattice architectures are pursued 
	\cite{DiVincenzoFTA4SQ} which require parallel control of all qubits simultaneous.
	For trapped ions, a two-dimensional lattice architecture has been proposed
	\cite{HugeIonTrap} as well as an architecture of an array of photonically coupled
	ion traps \cite{QC_Monroe_Kim}.	
	But as systems get bigger, the classical control in such massively-parallel
	approaches will become more complicated.  Eventually, mundane engineering challenges
	will limit the size of quantum computer.  For example, if one requires one
	wire per quantum bit, or short qubit, from room temperature to the cold stage 
	in a cryogenic quantum computer, such a quantum computer will require a massive 
	amount of cooling power when working with a million or more qubits, as a million
	wires would pose a huge thermal load on the cryostat.

	Although there are systems with long coherence times, like trapped ions
	\cite{BollingerBeMagnInsensitive,HighFidGateHarty,KihwanKimCoherence},
	most modern quantum computer architectures do not make use of the fact
	that QEC does not constantly have to be performed for fault-tolerant QC
	but one can have idle times.  These idle times could be used for a
	serialization of the computation.
	For superconducting circuit QED systems, a Quantum von Neumann 
	architecture has been demonstrated \cite{QuVonNeumannMartinis} which
	had the memory separate from the processing zone.

	Here, I propose a more general Quantum von Neumann architecture and
	explore what properties a system for this architecture needs to 
	have and what computation speed, one can expect with the resulting
	serialization of the quantum information processing (QIP).
	This architecture is for large scale systems and, thus, cannot be
	compared with the proof-of-principle experiment performed in
	a superconducting circuit QED system \cite{QuVonNeumannMartinis}.
	To illustrate the simplicity of the required hardware with this 
	general Quantum von Neumann architecture, I apply this concept on 
	a suitable system, trapped ions, and present a large-scale trapped 
	ion quantum computer based on Quantum von Neumann architecture.

	The manuscript is structured as follows: Section~\ref{sec: classical computer science}
	presents concepts from classical computer architecture which are applicable
	in a quantum computer architecture. In Section~\ref{sec: quantum computer},
	ideas for the required parallelism and achievable computation speed of
	quantum computer are discussed, before the Quantum von Neumann architecture
	is introduced in Section~\ref{sec: quantum von neumann}.  In
	Section~\ref{sec: quantum von neumann 4 trapped ions}, the Quantum von Neumann 
	architecture for trapped ions is derived which is the basis for the
	Quantum~4004, a trapped ion quantum computer with a simple hardware and
	Intel's~4004 as its blueprint, of Section~\ref{sec: quantum 4004}. 
	Finally, Section~\ref{sec: conclusion} sums up the presented content.

	\section{Introduction to classical computer science}
	\label{sec: classical computer science}

	This section covers the fundamentals of computer science that are most applicable 
	to quantum computers.  It is intended for physicists with no specialized
	knowledge in computer science. If one is already familiar with terms like
	von Neumann architecture, abstraction layers, pipelining, multiplexing, or Rent's
	rule, one can skip this section and continue with Section~\ref{sec: quantum computer}.

	A computer is a device that executes sequences of instructions \cite{Tanenbaum}. 
	A sequence to perform a certain task is called a program.  With such programs,
	one can solve mathematical problems by executing algorithms 
	\cite{MikeIke}.  How a computer solves problems and what types of problems 
	it can solve (efficiently\footnote{An algorithm is called efficient, if the
	required resources in time and memory scale polynomially with the system
	size \cite{MikeIke}.}) is given by the incorporated architecture.  Hence, the
	choice of computer architecture directly relates to how efficiently one can
	solve given problems.

	\begin{figure}[!htb]
		\begin{center}
			\includegraphics[width=10cm]{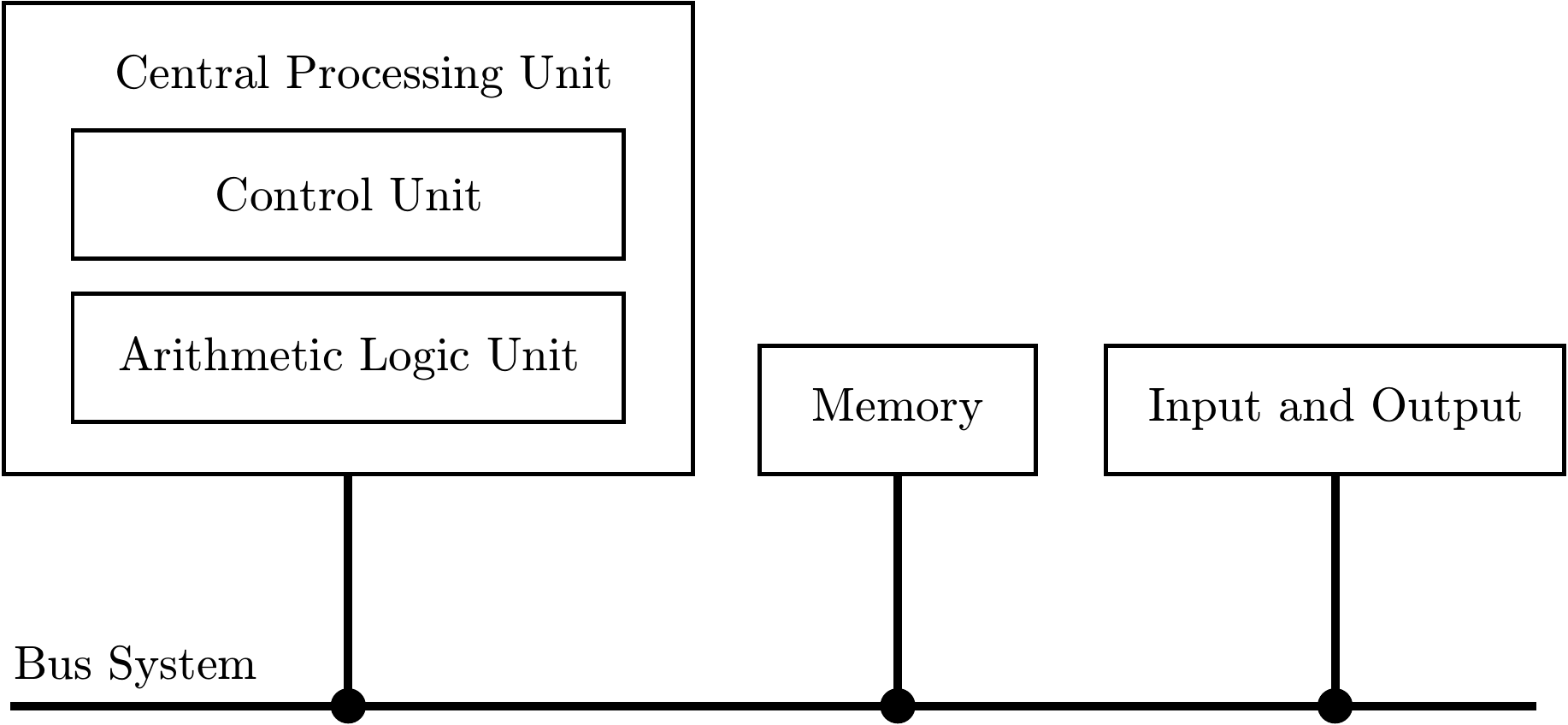}
			\caption{The von Neumann computer architecture.}
			\label{fig: von Neumann computer architecture}
		\end{center}
	\end{figure}

	Most modern computer architectures are based on von Neumann architecture \cite{vonNeumannArchitecture},
	which is depicted in Fig.~\ref{fig: von Neumann computer architecture}.
	The fundamental idea behind the von Neumann architecture is to divide the computer
	into individual parts: (1) arithmetic logic unit (ALU), (2) control unit (CU), (3)
	bus system, (4) memory, and (5) input and output (IO). The ALU
	can apply certain operations on its input registers.  These operations contain
	addition, subtraction, multiplication, and division and allow calculation with the 
	input registers.  Together with the CU, the ALU forms the central processing unit (CPU).  
	The CU controls the ALU by sending commands (e.g. when to perform what operation) as 
	well as a bus system which enables reading new commands and data from the bus into 
	the CPU and writing back the results %of the execution 
	to the bus.  A memory and an IO unit are attached to the bus so that one can
	read or store information, and react to external inputs by
	accessing the bus.

	To execute a command in a von Neumann architecture, the CU loads a command from
	the memory into the CPU where it is decoded and executed. The result of the execution
	can be access to the IO interface or storage in the memory.
	For example, if one wants to multiply two numbers, both numbers will be loaded
	into the CPU and stored in registers of the ALU, which are typically called
	Accumulator A and Accumulator B. The ALU performs the multiplication and writes the result
	back in the same registers where it can be used for further processing or it can 
	be written back into the memory.

	\subsection{Abstraction layers}
	\label{sec: classical - abstraction layers}

	A computer can only interpret a sequence of instructions to manipulate data.
	In order to simplify the usage of computers, the abstraction layers
	have been introduced to make the interface to the computers a more
	natural process for humans \cite{Tanenbaum}.  The idea behind these
	layers is that each layer hides the implementation details of the layer below 
	and offers simple functions for the layer above.  To execute a program
	at a higher level, the next-lower level translates, or converts, the program
	to instructions of this level which are then executed. This process is
	repeated until the program is physically executed.

	To discuss the concept of the abstraction layers with an example 
	which has five abstraction layers, let's assume a user wants to generate
	a program that reads a number from a file and displays on the screen
	whether the number is smaller than 1000, or not.  The user programs
	in the top layer, called the process layer, and the underlying layer,
	the operating system layer, offers functions e.g. for opening files, or 
	for comparing numbers.
	To execute the user's program, a compiler converts the functions
	of the operating system layer into assembler code for the assembler
	layer, the third layer.  In the firmware layer, which is the second
	layer from the bottom, these assembler commands are converted 
	into hardware commands that can be executed on this specific CPU.
	Finally, these hardware commands are executed on the hardware of
	the computer, the hardware layer.

	The concept of abstraction layers is fundamental in computer science.
	Without it, the usage of modern computers would not be as easy
	as it is today.

	\subsection{Parallel computing}
	\label{sec: classical - parallel computing}

\begin{figure}[!htb]
		\begin{center}
			\includegraphics[width=9cm]{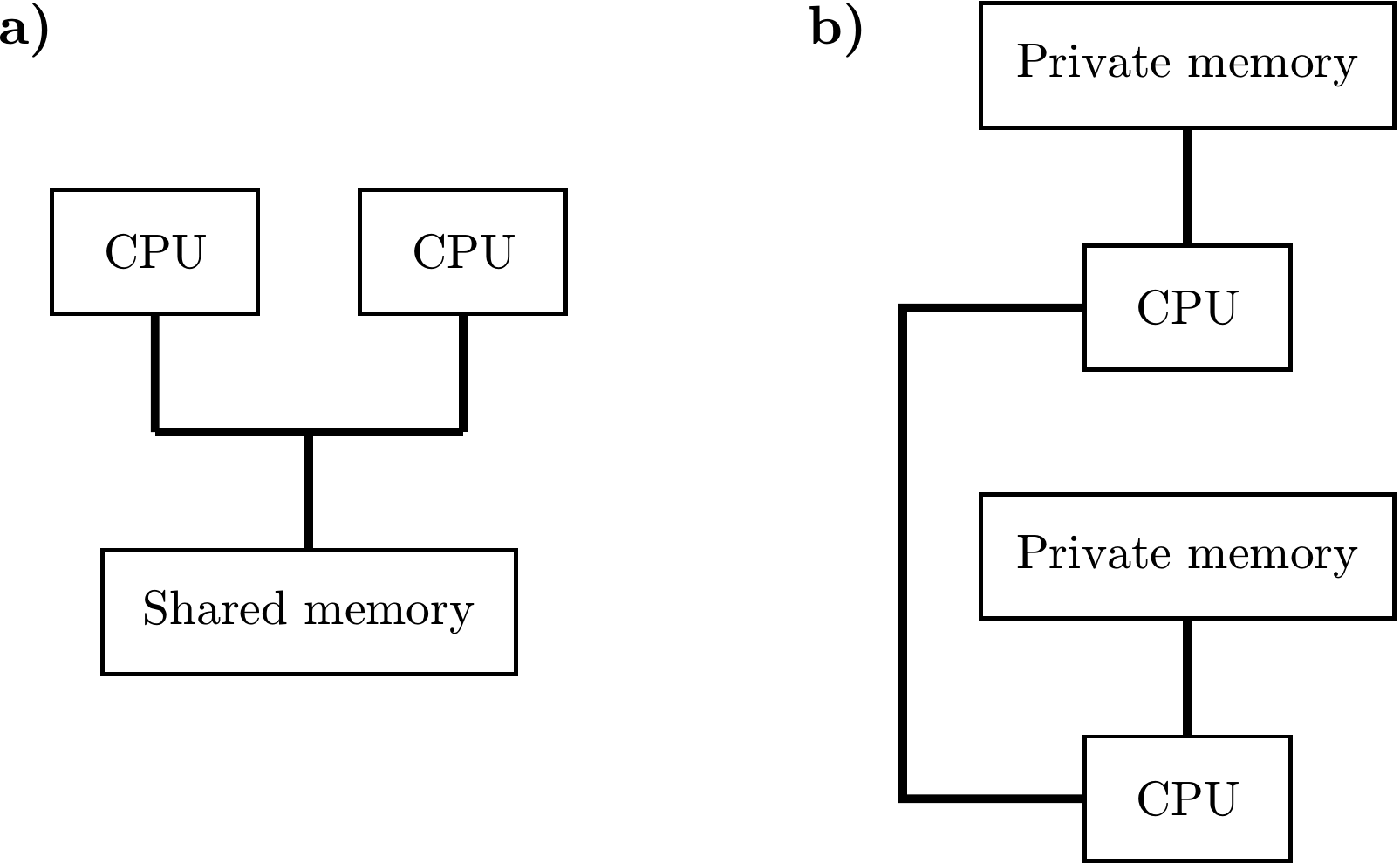}
			\caption{Parallelism in computer architecture.  Panel (a) shows
				a multiprocessor system with a shared memory. Panel (b) depicts
				a multicomputer.}
			\label{fig: von neumann bottleneck}
		\end{center}
	\end{figure}

	To speed up computation, one can either increase the clock speed
	of the CPU, or if this is not possible, one can execute
	multiple instructions simultaneously in parallel \cite{Tanenbaum}. 
	Parallel computers incorporate multiple CPUs
	in one computer, as shown in the multiprocessor system of 
	Fig.~\ref{fig: von neumann bottleneck}~a.
	Such a system illustrates a problem of the von Neumann architecture: data 
	processing in the CPU might be fast but all data has to come through
	the bus.  The clock of a bus is slower than the clock of a CPU.
	Hence, for fast CPUs (even for single processor systems), 
	the bus system is the limiting the performance of the computer.
	This is usually referred to as the "von Neumann bottleneck".

	Multiprocessor systems with shared memory have applications in for example
	graphics processing units (GPUs) \cite{Tanenbaum}. The algorithms behind 
	these applications have to be repeated many times (e.g. for each pixel
	on a computer screen).  %Modern GPUs can have thousands of single cores
	%allowing massive parallelism to solve these algorithms.
	Another way to increase parallelism is to build multicomputers, as 
	illustrated in Fig.~\ref{fig: von neumann bottleneck}~b. These
	computers contain multiple CPUs with their own private memories and buses.
	The CPUs are connected via special buses, making them
	very efficient when solving different problems in parallel.
	But as the different computers do not share memory, distributing single 
	algorithms usually involves additional overhead so that each CPU always
	works with correct data.
	%With the invention of the internet, computing grids emerged, like the
	%search for extraterrestrial intelligence (SETI) project \cite{Seti}.  
	%In such a grid, computers acquire a certain amount of data that they 
	%process independently.  Hence, grids are similar to multicomputers
	%for which the linking bus is replaced with an Ethernet connection.

	One thing that unites the different classical computer architectures 
	presented so far is that they separate the CPU from the memory.  
	CPUs require much more silicon area per bit of information\footnote{meaning 
	for example transistors} and produce much more heat per bit of information
	than memories. Thus, building the CPU is very hardware demanding and most 
	modern computers only have a few CPUs but a large memory.
	However in nature, brains compute massively parallel. %and asynchronously.
	As neurons in the brain both store information and are part of the computation, 
	people have tried to emulate the behavior of neurons with electrical circuits 
	to generate artificial neural networks (ANNs) \cite{NeuralNetwork}. 
	Such systems, also called neuromorphic systems \cite{NeuromorphicScience}, 
	map their input registers either in an analog or a digital way onto output 
	registers.  This mapping is performed in parallel and allows fast data 
	processing for specialized applications. To perform the desired task, these 
	mappings have to be "learned" with learning rules.  The applications of ANNs 
	are for example pattern recognition in images, video or audio. The field of 
	ANNs is still mainly research\footnote{e.g. DARPA's SyNAPSE project} but 
	companies like IBM and Qualcomm are building their first products based on 
	ANNs\footnote{e.g. Qualcomm Zeroth}.

	\subsection{Pipelining in CPUs}
	\label{sec: classical - pipelining}

	\begin{figure}[!htb]
		\begin{center}
			\includegraphics[width=13.5cm]{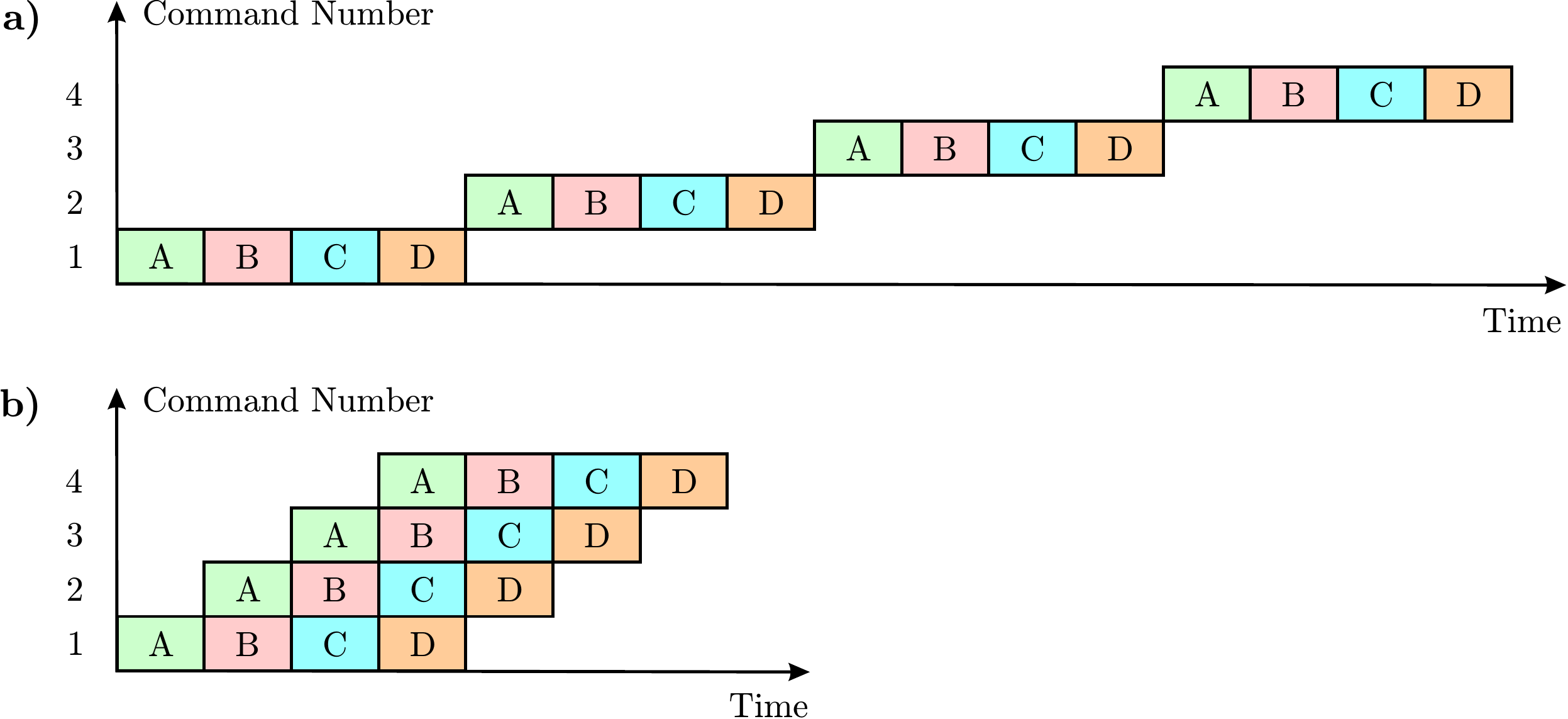}
			\caption{Execution of commands that can be split into four
				separate steps without (a) and with (b) pipelining.}
			\label{fig: pipelining}
		\end{center}
	\end{figure}

	Usually, the execution of a single command in a CPU can be divided into 
	multiple sub-operations. In the following example, a command
	can be divided into four parts.
	First, the command is loaded from the memory.  Second, the command
	is decoded to find out what the CPU has to do.  Third, the command
	is executed.  And fourth, the results are written back to the memory.
	In order to speed up processing, pipelining \cite{Hamacher,Tanenbaum}, illustrated in 
	Fig.~\ref{fig: pipelining}, is introduced where
	a CPU has the capability to execute all (four) different stages of a command
	in parallel.  Since the individual steps of a command have to be
	executed serially, the execution of one command takes the same
	time as without pipelining.  However, the tasks of the different
	stages of a pipeline can be executed simultaneously.  Hence, the
	execution of algorithms is speeded up by a factor of the number of 
	stages in the pipeline.  In the discussed example, that would be four.

	\subsection{Memory architecture - multiplexing}
	\label{sec: classical - multiplexing}

	Memory that can be read and written is called random access memory (RAM).
	In order to read data from or write data into a RAM, the CPU
	writes the RAM address, corresponding to the number of the memory segment 
	to access of the RAM, on the bus. A read/write
	flag indicates whether a read or a write process is being executed.
	As a result for a read process, the CPU expects that the data lines of the bus
	will contain the data that should be read.  For a write process,
	the CPU writes data on the same data lines of the bus.
	The number of memory registers that one can access scales as 2$^n$,
	where $n$ is the number of address lines on the bus which
	saves a lot of interconnections.

	Rent's rule is an empirical formula which is often used in the semiconductor
	industry as a scaling law.   It relates the number of pins $P$, 
	or external interconnections, of a system, with the number of logical 
	elements $B$ of this system, by a simple power law \cite{RentsRule1971,RentsRule2005}
	\begin{equation}
		P = K \cdot B^r \hskip 0.5cm ,
		\label{eq: rents rule}
	\end{equation}
	where $K$ is the Rent coefficient and $r$ is the Rent exponent.  The
	Rent coefficient is correlated to the average number of pins of a logical 
	element.  The Rent exponent sets the scaling with the systemsize, and 
	$r \leq 0.75$.  Hence, the number of external interconnections has to
	grow slower than the system size.  For example, a 1~MByte RAM chip must
	not have a thousand times more pins than 1~kByte RAM chip to obey Rent's rule.

	\begin{figure}[!htb]
		\begin{center}
			\includegraphics[width=13cm]{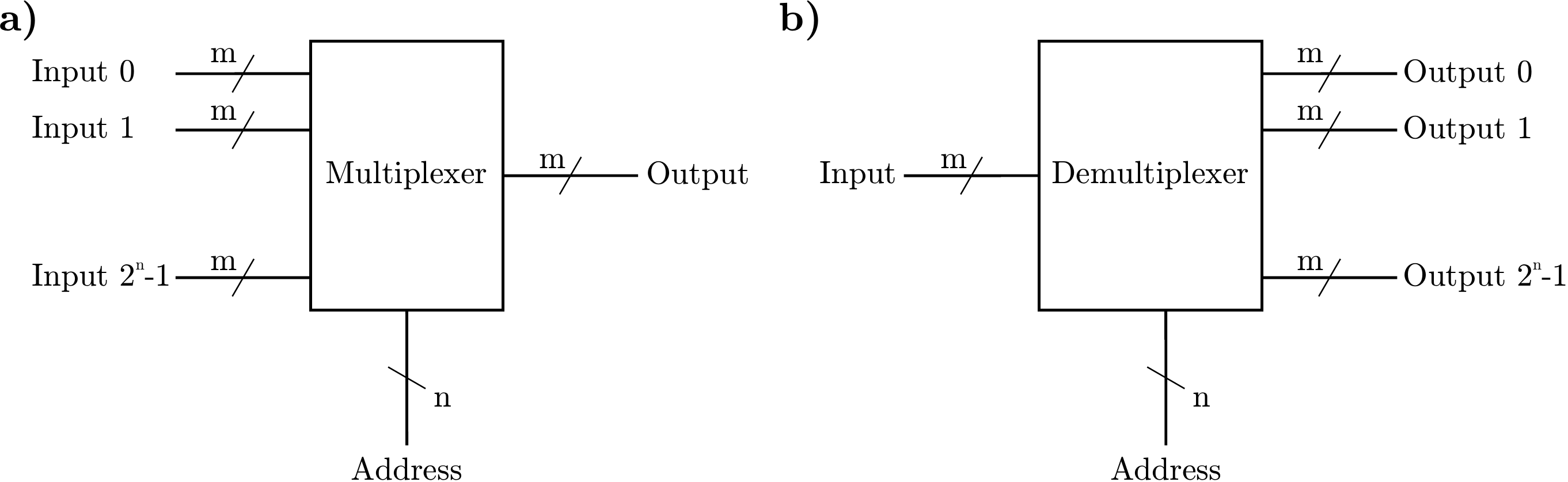}
			\caption{Multiplexer (a) and demultiplexer (b) circuits.}
			\label{fig: mux demux}
		\end{center}
	\end{figure}

	Multiplexers (Fig.~\ref{fig: mux demux}~a) are electronic
	circuits that have many input ports and only one output port \cite{Tanenbaum}.  
	An address applied to the multiplexer allows choosing which input is routed to the
	output.  Demultiplexers (Fig.~\ref{fig: mux demux}~b) are 
	the opposite of multiplexers and have only one input port but multiple
	output ports.  The address applied to the demultiplexer defines to which
	output port the input port is routed.  
	In a RAM, multiplexer and demultiplexer circuits are used to access (read or 
	write) a specific memory segment and require few pins to access a large
	number of memory cells.  This results in a Rent exponent $r$ of 0.12
	for static RAM (SRAM) \cite{RentsRule2005}.

	\section{Quantum computer}
	\label{sec: quantum computer}

	A quantum computer is a computer which executes sequences of instructions
	in a quantum mechanical system.  Quantum algorithms exploit quantum mechanical properties 
	like superposition and entanglement and can solve certain problems exponentially
	faster than classical computers \cite{ShorAlg,DeutschJozsa,GroverAlg}.
	The quantum information in quantum computers is stored in quantum mechanical
	two-level systems which are called quantum bits, or short qubits \cite{MikeIke}.
	As with any quantum mechanical system, the qubits are subject to noise-induced 
	decoherence.  Thus, quantum information cannot be stored infinitely long.
	Furthermore, quantum mechanical gate operations are analog operations
	which will introduce (small) errors in the computation in the Hilbert space spanned by the qubits. 
	To enable successful quantum computation (QC), quantum error correction (QEC) has to be
	performed repeatedly in the quantum computer \cite{MikeIke,ShorCode,SteaneCode}
	which corrects the errors induced by decoherence and gate operations.

	The concept of abstraction layers is fundamental in computer science and is
	applicable to quantum computers as well. In reference \cite{LayeredQCArchitecture},
	the authors propose a layered architecture for quantum computing with five layers.
	The bottom layer is the physical layer in which the physical gate operations
	are performed \cite{MonroeProgQC}. On top of that is the virtual layer in which one can use
	open-loop error cancellation for example with dynamical decoupling 
	\cite{DynamicalDecoupling1,DynamicalDecoupling2}.  In the third layer, QEC 
	is executed. And in the next layer, the logical layer, a substrate for universal 
	QC is constructed. Finally, the top layer is the application
	layer which provides an interface to the user who can input quantum
	algorithms there. In recent years, quantum assemblers \cite{Volckmarizer,MonroeAssembler,Estebanizer} 
	and quantum compilers \cite{MS_Liquid} are being investigated to generate 
	these sequences in an automated fashion similar to assemblers and compilers in
	classical computers.

	A physical quantum computer has to fulfill the following five criteria which 
	were proposed by DiVincenzo \cite{DiVincenzoCriteria}.  
	(1) A scalable physical system with well characterized qubits.
	(2) The ability to initialize the state of the qubits.
	(3) Long relevant decoherence times.
	(4) A universal set of quantum gates \cite{MikeIke}.
	(5) A qubit-specific measurement capability.
	Currently, many different technologies are investigated as possible candidates for a
	quantum computer, such as for example trapped ions \cite{QIPwithTIWineland,MonzShor}, 
	superconducting circuit quantum electro dynamics (QED) systems \cite{TransmonQubit,MartinisShor}, 
	quantum dots in silicon  \cite{SiliconTwoQubit,SpinInSilicon}, and ultra-cold atoms \cite{Saffman}.
	Finding a general architecture for a quantum computer, like von Neumann architecture
	for a classical computer, is challenging as the interactions for qubit manipulation
	vary vastly from technology to technology.  Therefore, the field of quantum computer
	architecture is still in its infancy \cite{VanMeterQC}.

	\subsection{Parallelism in quantum computation and classical hardware demand}
	\label{sec: parallelism in qc}

	There are two main sources of errors in QC. The first type is the storage error which is 
	caused by noise from the environment coupling to the quantum computer.  And the 
	second one is the error created by imperfect quantum gate operations.  QEC has to 
	correct these errors, and for arbitrarily long QC, these errors have to be lower than 
	a fault-tolerant limit\footnote{These thresholds assume systems that can grow
	arbitrarily in size.} \cite{KitaevFaultTolerance,KnillFaultTolerance,MikeIke}.  
	This means that errors during QC must be low enough for QEC to be effective.
	This fault-tolerant error threshold is assumed to be in the range between 10$^{-5}$
	and 3\% \cite{KnillFaultTolerantLimit}.

	For serial QC with just one processing zone, the coherence time of the qubits 
	would have to grow with growing number of qubits in the quantum computer to 
	maintain a constant memory error.  As this increase in coherence time is not 
	a realistic assumption, quantum computers with a large number of qubits will 
	require parallelism for QEC to be fault-tolerant \cite{SteaneParallelism}.

	On the path to minimize all errors in the system and thus to reach the fault-tolerant 
	threshold, one can decrease the memory error by making the quantum computer 
	massively parallel \cite{Parallelism_QC,DiVincenzoFTA4SQ,HugeIonTrap}.  If QEC is constantly 
	executed to "keep the qubits alive" and all necessary gate operations for QEC 
	are executed in parallel, the time required for QEC, and thereby the memory
	error, will be minimized.  The error threshold for fault-tolerant QC considers
	the memory and the gate error.  By minimizing the memory error, the threshold
	for the gate error will increase.  As there is no technology yet that has 
	shown fault-tolerance on a (small) set of qubits, a higher gate threshold
	will simplify reaching the fault-tolerant threshold of a given architecture.

	This massively parallel approach \cite{Parallelism_QC} is the most sensible one for small-scale systems.
	With growing system size, Rent's rule \cite{RentsRule1971,RentsRule2005} has to be 
	considered.  Otherwise, mundane engineering challenges like the complexity or 
	expenses will limit the fabrication of bigger systems \cite{VanMeterQC}.

	To estimate the possible size of such massively parallel architectures, one can
	look at classical architectures.  In such a massively parallel quantum architecture, 
	each qubit site has storage and calculation capability. The classical equivalent
	that is closest to such architecture is a neural network.  As there is no straight 
	forward way to build the hardware of ANNs yet, one can look at graphics processing 
	units (GPUs) which are often used to simulate ANNs in software because of their 
	massively parallel computation capability.
	Modern GPUs have thousands of ALUs. If one assumes that the hardware
	resources required for an ALU are comparable to the resources to control a qubit,
	one can assume that building a quantum computer in a parallel architecture with 
	hundreds or thousands of physical qubits will be feasible.

	The hardware demand of von Neumann type classical computers is a sum of the
	hardware demand for a CPU and the hardware demand for a memory. More memory will
	only affect the hardware demand for the memory and the bus connections of the CPU.
	As the CPU is very hardware demanding\footnote{In classical computers, hardware
	demand is often simplified as surface area on the silicon dye.} and one can use 
	multiplexing technologies in the memory \cite{Tanenbaum}, one can build classical 
	computers with large memories. In massively parallel quantum computer architectures, 
	which combine memory capability and computation capability on the same site, one 
	will require techniques from ANNs to reduce the hardware overhead for computation 
	or will otherwise be limited to small or medium size because of design considerations 
	like Rent's rule.  When following the classical approach for a large-scale 
	quantum computer architecture, one splits computation and memory into separate 
	regions \cite{Parallelism_QC}.  
	The computation region will be very hardware demanding and the memory region will 
	incorporate multiplexing technology to combine a big storage capability with low 
	hardware demand.

	In classical computer architecture, Rent's rule is the basis of models for power
	dissipation, interconnection requirements, or packaging \cite{RentsRule2005}.
	For quantum computers, one should find a similar model to estimate their
	hardware demand.  In general, one can state that typical parameters like
	price or power dissipation should scale as little as possible with the system
	size of (large-scale) quantum computers. For example, even small costs
	of the first qubit for computation can make hardware uneconomically expensive,
	if the scaling of the price with the number of qubits is linear and one
	wants to work with $10^8$ qubits\footnote{$10^8$ qubits are the approximated
	number of qubits required to factorize a number with about 10000 digits. See
	Section~\ref{sec: computation speed} for details.}.  However, if the cost scales 
	logarithmically with the number of qubits and the costs for the first qubit 
	are high, large-scale QC can still be economically sensible.

	For the estimation of the maximum number of physical qubits for fault-tolerant 
	QC per processing zone, the quantum computer must be built with a quantum 
	mechanical system, which allows moving quantum information from a memory 
	region to the processing zone and back, and as a first step, it will only 
	perform QEC but no additional QC to reach this maximum number of physical 
	qubits for fault-tolerant QC. 
	Typically, the coherence time of an experimental system is defined as a 
	decay time of a coherence measure, e.g. the \nicefrac{1}{e}-time of the 
	decay of the Ramsey contrast \cite{RamseyMeasurement}. In the following,
	the "QEC coherence time" is defined as the time when decoherence has caused
	a decay so big that QEC has to be performed and it can still be performed 
	with a fidelity high enough for fault-tolerant QC.  This time will depend 
	on the quantum computer and the incorporated QEC scheme but it will only 
	be a fraction of the stated coherence time.  To estimate the time of one cycle
	of QEC in a large-scale quantum computer, one can neglect the time for 
	detection and reinitialization because in a serial architecture with one 
	processing zone, detection and (re-)initialization can be performed in 
	different locations\footnote{An additional assumption here is that the
	detection time is so small that it can be neglected when compared to
	the "QEC coherence time".}.
	Thus, only the sum of all gate times will matter if the time, which is 
	required to move the quantum information in and out of the processing zone,
	is part of the gate time.  As an example, for a 
	syndrome measurement with a 7-qubit Steane code \cite{SteaneCode}, one 
	needs 4 entangling gates per syndrome, 3 syndrome measurements for bit 
	flips and 3 for phase flips.  This sums up to 24 entangling gates for 
	7 qubits.  Hence for a 7-qubit Steane code, the time for a single cycle 
	of QEC per physical qubit is \nicefrac{24}{7} times the average time of 
	an entangling gate operation plus \nicefrac{1}{7} of the time to perform 
	the correcting gate operation.  For higher level logical qubit encoding, 
	this value can be higher. With these assumptions, the maximum number of 
	physical qubits per processing zone $\kappa$ can be defined as
	\begin{equation}
		\kappa = \frac{\textrm{"QEC coherence time"}}{\textrm{time for a single cycle of QEC per physical qubit}}
		\label{eq: max number qubit per zone} \hskip 0.5cm ,
	\end{equation}
	where $\kappa$ must be greater than 1 for fault-tolerant QC. 
	The $\kappa$ value states how much serialization is possible in a quantum
	computer.  If this serialization can be implemented with multiplexing
	technologies, and thus much lower hardware demand than for computation, 
	high $\kappa$ values will indicate a potential for large-scale QC with serialization.

	Two of the most promising technologies to fabricate a quantum computer are
	superconducting circuit QED systems and trapped ions.  To estimate the (rough) $\kappa$ 
	values of these systems, let us assume in the following that in the chosen 
	logic qubit encoding, it takes the time equivalent of 10 physical entangling 
	gate operations to perform QEC\footnote{10 was chosen for simplicity not because
	of a specific qubit encoding.}. Furthermore, for simplicity the "QEC coherence time" 
	is chosen between 1 and 10~\% of the experimentally measured coherence time.
	For superconducting circuit QED systems \cite{QuVonNeumannMartinis}, the coherence 
	times are on the order of 100~$\mu$s \cite{SupercQubitsOutlook,SupercQubitsCoherence} 
	and entangling gates require about 100~ns \cite{SupercQubitsEntanglement,SupercQubitsEntanglement2}.  
	The resulting $\kappa$ value is between 1 and 10, which suggests a parallel 
	architecture for circuit QED systems\footnote{As circuit QED systems with
	such coherence times do not allow for large-scale QC with a single processing
	zone, one would have to include the detection time in the "QEC coherence time" 
	to estimate $\kappa$ more precisely.}.
	In trapped ion experiments, coherence times on the order of 100~s 
	\cite{HighFidGateHarty,KihwanKimCoherence} and entangling gate times of
	about 100~$\mu$s \cite{HFentangling} have been demonstrated, which
	yields a $\kappa$ value between 1000 and 10000.  This $\kappa$ value
	classifies trapped ions as a system which can be used for parallel 
	processing as well as for serial processing.  Since the qubits in
	this system can be moved to a processing zone \cite{Kielpinski} and multiplexing is
	possible, the (classical) hardware demand for large-scale trapped ion 
	systems can be reduced drastically by serialization of the computation, 
	see Section~\ref{sec: quantum von neumann 4 trapped ions} and 
	\ref{sec: quantum 4004} for details.

	\subsection{Computation speed}
	\label{sec: computation speed}

	In the previous section, serialization in quantum information processing
	(QIP) was introduced as a method to reduce the hardware demand in
	large-scale systems.  This serialization implies that fewer physical gate 
	operations can be performed per unit of time compared to a fully parallel
	architecture.  But this serialization allows working with more qubits
	than in a parallel architecture with similar hardware resources, which 
	will lead to a speed-up as discussed later in this section.

	The comparison of the computation speed of different architectures heavily
	depends on the quantum algorithm that has to be executed.  In this section,
	Shor's algorithm \cite{ShorAlg}, which allows fast factoring of large numbers, 
	is used as a benchmark for computation speed, as discussed in Van Meter's PhD 
	thesis \cite{VanMeterPhD}.

	With Shor's algorithm, one can factorize a number $N$ which has a binary bit 
	length of $n$.  In his thesis, Van Meter introduces different architectural 
	models to execute Shor's algorithm and relates them to their execution time.
	For this, he requires a logical clock speed of the hardware, which states
	how many operations can be executed on the logical qubits per time and implies
	that QEC is executed in between the operations on the logical qubits.

	One architectural model used by Van Meter is Beckman-Chari-Devabhaktuni-Preskill 
	(BCDP), which requires next-neighbor-only interaction.  
	It uses ($5n+3$) qubits to factorize a number with $n$ bit length
	and its execution time scales as $\sim 54n^3$.  Another model is the 
	Neighbor-only, Two-qubit-gate, Concurrent (NTC) architecture, which is executed 
	with next-neighbor-only interaction in a $2n^2$ qubit space and its execution 
	time scales as $\sim 20n^2 \textrm{log}_2(n)$.
	The last model is the Abstract Concurrent (AC) architecture, which allows
	two- and three-qubit interactions between arbitrary qubits.
	This model is executed in a $2n^2$ qubit space and its execution 
	time scales as $\sim 9n \textrm{log}_2^2(n)$.

	As the qubit demand in the BCDP architecture only scales linearly with
	the length of the number to factorize, BCDP is the obvious choice for
	a massively parallel system, typically systems with low $\kappa$ values.  
	Whereas, the other two architectures are executed in a $2n^2$ qubit space,
	which suggests some kind of serialization to reduce the classical
	hardware demand for factorizing reasonably big numbers. Hence, such
	models are executed on hardware with a high $\kappa$ value, e.g. 
	$\kappa > 1000$.

\begin{figure}[!htb]
		\begin{center}
			\includegraphics[width=12cm]{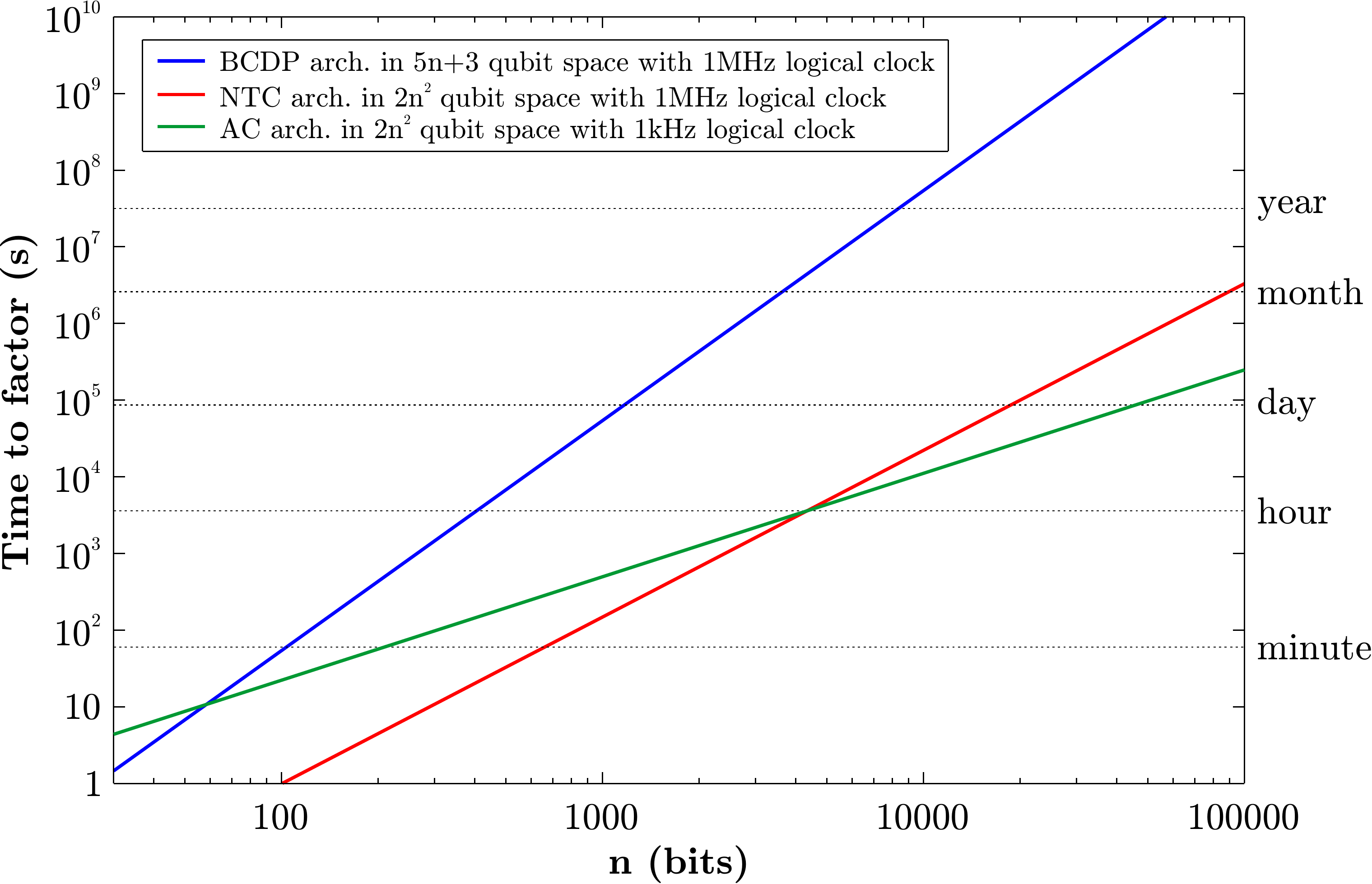}
			\caption{Time to factor a number $N$ with $n$ bit length with Shor's
				algorithm for different architectures.}
			\label{fig: shor time}
		\end{center}
	\end{figure}

	In the following example, let us assume that hardware allows for
	1~MHz logical clock speed in a massively parallel approach used
	in the BCDP architectural model.  In the NTC model, it is possible
	to compute with next-neighbor-only interaction even in a $2n^2$ 
	qubit space with a 1~MHz logical clock speed.
	And due to serialization, the AC model can only be executed with a 
	1~kHz logical clock speed.  The execution times of a factorization with Shor's
	algorithm are depicted in Fig.~\ref{fig: shor time} in dependence
	of $n$.

	For short numbers ($n < 50$), the higher logical clock speed in
	the BCDP and NTC architectures allows for the fastest (quantum) computation.
	NTC is faster than BCDP and is less dependent on $n$ than BCDP.
	Hence, a bigger computation space allows for faster computation.
	As $n$ grows, the AC architecture become faster than BCDP and NTC.  
	This illustrated that for large-scale systems, interaction between 
	arbitrary qubits speeds up computation more than a fast logical
	clock.

	It is not possible to generalize the following statement for all quantum algorithms but
	typically one can state: small- and medium-scale QC is best executed 
	with a fast logical clock speed for shortest computation time, suggesting 
	a massively parallel hardware. However, the bigger the computation, the 
	more it makes sense to switch to hardware which allows interaction 
	between arbitrary qubits and temporarily storing quantum information 
	in ancilla qubits.  To outperform massively parallel architectures,
	such hardware requires a big total number of qubits which suggests some 
	kind of multiplexing to reduce the classical hardware demand (Rent's rule). 
	Then, systems with high $\kappa$ values allow not only computation with 
	more qubits but also faster computation.

	\section{Quantum von Neumann architecture}
	\label{sec: quantum von neumann}

	When building a novel type of computer, like a quantum computer, 
	one can use an architecture that is based on one of the 
	(presented) classical architectures \cite{Tanenbaum} to avoid 
	having to design a fundamentally different kind of architecture.
	Furthermore, facilitating scalability in quantum computer hardware is one of 
	the most challenging tasks of quantum computer architectures.
	In this section, the quantum von Neumann architecture is introduced
	which combines the classical von Neumann architecture with the requirements
	of the DiVincenzo-criteria in QC \cite{DiVincenzoCriteria} resulting
	in quantum hardware which incorporates scalability.

	In a massively parallel quantum computer, the DiVincenzo criteria
	have to be fulfilled at every site that holds a qubit. 
	In order to simplify the hardware of a quantum computer, one can 
	fabricate hardware specialized on only one criteria and move the
	quantum information between these specialized hardware components
	to perform QC \cite{Parallelism_QC}. 

\begin{figure}[!htb]
		\begin{center}
			\includegraphics[width=12cm]{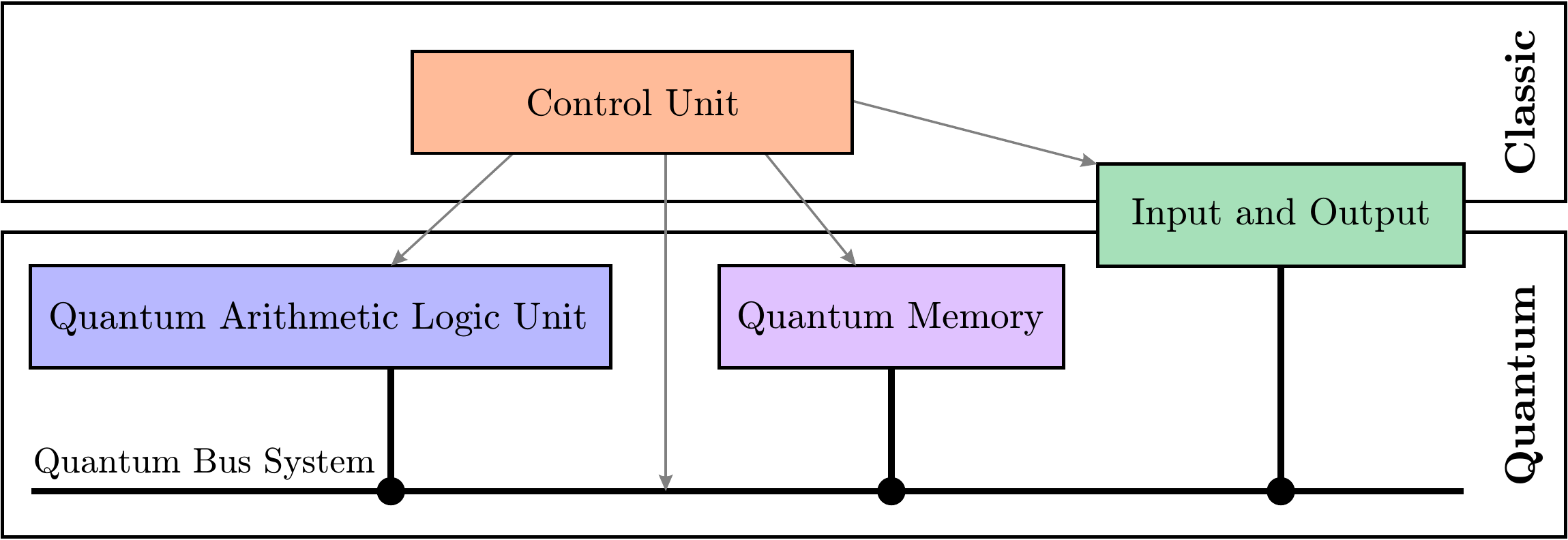}
			\caption{The Quantum von Neumann architecture.}
			\label{fig: quantum von Neumann computer architecture}
		\end{center}
	\end{figure}

	The schematic diagram of the quantum von Neumann architecture is depicted in 
	Fig.~\ref{fig: quantum von Neumann computer architecture}.
	Like any quantum computer, it will require a classical control
	unit which controls the quantum computer.
	A quantum bus system allows moving quantum information
	between the different parts of the quantum computer.
	The manipulation of the quantum information is executed in 
	the quantum arithmetic logic unit (QALU)
	which is the most hardware demanding\footnote{For the hardware demand
	of a quantum computer, one has to consider the required classical 
	hardware for quantum state manipulation (per qubit) and the
	complexity for entangling operations.  This hardware can include
	RF sources or laser sources but can also include the heat impact on
	a cryostat cause by the wiring required per qubit. For the complexity
	of the hardware, one has to consider things like for example the scaling of RF crosstalk
	between different RF lines in dependence of the system size.} 
	part of the quantum computer as quantum gate operations are executed here.
	The quantum information is stored in the quantum memory
	which should rely on multiplexing technology for large storage
	capability.  Furthermore, an input and output region acts as 
	an interface to the classical world in which the qubit
	state is initialized and/or detected.

	The operation principle of the quantum von Neumann architecture 
	is similar to that of classical von Neumann architecture.
	A quantum von Neumann machine executes a series of
	quantum gate operations by loading the qubits which should
	be manipulated into quantum registers of the QALU.  
	The quantum register length can be arbitrarily long. %, starting at one qubit.
	In order to entangle two qubits from arbitrary positions in the 
	memory, the two qubits are loaded from the quantum memory into the 
	QALU where the gate operations are performed.  After the quantum
	gate operations, the qubits can stay in the QALU for further 
	processing or the qubits are moved back into the quantum memory.
	To detect quantum states, the required quantum information
	can be moved to an output, or detection, region.
	As this detection region works independently from the QALU,
	QIP in the QALU and detection in the output region can
	be performed simultaneously.  After quantum state detection,
	the qubits can be moved to an input region for initialization
	into one specific state before the (now initialized) quantum 
	information can be moved back into the quantum memory. 
	If a desired initial state is a more complex state, e.g.
	an entangled state, the initialized qubits can be moved
	into the QALU which performs quantum gate operations
	to generate the desired initial quantum state.

	\subsection{Quantum memory region}
	\label{sec: quantum memory}

\begin{figure}[!htb]
		\begin{center}
			\includegraphics[width=12cm]{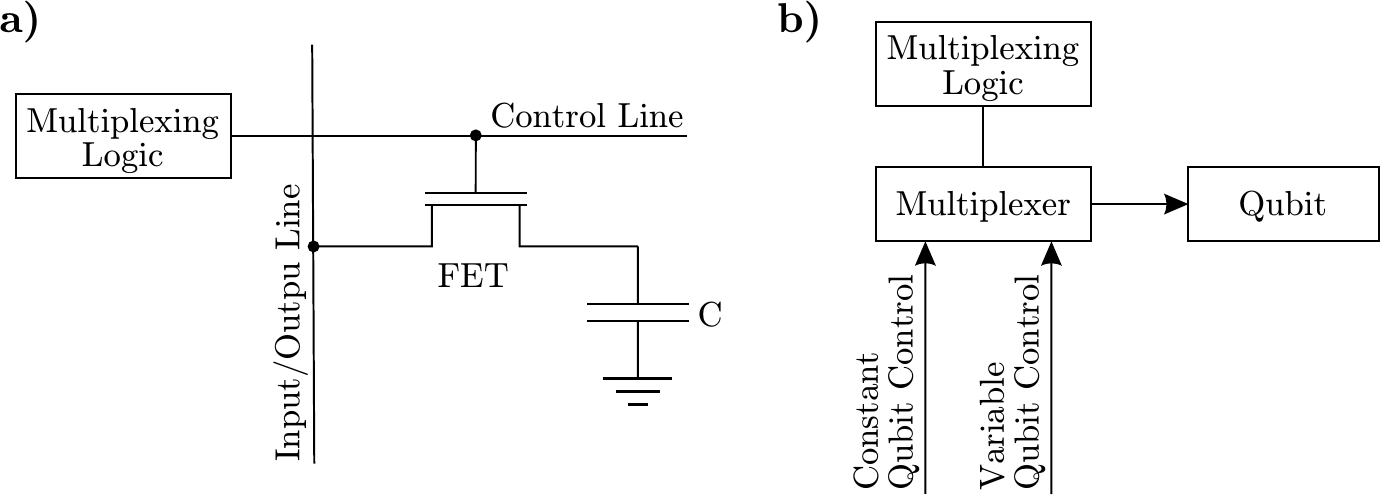}
			\caption{A memory cell in a DRAM (a) and in a
				multiplexing quantum memory (b).}
			\label{fig: quantum memory architecture}
		\end{center}
	\end{figure}

	In a classical computer, the components of a dynamic RAM (DRAM)
	needed to store one bit of information are a field effect 
	transistor (FET) and a capacitor, as depicted in 
	Fig.~\ref{fig: quantum memory architecture}~a. Digital multiplexing logic
	controls the FET to access the DRAM cell.  This low hardware
	demand per bit results in big data storage capacities on DRAM 
	chips \cite{Tanenbaum,RentsRule2005}.

	In large-scale quantum computers, one has to achieve big quantum
	data storage capacities with low (classical) hardware demand in
	the quantum memory.  If the hardware demand scaled linearly
	with the number of qubits that were stored, it would not obey
	Rent's rule and the control hardware would get too complicated
	and too expensive for large-scale QC with thousands or millions 
	of qubits.   

	Reducing the hardware demand can be achieved with 
	multiplexing circuits, as depicted in Fig.~\ref{fig: quantum memory architecture}~b.  
	Therefore, one needs to have the ability to store quantum 
	information with a set of constant parameters.  
	For example, to store an ion chain in a segmented
	Paul trap, one only needs a negative DC voltage at the position of the
	ion string and positive DC voltages surrounding it which form an 
	axial confinement for the ion string.  These few
	voltages can in principle be used to store arbitrarily many
	ion strings.  During storage, this set of parameters (for
	trapped ions that would be a set of DC voltages) is applied
	to all qubits in the quantum memory. In order to access
	a specific memory cell, multiplexing technology allows a change of
	this set of parameters to another set which can be controlled
	independently.  This independent set of parameters enables
	movement of the quantum information of an arbitrary memory 
	cell out of the quantum memory.

	\subsection{Quantum information transport}
	\label{sec: qi transport}

	One of the most critical features of this quantum von
	Neumann architecture is the quantum bus system for
	quantum information transport, which has to be performed
	with high fidelity to allow fault-tolerant QC.
	As quantum information cannot be copied \cite{NoCloning},
	quantum information can only be transported by
	physically moving the qubits, quantum teleportation
	\cite{QuantumTeleportation} or via coupling to photons
	\cite{FlyingQubits,CQEDcoupling}. 

	Atomic or molecular qubit systems enable quantum information
	transport via physical movement of the qubit from one 
	location in space to another. For example in trapped ion 
	systems with segmented Paul traps \cite{Wineland1998,Kielpinski}, 
	ions or ion strings can be moved by changing the confining
	axial DC potential.
	In solid state systems, such movement is not possible in
	general.  However, there are solid state systems which allow
	qubit movement, like spins in silicon 
	\cite{SiliconTwoQubit,SiliconQCoverview,SiliconQCarch}.

	Quantum teleportation \cite{QuantumTeleportation}
	requires an entangled qubit pair, of which one qubit
	is at the location from where the quantum information is taken
	and the second is the qubit at the destination. Furthermore,
	it requires a qubit measurement with a classical channel
	to the destination where a conditional quantum gate has
	to be performed.  In order to store and read quantum
	information in the quantum memory, it implies read-out-
	and quantum-gate-capability at every site in the memory.
	This is in contradiction to a specialized hardware
	for each DiVincenzo criteria and, thus, more hardware
	demanding than physical movement. But it could be a strategy 
	in many solid-state systems. 

	Mapping qubits to photons was demonstrated in 
	atomic or molecular qubit systems \cite{MonroeRemoteEntanglement,FlyingQubitIon}
	as well as in solid-state systems \cite{CQEDcoupling}.
	Like in quantum teleportation, this approach requires 
	quantum logic at every site in the memory and is therefore
	hardware demanding.

	Quantum information transport with quantum teleportation or
	mapping to photons have one advantage over qubit movement:
	it is possible to change from one qubit system to another. 
	For example, QIP could be performed with superconducting 
	circuit QED systems \cite{TransmonQubit,MartinisShor} and for 
	long storage in the quantum memory, one could use nitrogen 
	vacancy centers in diamond \cite{NVoverview}. 
	The disadvantage of these technologies compared to
	systems, which allow qubit movement, is the high hardware
	demand in the memory, as quantum gate operations and 
	quantum state readout are required at every site in the quantum memory.
	If this cannot be overcome, quantum teleportation and mapping to photons
	will only be applicable to small- and medium-scale systems.
	Large-scale systems with low hardware demand per stored 
	qubit may have to move the qubits in the quantum computer \cite{Parallelism_QC}.

	\subsection{Parallelism in quantum von Neumann architectures}
	\label{sec: parallelism in qvNa}

\begin{figure}[!htb]
		\begin{center}
			\includegraphics[width=13.5cm]{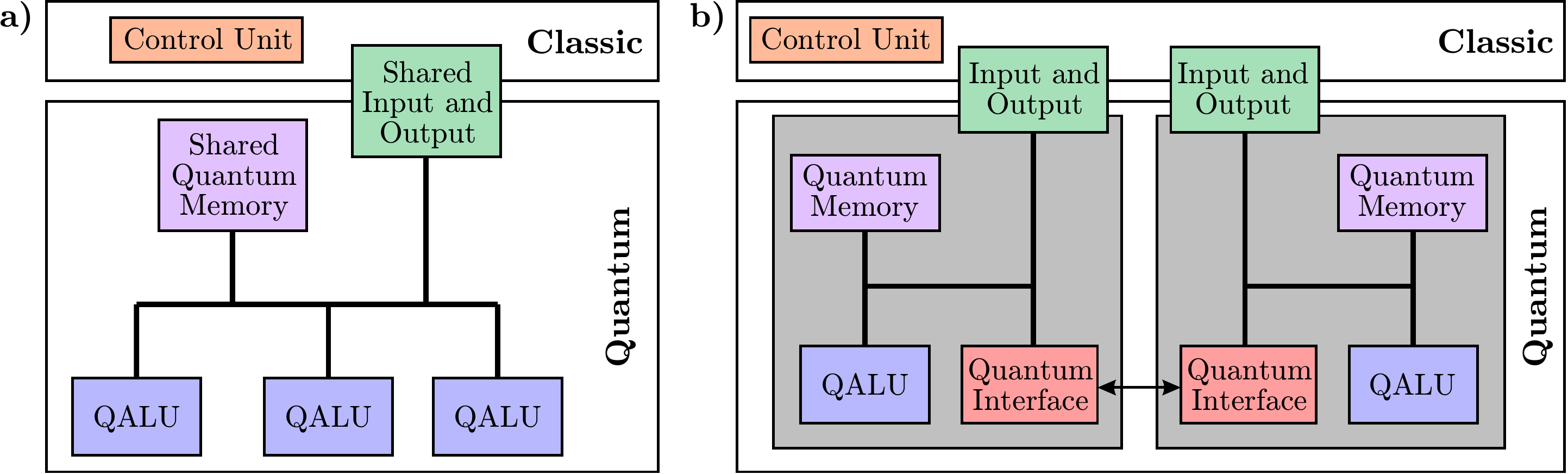}
			\caption{Parallelism in Quantum von Neumann architecture.  
				Panel (a) shows a multiquantumprocessor system with a 
				shared quantum memory. Panel (b) depicts a 
				multiquantumcomputer which couples multiple quantum computers
				via a quantum interface.}
			\label{fig: parallelism quantum von neumann}
		\end{center}
	\end{figure}

	In order to work with an increasing number of qubits in a quantum
	von Neumann architecture, one has to increase the $\kappa$ value
	to compensate decoherence in the quantum memory.
	Therefore, one can either increase the coherence time or decrease 
	the time per quantum gate operation.  If both options are not
	feasible, one has to parallelize QIP.  Similar to classical
	multiprocessor systems, one can use multiple QALUs in one
	quantum computer, as depicted in Fig.~\ref{fig: parallelism quantum von neumann}~a.

	Another option, illustrated in Fig.~\ref{fig: parallelism quantum von neumann}~b,
	is to couple multiple quantum computers via quantum interfaces,
	which is the quantum equivalent of a multicomputer system.
	These quantum interfaces can either be implemented by actual
	physical qubit exchange (qubit movement), quantum teleportation 
	\cite{QuantumTeleportation}, or mapping to photons 
	\cite{FlyingQubits,CQEDcoupling}.  Here, the hardware demand
	for the interface can be higher because it is not needed at
	every storage site in the quantum memory but only once
	for the interface.

	\subsection{Possible technologies}
	\label{sec: possible technologies}

	A technology to implement the quantum von Neumann architecture
	has to have a high $\kappa$ value and needs to be capable of
	quantum information transport, ideally by physically moving the
	qubits.  Both criteria are fulfilled in trapped ion experiments
	and details on the implementation of a quantum von Neumann architecture
	for trapped ion QC are presented in Chapter~\ref{sec: quantum von neumann 4 trapped ions}
	and \ref{sec: quantum 4004}.

	Another technology suitable for a quantum von Neumann architecture
	is QC with ultracold atoms \cite{Saffman}.  There, atoms can be stored
	in optical lattices.  Hence, quantum memories with low hardware
	demand are feasible.  Micro-electro-mechanical systems (MEMS) technology
	enables beam steering which, in combination with optical tweezers
	\cite{OpticalTweezers}, can be used to move atoms or qubits in
	QC with ultracold atoms.

	A promising candidate for a solid-state system with high $\kappa$ values and capability
	to move qubits is QC with spins in silicon \cite{SiliconQCoverview,SiliconQCarch},
	where coherence times of 28~ms have been demonstrated \cite{SiliconQCcoherence}.
	There, the qubits can be moved in an electric field by changing 
	DC voltages \cite{SiliconTwoQubit}.

	\section{A quantum von Neumann architecture for trapped ion quantum computation}
	\label{sec: quantum von neumann 4 trapped ions}

	This section covers how one can build the different parts of a 
	quantum von Neumann architecture in a trapped ion system. 
	The next section will combine these individual parts to build
	a model trapped ion quantum computer based on
	quantum von Neumann architecture called Quantum 4004.

	The guideline for development of the architecture is as follows: 
	(1) Trapped ions QC was chosen because ion traps are a technology
	with high $\kappa$ values.
	(2) If it is possible, one should only use operations 
	that have already been demonstrated with high fidelity.
	(3) This document covers only the hardware of the quantum computer.
	(4) Simplicity of the hardware, especially for scaling of the
	quantum computer, is favored over optimization for higher abstraction
	layer tasks, containing things like QEC or quantum algorithms, throughout this section.	
	(5) As there is no functioning fault-tolerant quantum computer
	yet, one cannot expect a first-generation quantum computer to work
	with high computation speed. Thus, computation speed has low priority
	in the development of the architecture.
	If a fault-tolerant quantum computer can be built and if Moore's law 
	is applicable to quantum computer development, the computation
	speed (and quantum memory size) will increase exponentially
	over time.

\begin{figure}[!htb]
		\begin{center}
			\includegraphics[width=6cm]{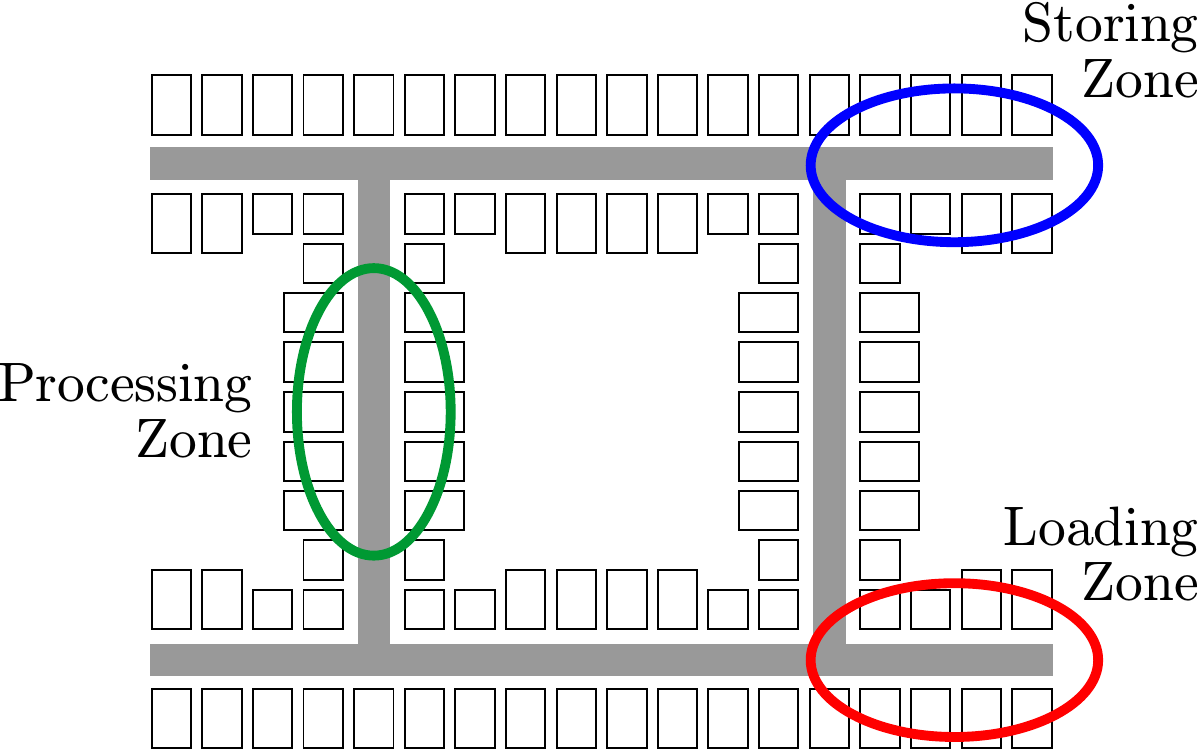}
			\caption{The QCCD principle in which ions (charges) are moved along 
				(gray) RF rails to different parts of the trap.}
			\label{fig: QCCD}
		\end{center}
	\end{figure}

	In the quantum charged coupled device (QCCD) principle \cite{Kielpinski},
	shown in Fig.~\ref{fig: QCCD}, a segmented ion trap is used to move ions 
	to different positions on the trap by changing the axially confining DC voltages.
	This allows using one part of the trap as a quantum memory and another
	part as a processing zone, or QALU.
	The QCCD \cite{Kielpinski} is a general concept for trapped ion QC and
	resembles a quantum von Neumann architecture, as there are separate
	regions for the different DiVincenzo criteria \cite{DiVincenzoCriteria}
	and it enables qubit movement (along the RF rails of the segmented Paul trap).
	In the following, the QCCD is used as the underlying principle of a 
	quantum von Neumann architecture for trapped ion QC.

	For the QCCD, ideas for QEC and higher level architectures\footnote{
	Higher level means in abstraction layers above the hardware level.}
	have been proposed \cite{QuantumLogicArrayMicroarchitecture}.
	However, trapped ions offer a variety of different gate operations and, 
	thus, it makes sense to adapt at the abstraction layer scheme for 
	trapped ion QC.  Gates can be performed using local RF
	fields \cite{Wunderlich_RF,HighFidGateHarty,MicrowaveAddressing,EntanglingRFgate},
	global RF fields \cite{Hensinger_Global_Field} or 
	optical fields \cite{PhilippNJP,HFentangling}.  Even for optical entangling
	gates, there are multiple types of gates \cite{Cirac-Zoller,MS-Gate,EntanglingPhasegate}.
	Similarly, multiple procedures for efficient ion movement have been
	demonstrated \cite{FirstShuttling,FastShuttling1,FastShuttling2}.
	In the following, the lowest abstraction layer of the scheme described
	in Section~\ref{sec: quantum computer} (and reference \cite{LayeredQCArchitecture}) 
	is split into two.  The new lowest level is then called the hardware layer, 
	and on top of the hardware layer, a firmware layer is inserted.  
	This new layer contains the firmwire, such as the type of quantum gates 
	and ion movements.  Since the top layers have only weak hardware dependence,
	they do not need to be adapted.  

	The quantum von Neumann architecture, presented in this section, covers 
	only the hardware abstraction layer. % and gives a few suggestions for the 
	%firmware layer and the virtual layer.  
	As the exact performance of the hardware is not known and, thus, the 
	optimum QEC scheme cannot be identified\footnote{Different QEC schemes have 
	for example different fidelity thresholds
	or require a different amount of ion movement during computation.  Hence, one has
	to find the best QEC scheme for a given architecture by evaluating the parameters
	of this architecture.}, it does not make sense to discuss the higher abstraction
	layers for this architecture at this point.

	The different design challenges for such a quantum von Neumann architecture 
	with trapped ions are
	\begin{itemize}
		\item vacuum pressure,
		\item decoherence in the quantum memory,
		\item multiplexing to enable large quantum memories,
		\item quantum gates,
		\item read out and initialization, and
		\item choice of qubits,
	\end{itemize}
	which are discussed in the following subsections.

	\subsection{Reduce collisions with background gas}
	\label{sec: collisions with background gas}

	In order to maximize the coherence time in trapped ion systems, collisions 
	with background gas should not limit the coherence times, as they can lead 
	to ion loss or to loss of quantum information in the ion chain.  
	Although these losses can be corrected with QEC, it is advisable to suppress 
	such collisions as much as possible.   In room temperature 
	setups, collisions with residual background gas occur roughly once per hour 
	per ion at typical UHV pressures of 10$^\mathrm{-11}$~mbar \cite{MonroeKim}.
	That means when working with, for example 3600, ions in room temperature setups, 
	one will have approximately one collision per second.  In a cryogenic ion trap
	experiment at a temperature of 4~K, a residual background pressure of 
	10$^\mathrm{-16}$~mbar has been observed \cite{CryogenicVacuum}.  
	Such pressures reduce the collision rate by 5 orders of magnitude compared to 
	room temperature setups.  Hence in cryogenic experiments, one can work with 
	more ions than in room temperature setups while at the same time reducing the 
	collisions with background gas. This suggests that large-scale QC with trapped 
	ions will have to be performed in a cryogenic environment.

	Ideally, one wants to be able to neglect collisions with background gas
	as a source of qubit loss or decoherence.  Therefore, one can look at the
	two elements with the lowest boiling point (or triple point), hydrogen and helium.
	As the exact vacuum pressure in an experiment strongly depends on the used
	materials, whether they were baked before, and so on, one can only perform
	a worst-case analysis by looking at the vapor pressure of hydrogen and helium.
	For the vapor pressure, one assumes at the whole vacuum chamber is covered
	with at least one monolayer of the element in question.  Hydrogen has a 
	sublimation equilibrium pressure of 10$^{-6}$~mbar at a temperature of 4.2~K, 
	and 10$^{-12}$~mbar at a temperature of 2.6~K \cite{CryoVacuumPressures,H2SaturationPressure}.
	Hence at a temperature around 2~K, hydrogen can no longer sublimate and
	will definitely be frozen out.
	If helium is also a source for collision with the ions, one will have to
	cool even further, as $^4$He has a sublimation equilibrium pressure of 
	10$^{-6}$~mbar at a temperature of 0.46~K, and 10$^{-12}$~mbar at a 
	temperature of 0.24~K \cite{CryoVacuumPressures}. $^3$He shows a sublimation 
	equilibrium pressure of 10$^{-6}$~mbar already at a temperature of 0.22~K, 
	and 10$^{-12}$~mbar at a temperature of 0.1~K \cite{CryoVacuumPressures}.
	This does not imply that the whole experiment has to be performed
	at a temperature of 0.1~K to not be limited by collisions with $^3$He.
	But at least one surface in the cryostat will have to be that cold to 
	exclude collisions with background gas from the sources of qubit loss 
	or decoherence.

	\subsection{Decoherence in the quantum memory and magnetic shielding}
	\label{sec: magnetic shielding}

	In trapped ion QC, the qubit can either be encoded in an optical qubit
	\cite{OldOpticalQubit,PhilippNJP} or a ground state qubit 
	\cite{QubitTypes,MainzCoherenceTime,HyperfineQubit}. In the optical
	qubit, one state of the qubit is a meta-stable D-state of the ion
	whereas the other one is in the ground state.  The qubit transition 
	frequency is in the optical regime and thus it is called optical qubit.
	As the live-time of this qubit is limited by the life-time of the
	meta-stable state, which is typically on the order of 1~s \cite{QubitTypes},
	the coherence time will ultimately be limited by its spontaneous decay.
	Therefore, to achieve a long coherence time and a high $\kappa$ value 
	of the system, the qubit has to be encoded in the ground state of the ion,
	which does not suffer from such decoherence.

	For ground state qubits, the main source of decoherence is magnetic 
	field fluctuations.  Therefore, generating a constant magnetic field
	and magnetic shielding are the most critical challenges to achieve
	long coherence times in trapped ion systems.  Decoherence sources like 
	spin-spin interaction \cite{OzeriSpinSpin} must be suppressed, e.g. by the
	choice of an $\left|F_1,M_F=0\right\rangle$ to $\left|F_2,M_F=0\right\rangle$ 
	transition qubit, or by the choice of a qubit at a 'clock transition',
	for which the energy separation does to first order not depend on the
	magnetic field \cite{QubitTypes}.  Other decoherence sources like 
	leakage of resonant light must be reduced such that they can be neglected 
	in the quantum memory, which is discussed in 
	Section~\ref{sec: quantum gates - trap constraints}.

	Quantum gate operations in trapped ion systems take between 10 and 
	100~$\mu$s \cite{HighFidGateHarty,HFentangling}.  Experimentally,
	coherence times of more than 100~ms have been shown with mu-metal magnetic
	shielding \cite{MainzCoherenceTime} and dressed states \cite{DressedStates}.
	All trapped ion experiments with coherence times of more than 100~s
	\cite{BollingerBeMagnInsensitive,KihwanKimCoherence} were performed
	with hyperfine qubits at a clock transition \cite{QubitTypes}
	and without external magnetic shielding.  
	Hence with appropriate magnetic shielding, one should be able to
	increase the coherence by several orders of magnitude. This results in
	coherence times of hours or days and $\kappa$ values\footnote{As previously
	discussed, one will only be able to use a small fraction of the stated
	coherence time, which is typically a \nicefrac{1}{e} value of a 
	coherence measure. Although the exact $\kappa$ value depends on the
	logical qubit encoding, a stated coherence time of about 10$^5$~s
	with gate times of about 10 to 100~$\mu$s may lead to 
	$\kappa \approx 10^6$.} greater than 10$^6$.

	As the main magnetic field noise in a laboratory environment is from
	alternating current (AC) sources, one way to shield against AC magnetic
	field is using skin-effect in a highly conducting material surrounding
	the experiment \cite{MyRSI}.  Another way is to encapsulate the 
	experiment in a mu-metal shield \cite{BlakestadMuMetal}, which
	provides shielding against AC and DC magnetic fluctuations.
	However, slow magnetic field drifts such as changes in earth's 
	magnetic field \cite{EarthMagneticField} still penetrate a magnetic
	shield made out of a highly conducting material or mu-metal\footnote{A
	typical value of the DC attenuation of magnetic field of a mu-metal
	shield is about 30~dB.}.  Thus, these simple magnetic shielding 
	schemes will not allow the desired coherence times of hours or days.

	A consequence of Meissner effect \cite{MeissnerEffect} is that
	superconductors are perfect diamagnets and thus perfect magnetic shields.  
	Inside a hollow superconductor, the magnetic field is constant and 
	shielded from external magnetic fields.  When placing the ion trap 
	(equivalent to the whole quantum computer) in such an environment,
	the desired coherence times should be feasible with clock transitions in 
	hyperfine qubits.

	In practice, it is not straightforward to define a certain magnetic 
	field strength inside a superconductor, as required for clock transitions
	in hyperfine qubits.  During the phase transition into the superconducting 
	regime, local magnetic flux can get pinned\footnote{This local flux can 
	get pinned to grain boundaries, strains, etc. inside the superconductor 
	which alters the magnetic field inside the superconductor 
	\cite{SuperconductingShielding1970}.
	The measured magnetic field strength inside a superconducting cylinder can 
	reach 100 times the externally applied magnetic field strength.}.
	To avoid this pinning, the suggested solution is to have the superconductor 
	undergo the phase transition in a zero-field environment 
	\cite{SuperconductingShielding1970}.
	For such a cool-down, the experiment has to be located inside a 
	magnetically shielded room (MSR) \cite{MuMetalRoom}.
	Once the entire shield is superconducting, external magnetic field 
	changes will no longer be able to penetrate the shield.  Cables, fibers, etc. 
	to operate the Paul trap will have to enter the superconducting shield 
	through holes to which superconducting tubes should be attached.  
	The shielding of such superconducting cylinders depends 
	exponentially on its length (for a given diameter) \cite{MShieldingCylinder}.
	Hence, long and thin cylinders are desired for high shielding 
	against the environment.

\begin{figure}[!htb]
		\begin{center}
			\includegraphics[width=13.5cm]{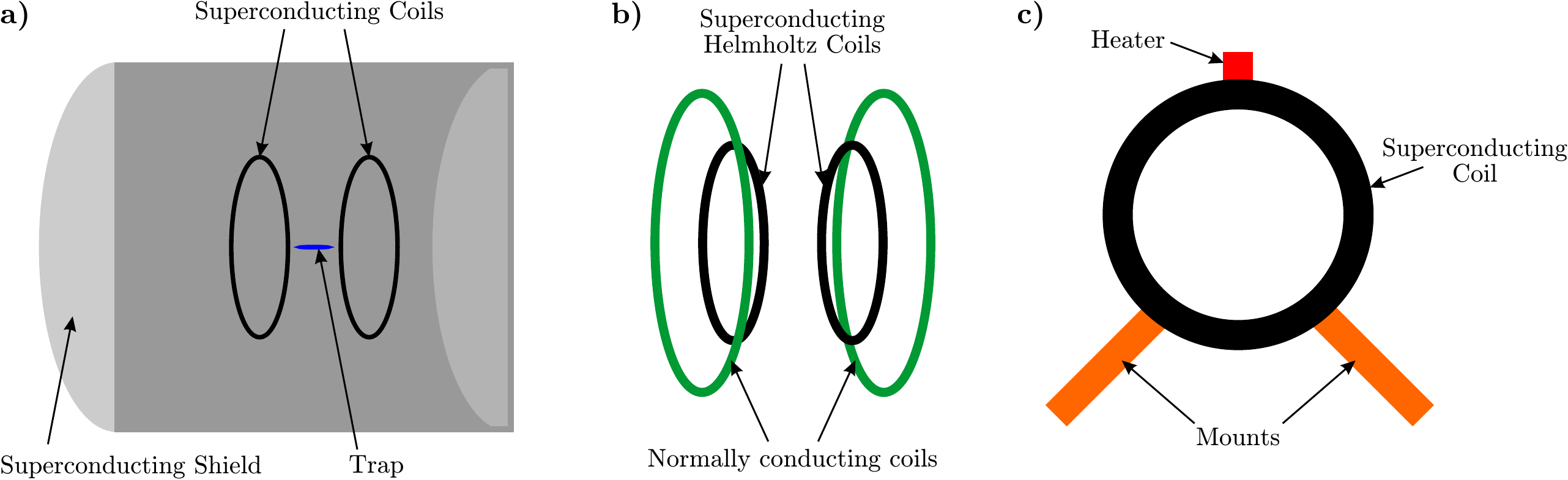}
			\caption{Panel (a) shows a cutaway view of a superconducting shield
				with the coils and the trap.  Panel (b) illustrates a setup
				of two coil pairs required to generate a persistent current in the
				superconducting coils. Panel (c) depicts a superconducting
				coil with a heater.}
			\label{fig: superconducting shield}
		\end{center}
	\end{figure}

	The bias magnetic field at the position of the ions can be generated by 
	superconducting coils inside the magnetic shield, as depicted in
	Fig.~\ref{fig: superconducting shield}~a.  During the cool-down in a 
	zero-field environment, the superconducting coils do not contain persistent 
	current \cite{PersistentCurrent}.  With additional normally conducting coils, 
	one can generate a magnetic field inside in the shield, shown in 
	Fig.~\ref{fig: superconducting shield}~b.  When the superconducting 
	coils are heated locally, as illustrated in Fig.~\ref{fig: superconducting shield}~c, 
	the generated magnetic field can penetrate the superconducting coils.  
	After they are cooled back down into a superconducting regime, the magnetic 
	field produced by the normally conducting coils can be switched off.  
	The resulting persistent current in the superconducting coils will generate an 
	ultra-stable magnetic field inside the superconducting shield.

	The zero-field environments in MSRs with a residual field of less than 
	1.5~nT have been demonstrated \cite{MSR}.  If the pinning of magnetic flux 
	in the superconductor were to increase the residual magnetic field by a 
	factor of 100, the magnetic field strength would be on the order of 100~nT.  
	The magnetic field strength for clock transitions in hyperfine qubits is 
	generated by the superconducting coils inside the superconducting shield
	and is on the order of 10~mT \cite{BeClockQubit,HighFidGateHarty}.
	Hence, the pinned magnetic field in the center of the superconducting coils 
	(at the position of the ion trap) can only produce a relative offset of 
	10$^\mathrm{-5}$ of the total magnetic field, which 
	leaves hyperfine qubits safely in the regime with only quadratic Zeeman shift.

	Such a setup will provide a temporally stable magnetic field to reach
	the desired coherence times of hours or days.  In general, it is not 
	necessary to completely eliminate magnetic gradients in trapped ion QC.  
	If the magnetic field at each storage point is well known, one can calculate 
	the phase evolution of all qubits.  However, techniques like decoherence free 
	subspace (DFS) encoding \cite{DFStheory} require the same magnetic field for multiple ions.  
	Therefore, it is desirable but not necessary to have high homogeneity.
	Spatial homogeneity is discussed in the appendix in 
	Appendix~\ref{sec: magnetic field homogeneity} in more detail.

	\subsection{Local oscillator stability}
	\label{sec: clock ions}

	The transition frequencies of hyperfine transitions are typically on 
	the order of 1-10~GHz \cite{YbClockQubit,BeClockQubit,HighFidGateHarty}.  
	If one wants to achieve coherence times of up to days, a frequency 
	reference with a stability of about 10$^\mathrm{-15}$ will be required.
	In order to achieve the required stability of the reference clock, 
	one can sacrifice some ions of the quantum computer to act as a precise 
	long-term frequency reference.  Although one has to remove some ions 
	from QIP for the clock signal generation, such a scheme allows 
	stabilizing the local oscillator.  It will even enable using the 
	quantum computer for atomic clock measurements.

	\subsection{Multiplexing: ion storage and movement}
	\label{sec: multiplexing ion storage and movement}

	An idling ion string accumulates on the order of 10 quanta/s and thus 
	about a million phonons during a day of uncooled storage.  These high 
	phonon numbers will cause a melting of the ion crystal and the order 
	in the ion string will be lost after a refreeze.  Hence, ion storage 
	times of hours or days require sympathetic cooling with a second ion 
	species \cite{SympatheticCooling,SympatheticCooling2}.

	\begin{figure}[!htb]
		\begin{center}
			\includegraphics[width=14cm]{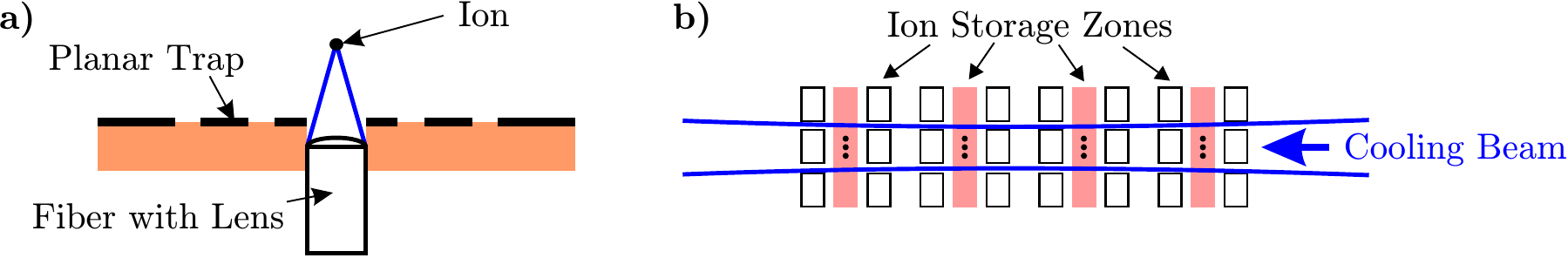}
			\caption{Setups for sympathetic cooling with integrated fiber 
				optics at each ion storage zone (a) or with beams along 
				the surface for multiple ion storage zones (b).}
			\label{fig: sympathetic cooling setup}
		\end{center}
	\end{figure}

\begin{figure}[!htb]
		\begin{center}
			\includegraphics[width=12cm]{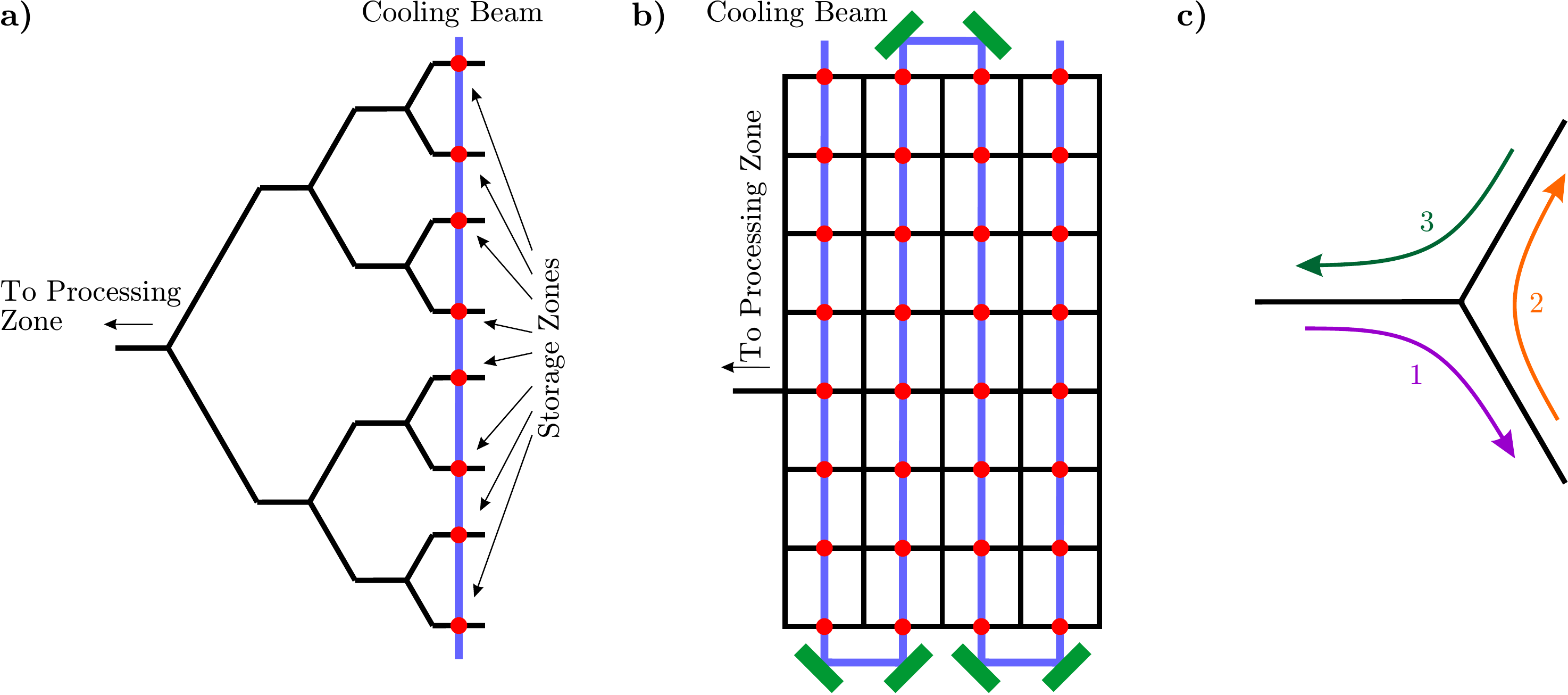}
			\caption{Memory zone either structured as a branching with
				Y-junctions (a) or a grid generated with X-junctions 
				(b). Panel~(c) depicts the series of movements
				in a Y-junction to rotate an ion string. \\
				The RF rails along which ions can be moved are depicted
				by black lines.}
			\label{fig: memory zone}
		\end{center}
	\end{figure}

	Besides working with two ion species, sympathetic cooling implies that 
	one needs cooling beams at each storage position.  The illumination of 
	each storage zone can either be accomplished by integrating fibers into 
	the trap \cite{FiberCouplingDetection} or by illuminating multiple 
	storage zones with a beam parallel to the trap surface, as displayed in 
	Fig.~\ref{fig: sympathetic cooling setup}.  Integrated fiber optics
	facilitate cooling of ion strings.  However, one fiber per storage position
	will complicate the trap design whereas cooling multiple storage zones with
	a single beam will simplify the optical setup.
	These beams along the surface can even be reused by reflecting the light 
	from one line of storage zones to the next line of storage zones, similar 
	to the ideas discussed in reference \cite{KimFiberSwitch} and shown in 
	Fig.~\ref{fig: memory zone}~b.

	One thing that has to be kept in mind when designing a large-scale
	quantum computer in a cryogenic environment is the heat load.
	Large-scale QC will require thousands of storage sites.
	If light is coming from a fiber at every storage site, it will
	be hard to couple the light back into fibers to avoid heating the
	cryostat due to the light absorption.  Whereas, light parallel
	to the surface cools multiple sites and is easier to couple back
	into a fiber.

	\begin{figure}[!htb]
		\begin{center}
			\includegraphics[width=12cm]{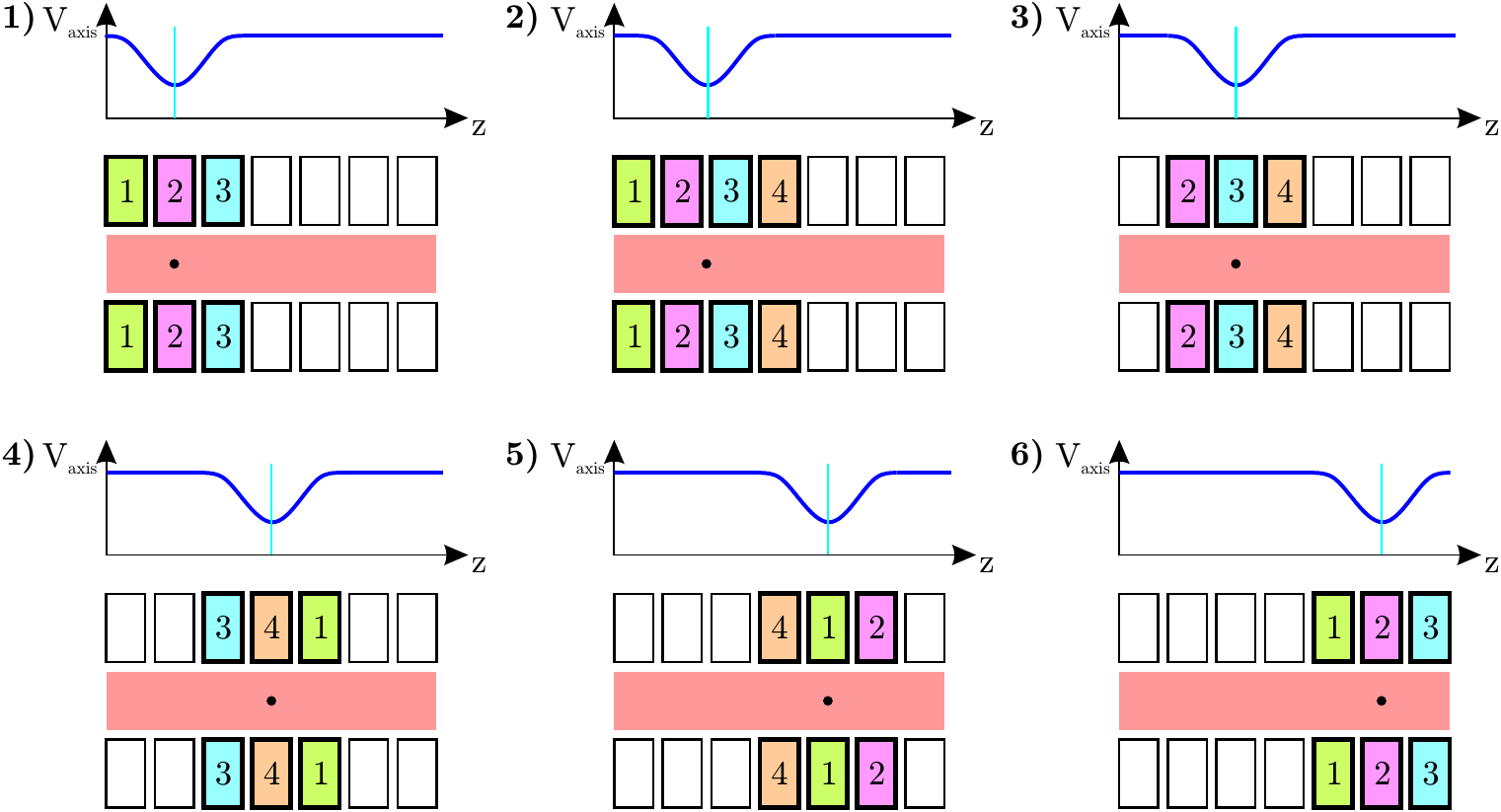}
			\caption{Moving an ion (string) from left to right. Instead of 
				controlling all segments individually, it is sufficient to control 
				at maximum four segment pairs. \\
				A number on a segment corresponds to the DAC pair number that 
				controls the voltages on this segment pair.}
			\label{fig: segment mux}
		\end{center}
	\end{figure}

	A trap suitable for QIP with thousands of ions will require the 
	control over thousands of segments and thus over thousands of 
	voltages with digital-to-analog converter (DAC) channels.  
	In order to reduce this hardware demand, one can use analog 
	multiplexers.  By employing such analog switches, one DAC channel 
	can control multiple segments.  An example of how this can be 
	incorporated in ion movement is shown in Fig.~\ref{fig: segment mux}.  
	At first, the ion is stored on the left side by controlling three 
	segment pairs.  During the shuttling, one has to control at maximum 
	four segment pairs.  When moving the ion right, the control of an 
	unused segment pair on the left can be exchanged to control over 
	the next segment pair on the right\footnote{For better illustration,
	please refer to IonMovement.mp4 in the supplementary data.}.  
	Hence, with this multiplexing 
	scheme, it is possible to move ions in an arbitrarily big segmented 
	trap with DC control over only four segment pairs and digital 
	multiplexing logic.  Furthermore, it is possible to adapt the 
	voltage ramps for each segment individually which enables the 
	compensation of stray fields on all parts of the trap.

	\begin{figure}[!htb]
		\begin{center}
			\includegraphics[width=13.5cm]{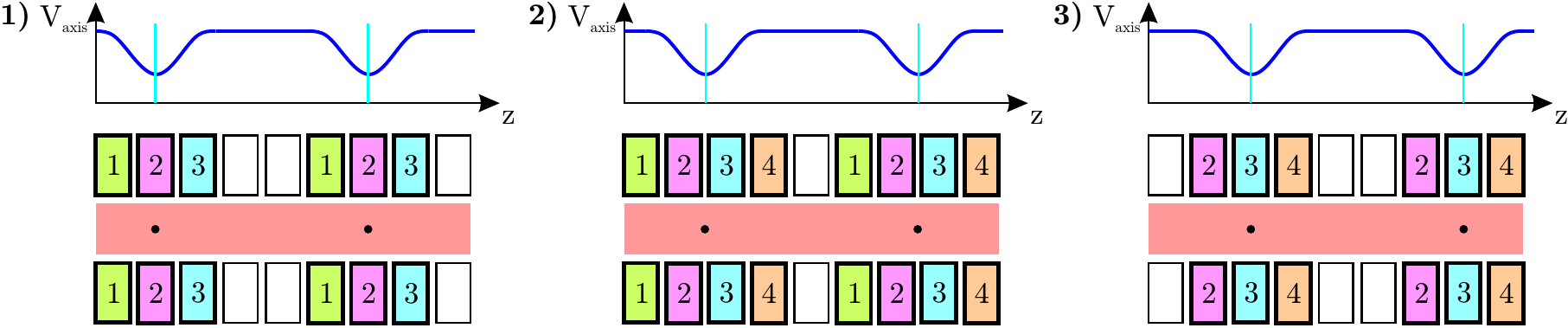}
			\caption{Moving multiple ion strings from left to right by 
				controlling only four DAC pairs.}
			\label{fig: segment mux multiple}
		\end{center}
	\end{figure}

	The digital multiplexing logic circuits have to contain at least 
	as many digital outputs as there are segment pairs on the trap.  
	As traps for large scale QC will contain thousands of segment 
	pairs, this will require thousands of interconnects, and it is 
	advisable to place both the analog switches and demultiplexer 
	circuits close to the trap or even integrate it into the trap chip.

	If the digital multiplexing logic allows the control of multiple 
	segments with just one DAC, one can generate multiple confining 
	potentials on the trap with the same DACs.  As illustrated in 
	Fig.~\ref{fig: segment mux multiple}, these multiple confining 
	potentials can be moved on the trap the same way as a single one, 
	allowing the transfer of multiple ion strings simultaneously
	with the same DAC channels\footnote{For better illustration,
	please refer to DualIonMovement.mp4 in the supplementary data.}.
	As the same confining potential is used for multiple ion strings,
	stray fields on individual ion strings cannot be compensated
	separately. Hence, this scheme can only be used in regions where
	micromotion does not influence the operation of the quantum computer.

	\begin{figure}[!htb]
		\begin{center}
			\includegraphics[width=8cm]{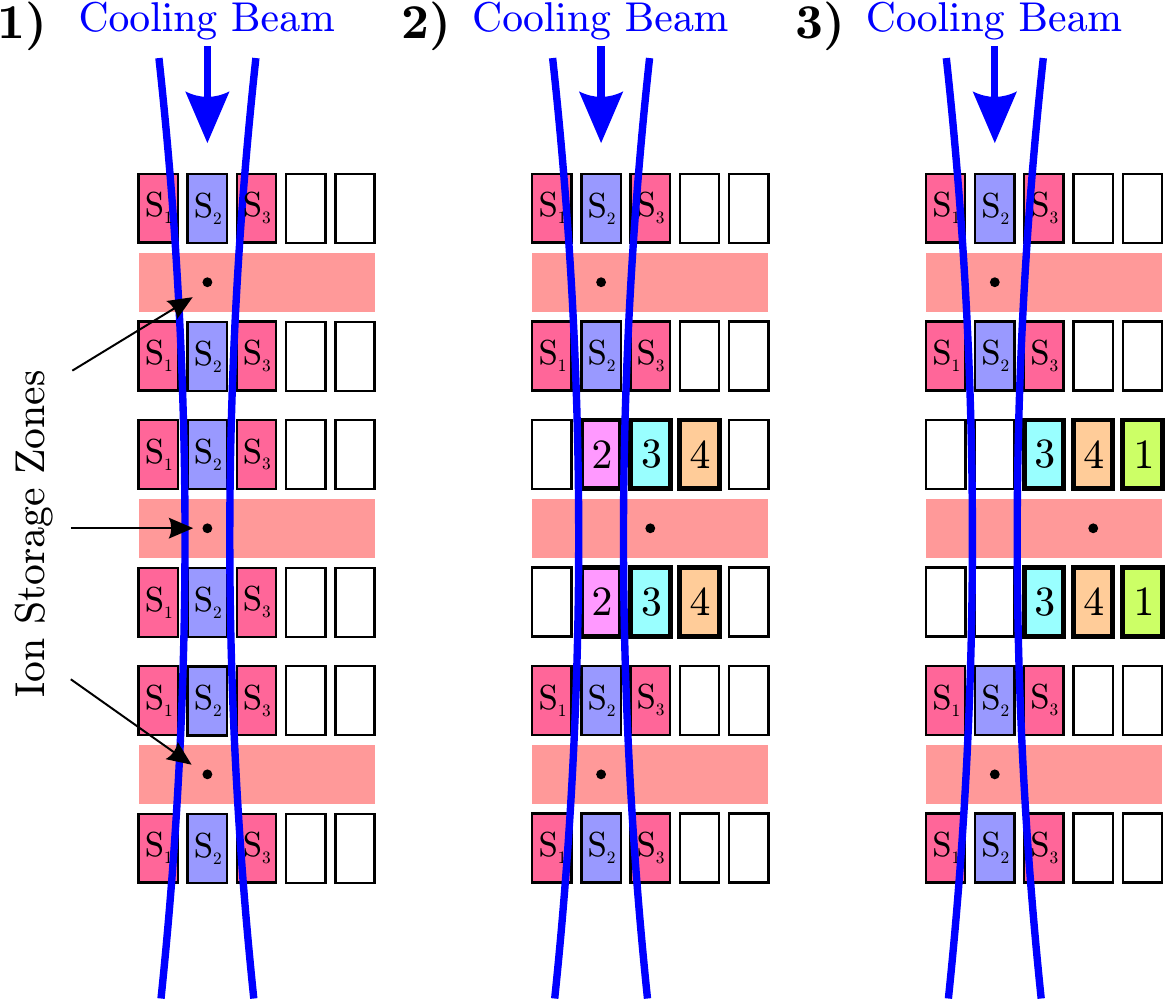}
			\caption{Multiplexing in the memory region.  All ions can be 
				confined by controlling the voltages of three segment pairs 
				(S$_\mathrm{1}$, S$_\mathrm{2}$, S$_\mathrm{3}$).  
				To access an ion string, analog multiplexers connect the 
				segments to the individually controlled DACs (1-4).}
			\label{fig: memory mux}
		\end{center}
	\end{figure}

	Following the ideas of the hardware requirements for storage 
	in large-scale Quantum von Neumann setups in Section~\ref{sec: quantum memory}, 
	the ions in all storage zones can be confined 
	by a small set of static voltages\footnote{This implies that 
	stray fields on the trap have to be small enough to allow 
	efficient sympathetic cooling in the storage zones.}, as shown
	in Fig.~\ref{fig: memory mux}~(1).  To access a quantum memory cell, 
	a digital signal from the multiplexing logic switches the voltage 
	from the set of static voltages to a set of voltages controlled 
	by DACs.  With the control over the confining voltages of the 
	single memory cell, one can move the stored quantum information 
	from the quantum memory to another region of the trap for further
	processing\footnote{For better illustration, please refer to 
	MemoryMultiplexing.mp4 in the supplementary data.}, 
	as depicted in Fig.~\ref{fig: memory mux}.
	Should splitting of the qubits and the cooling ions be required, 
	it can either be performed in the quantum memory region or in 
	the QALU before QIP is performed.

	The movement through X- and Y-junctions 
	\cite{ShuttlingYJunction,ShuttlingXJunction} may require the control 
	over more than four segment pairs to block the ions from entering 
	a wrong arm in the junction.  But other than possibly having to 
	control more voltages, there is no reason why this multiplexing 
	architecture cannot be used to move ions through junctions.
	This allows structuring the memory region by branching with 
	Y-junctions or by generating a grid with X-junctions, as depicted 
	in Fig.~\ref{fig: memory zone}~a and b.

	If one wants to use a qubit encoding scheme that is sensitive to
	magnetic field gradients, like DFS encoding, even tiny magnetic field 
	gradients along the trap axis will cause different phase evolution 
	in the different ions over long storage times of hours or days.
	If the gradient is linear, one could rotate the ion string in the 
	middle of the storage time or repeatedly after a certain time interval to 
	cancel the effect of the magnetic field gradient.  Rotation of 
	an ion string is easiest in a junction by moving the ions from 
	arm~1 to into arm~2, from there into arm~3 and then back into 
	arm~1, as illustrated in Fig.~\ref{fig: memory zone}~c. 

	\subsection{Quantum gates}
	\label{sec: quantum gates}

	\subsubsection{Heating}
	\label{sec: quantum gates - heating}

	A major problem with entangling gates which use Coulomb interaction 
	\cite{Cirac-Zoller,MS-Gate}, thus phonons in an ion crystal, is 
	motional heating \cite{Heating-Rates}.  A lot of effort has been 
	made to characterize heating \cite{BrownnuttHR}, especially its dependence
	on the distance of the ion to the surface of the trap, and it has been 
	shown that the heating rate is reduced in cryogenic environments
	\cite{CryoHeatingRates,CryoHeatingRates2}.  Experimentally, heating
	rates as low as 0.33~ph/s have been observed in surface traps 
	\cite{MichiSilicon}.

	Due to sympathetic cooling in the memory region and short transport 
	times between the quantum memory region and the QALU of less than
	about 1~ms, heating only affects QIP in the QALU.
	The quantum computer based on this quantum von Neumann architecture 
	for trapped ions has to be operated in a cryogenic environment and, 
	thus, heating rate should be low enough to allow for fault-tolerant QC. 

	\subsubsection{RF or optical drive fields}
	\label{sec: quantum gates - optical vs RF}

	RF fields enable qubit operations with the lowest infidelity in 
	trapped ion systems to date \cite{HighFidGateHarty}.
	Entangling operations via Coulomb interaction require high
	field gradients due to the low Lamb-Dicke parameter of RF fields.
	These high field gradients are typically generated with high
	RF amplitudes. If the QALU is surrounded by memory zones, one
	must protect the qubits in the quantum memory from the resonant
	and near-resonant RF fields. There is research on minimizing RF
	surrounding the processing zones of traps.  However, it is unclear how
	well this RF field suppression would work for a large-scale
	quantum computer with tens of thousands of qubits or more
	surrounding the QALU.  Experimentally, one has to stabilize
	the phase of the RF in QALU for high fidelity operation such 
	that the length between the RF source and the ion does not 
	fluctuate on a (tens of) micrometer scale.
	
	On the other hand, high fidelity quantum operations can be
	performed with optical drive fields as well \cite{HFentangling,MonroeProgQC}.
	There, the demonstrated infidelity is about one order of magnitude
	worse than with RF fields.  However, unwanted 
	fields can be avoided by inhibiting direct line of sight between 
	the quantum memory and surfaces of the QALU that scatter light, 
	see the Section~\ref{sec: quantum gates - trap constraints} for details.  
	Experimentally, the most challenging part is amplitude and phase 
	control of the light field at the position of the ion.
	Given that optical frequencies are much higher than the RF frequencies,
	one has to stabilize the phase of the light with sub-nanometer 
	precision. Suggestions on the phase stabilization is given
	in Appendix~\ref{sec: optical setup}. Furthermore,
	in order to avoid long distances between the quantum memory region 
	and the QALU, the processing zone will be in the center of the trap.  
	Single ion addressing with laser beams will require a numerical
	aperture (NA) of 0.2 or higher for ion-to-ion distances of about 
	5~$\mu$m.  Therefore, the trap needs to be slotted in the region
	of the QALU to allow high NA addressing perpendicular to the trap 
	surface. 

	\subsubsection{Physical requirements for the gate operations}
	\label{sec: quantum gates - physical requirements}

	So far, this architecture requires (at least) two ion chains
	to be loaded into the QALU for QIP, where gate operations are performed.
	The type of gates \cite{MS-Gate,EntanglingPhasegate,Wunderlich_RF,HighFidGateHarty} 
	that are executed is defined in the firmware layer of the architecture.
	For a full set of quantum operations, single ion addressing
	capability is required.

	As the length of path between the drive field's source
	and the ions should not fluctuate for a stable phase reference,
	vibration isolation of the superconducting magnetic shield
	will be required, e.g. by suspending the shield with ropes from
	the vacuum chamber.  Please, refer to Appendix~\ref{sec: cryogenic system}
	for more details.  

	With RF gates, the trap can be used as part of the transmission line
	which simplifies the setup.  For optical gates, light can be 
	guided via fibers into the magnetic shield and optical alignment 
	in the shield will enable enough optical access to perform the 
	required gate operations.  Furthermore, vibration isolation will 
	reduce beam pointing instabilities and thus undesired varying 
	optical crosstalk between the ions.
	The tight focusing, required for single ion addressing, results 
	in a high local light intensity at the position of the ion.  
	In reference \cite{SpontaneousPhotonScattering}, the authors 
	state that between 1 and 10~mW optical power is required for 
	single qubit gates with a gate infidelity of 10$^\mathrm{-4}$ 
	employing Raman transitions.  Moreover, between 100~mW and 1~W 
	optical power is required for entangling gates\footnote{Their 
	experimental parameters are: the duration for a single qubit 
	$\pi$ rotation is 1~$\mu$s; the gate time of the entangling gate
	is 10~$\mu$s and the gates are performed on a radial mode with 
	$\omega_\mathrm{rad}~=~2\pi\cdot5$~MHz.} using a Gaussian beam 
	with w$_0$~=~20~$\mathrm{\mu}$m.
	If all gates in a quantum von Neumann setup are performed with 
	highly focused Gaussian beams with w$_0$~$\approx$~1~$\mathrm{\mu}$m, 
	the required total optical power will drop by a factor of 400 
	compared to their stated values.  This lower optical power reduces 
	problems like bleaching of fibers, which is worse at higher powers.
	
	The crosstalk onto neighboring ions is a coherent process and 
	thus can be eliminated by calibration and composite pulses 
	\cite{PhilippNJP}.  If the crosstalk on all ions is known, one 
	can construct a pulse sequence that performs all single qubit 
	operations required by the quantum algorithm and at the same time 
	corrects for the crosstalk \cite{Estebanizer}.
	Such calibration requires precise control over the amplitude of 
	the driving field at the position of the ion. 
	For operation with RF gates, this implies a clever segment
	structure of the trap. For operation with optical gates, beam pointing
	instabilities and imperfections in amplitude and timing control 
	must be negligibly small.  Thermal drifts might still cause spatial 
	drifts on time scales of seconds or minutes.  Therefore, it might 
	be necessary to regularly place "calibration ions" in 
	the QALU to track drifts of the crosstalk.

	In order to protect idling qubits, it is possible to shelve 
	populations from the clock state to other states in the Zeeman 
	manifold \cite{PhilippNJP} in which the QIP is performed.  With this 
	scheme, gate operations are not resonant with the clock transition 
	in which quantum information is stored in the quantum memory. 
	However, the imperfect shelving operations introduce leakage from 
	the qubit states which needs to be considered in the employed QEC.

	\subsubsection{Pipelining}
	\label{sec: quantum gates - pipelining}

	Since ions which arrive in the QALU from the quantum memory are 
	only Doppler cooled, they have to be groundstate-cooled for high 
	fidelity QIP.  If cooling and QIP are executed in the same 
	processing zone of the QALU, the processing cycle will be slowed 
	down by the required initial cooling.  Following the pipelining 
	approach from classical computer science, one can use separate
	regions in the processing zone for the individual tasks required for
	efficient QIP.  These tasks could be:
	\begin{enumerate}
		\item Combining two (or more) ion strings to a single string
		\item Sympathetic Doppler cooling
		\item Sympathetic ground state cooling: e.g. by using electromagnetically-induced-transparency (EIT) cooling
		\item Qubit decoding, e.g. map from a DFS basis back to a single ion basis
		\item Map ions for QIP from clock states, or dressed states, to processing states
		\item Perform QIP
		\item Map ions back from processing state to clock states, or dressed states
		\item Qubit encoding, e.g. map the single ion basis back into DFS basis
		\item Split the ion string into multiple parts for storage
	\end{enumerate}

	\begin{figure}[!htb]
		\begin{center}
			\includegraphics[width=13cm]{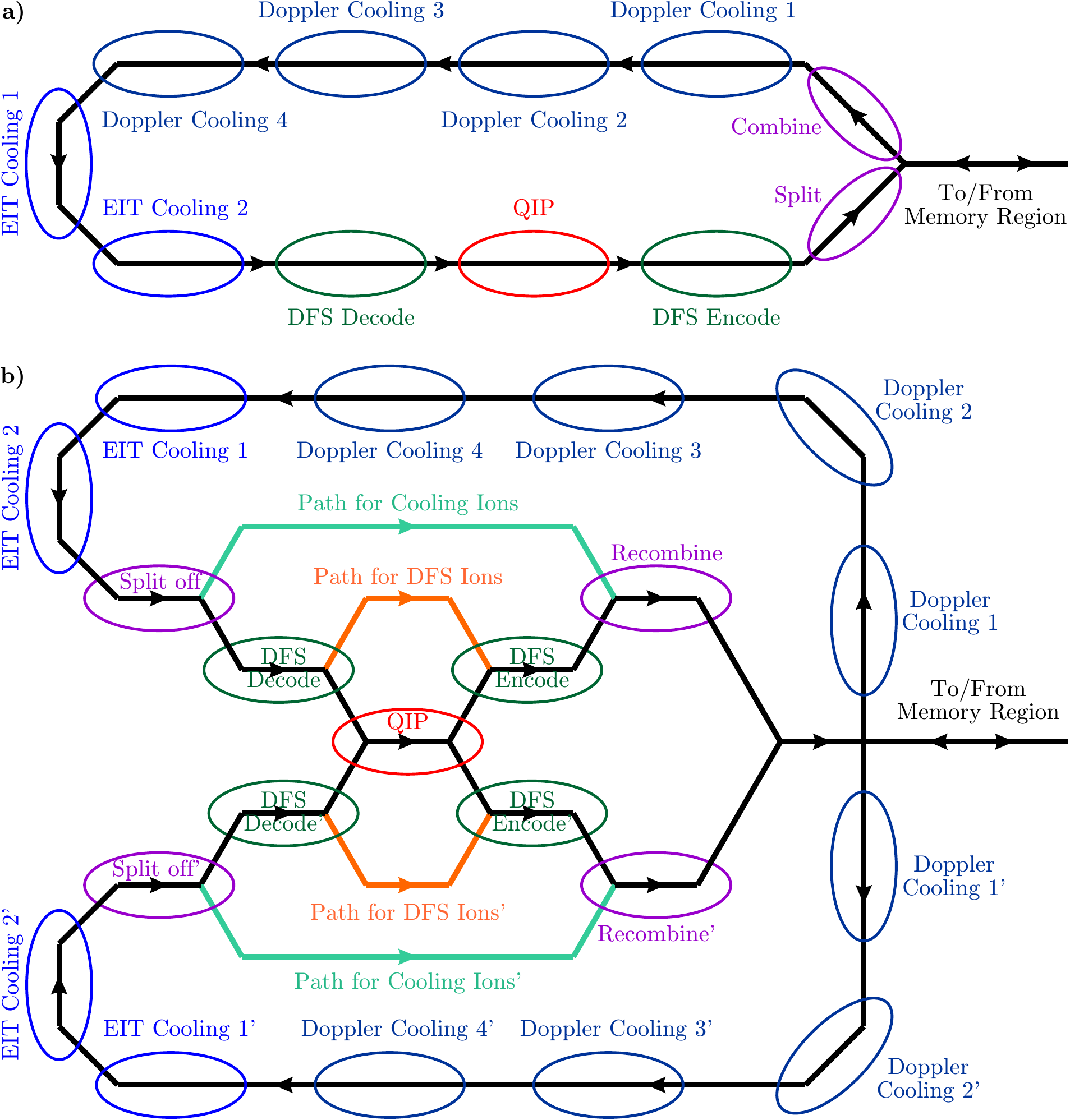}
			\caption{Pipelining in the processing zone.  Each task 
				required in the QALU gets its own region, and the ions 
				are shuttled from region to region. Panel~(a)
				shows a version in which two ion strings are combined.
				In this case, cooling and QIP are performed on the long
				combined string. Panel~(b) depicts an advanced
				pipelining architecture which requires efficient
				splitting operations of ground state cooled ion strings,
				as the two ion strings from the memory are cooled separately.
				After ground state cooling, they are combined and QIP 
				can be performed.}
			\label{fig: pipelining processing zone}
		\end{center}
	\end{figure}

	Fig.~\ref{fig: pipelining processing zone} depicts such
	a pipelining approach which enables the execution of multiple 
	tasks on multiple ion strings simultaneously.  The thick
	black lines illustrate the RF rails along which ion strings
	can be moved. Fig.~\ref{fig: pipelining processing zone}~a
	shows a QALU architecture for which two (or more) ion strings, which 
	shall interact during QIP, are loaded from the quantum memory
	and combined to a single ion string.
	As Doppler cooling typically lasts milliseconds, whereas QIP 
	is performed in tens of microseconds, the ion string passes through 
	multiple stages of sympathetic Doppler cooling to ensure that 
	the ions are at the Doppler limit before further processing 
	is performed.  After Doppler cooling, the ion string is 
	ground state cooled with sympathetic EIT cooling.
	In the next step, the quantum information is decoded for
	example by transferring from the DFS encoding to the bare physical qubit.
	After qubit decoding, QIP is performed on the ion string.	 This enables interaction 
	between arbitrary qubits of the quantum memory.  After QIP, 
	the ion string is encoded, e.g. with DFS encoding. 
	At last, the long ion string is split into multiple ion strings
	which can then be sent back to the quantum memory.

	Another QALU architecture is depicted in 
	Fig.~\ref{fig: pipelining processing zone}~b.  It has the
	same cooling and QIP procedure as the previous one.  However,
	the different ion strings loaded from the quantum memory 
	are not combined in the first pipeline step but cooled individually.
	After ground state cooling, the cooling ions can be separated
	from the qubit ions. This simplifies the mode structure of the
	ion crystal but requires efficient ion splitting of ground state
	cooled ion strings.  To simplify the mode structure even further,
	the qubit ions used only for DFS encoding are separated from the 
	ones containing the quantum information after
	DFS decoding. In the QIP region, the two ion strings are combined
	and QIP can be performed. For DFS encoding, the ions that were
	split off can be reused. At last, the qubit ions are recombined
	with the cooling ions before ion strings can be sent back to
	the quantum memory.

	Qubit encoding/decoding and QIP require single ion addressing.  
	With optical gates, if there is not enough optical access to 
	perform single ion addressing at multiple locations, these tasks 
	may have to be performed at different positions on the trap.

	Having different regions for the different parts required 
	for QIP is not yet pipelining. In the pipelining process, an ion 
	string is moved from one processing region to the next, while
	the next ion string is moved into the previous processing 
	region\footnote{For better illustration, please refer to 
	Pipeline.mp4 and DualPipeline.mp4 in the supplementary 
	data.}. Thus, the number of 
	processing regions defines the depth of the pipeline. 
	The parameters of the cooling and processing time have to 
	be chosen such that they can be synchronized.  The time 
	of one execution cycle defines the speed of the processing.  
	The distance between the different processing regions 
	should be short so that ion movement does not increase 
	the execution time of one pipeline step considerably.

	In the processing regions, micromotion \cite{NISTMicromotion} 
	has to be compensated for effective cooling and QIP.  This 
	will require many independently controlled voltages in the 
	processing zone.  However, in the shuttling regions between 
	the processing regions, micromotion is not crucial and one 
	can use multiplexing, as shown in Fig.~\ref{fig: segment mux multiple}, 
	to reduce the number of DC voltages which need to be controlled
	in the QALU.

	In this general pipelining approach, there are no restrictions
	on the ion strings processed in the QALU.  In order to keep
	the vibrational mode structure of the ion strings simple, one
	has to limit the length of the ion strings.  By choosing
	the ion string loaded from the quantum memory such that
	the qubit ions are surrounded by the cooling ions, one
	can detect ion loss during Doppler cooling.  For this, 
	one uses a camera to detect the number and positions of the 
	cooling ions.  From the spacing between the cooling ions, 
	the number of processing ions can be inferred.  Ion loss can
	be compensated by adding ions to the ion string either in the 
	QALU or in a special zone outside the QALU. 

	\subsubsection{Trap constraints}
	\label{sec: quantum gates - trap constraints}

	In order to minimize axial micromotion (which cannot be 
	compensated), it is imperative to design the trap in the 
	processing zone as symmetric as possible, e.g. as 
	illustrated in Fig.~\ref{fig: pipelining processing zone}~a and b.

	In both the quantum memory and the QALU, the tracks along 
	ions can be shuttled will form loops.  Hence, inter-layer 
	connectivity (vias) will be required for the fabrication of 
	such trap structures.
	Modern traps with vias\footnote{e.g. Sandia's high optical 
	access (HOA) 2 trap (Sandia National Laboratories, New Mexico, 
	P.O. Box 5800, Albuquerque, NM 87185, USA)} typically route 
	the signal lines underneath the trap surface to the segment.
	These traps use vias to connect the actual segments with the 
	routing tracks.  This enables placing ground planes at areas 
	on the trap surface which are not used for electrodes.  
	These ground planes shield against electric fields from the 
	lower lying routing layer, thereby, reduce the cross-talk 
	between segments.

	\begin{figure}[!htb]
		\begin{center}
			\includegraphics[width=11cm]{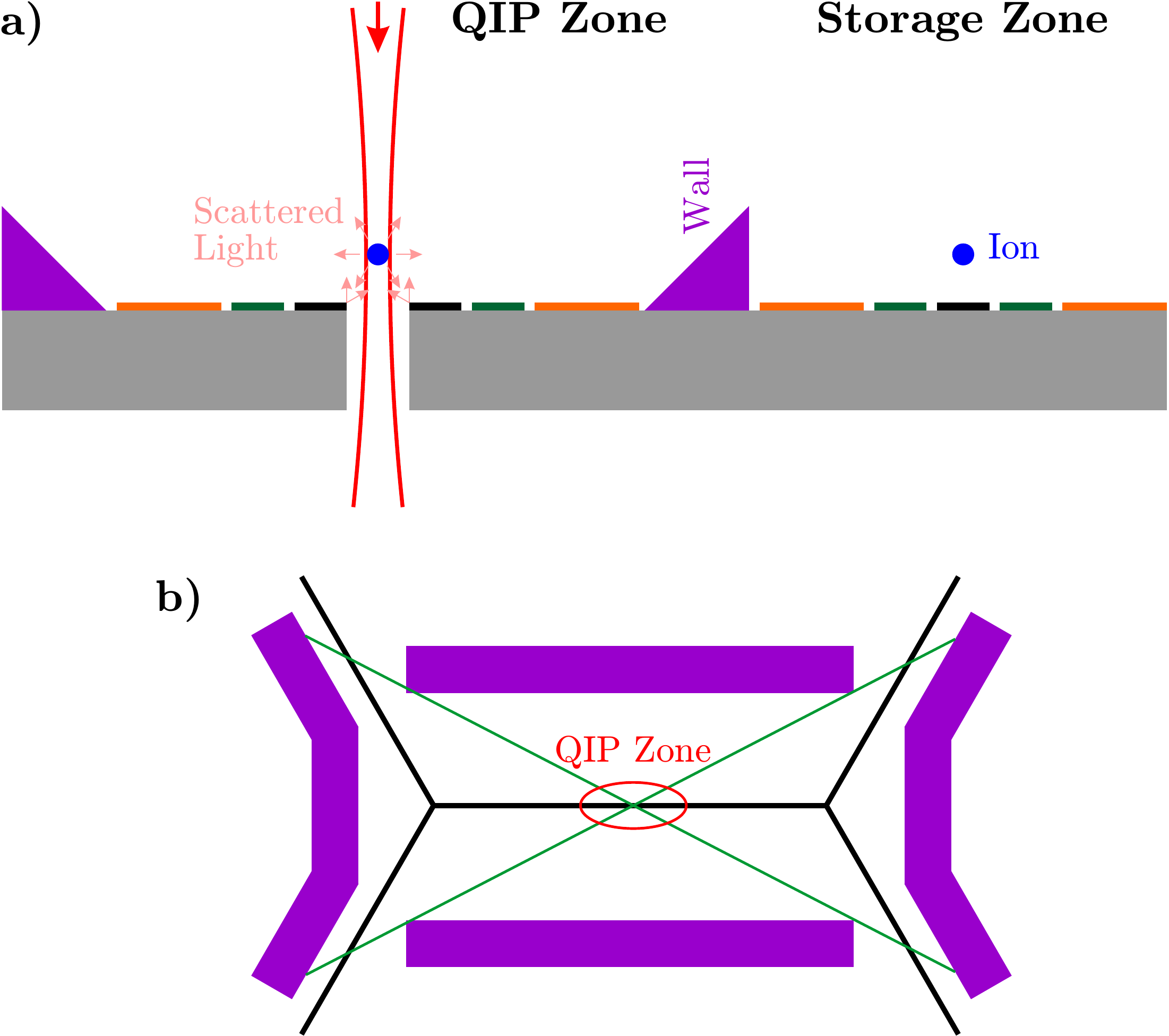}
			\caption{Block line of sight between the quantum memory and the
				QALU by placing walls on the trap which are higher 
				than the ion height above the trap surface. Panel (a)
				shows a cross-section of a trap with a memory zone next to 
				a QIP-zone separated by a wall. Panel (b) depicts a top view 
				of a trap on which the QIP-zone is completely surrounded by
				walls.}
			\label{fig: wall on trap}
		\end{center}
	\end{figure}

	For the operation with optical gates, stray light that is 
	(near-)resonant to a qubit transition 
	is a serious problem for the long coherence times required
	for a quantum von Neumann architecture. A main source of 
	stray light is light scatter at the slot, required for QIP 
	with high NA, in the QALU, and it can be minimized by 
	blocking direct line of sight between the quantum memory 
	and the QALU, for example with walls on the segmented traps, 
	as depicted in Fig.~\ref{fig: wall on trap}.
	The height of the walls should be higher than the distance 
	from an ion to the surface of the trap.  
	These walls should not be perpendicular to the trap surface 
	but under an angle so that reflections on the wall's surface 
	reflect the light away from the trap surface.  If the walls 
	are made of a conducting material, they can be grounded and 
	will have little impact on the trapping potentials.
	For the operation with RF gates, blocking of stray fields 
	is not possible. It can only be minimized by clever segment
	structures.

	\subsection{Detection and initialization}
	\label{sec: detection and init}

	For QIP with optical gates on ground state qubits, Raman 
	transitions are incorporated to couple the quantum states 
	\cite{SpontaneousPhotonScattering}.  These transitions are 
	off-resonant with a typical detuning in the GHz or low THz regime.
	Because of this large detuning, a single photon is very unlikely 
	to affect a qubit.  Thus, it is safe it assume that reflections 
	somewhere in the vacuum chamber can be neglected and it is 
	sufficient to place a wall around the QIP zone in the QALU
	to shield the quantum memory from stray light.
	However, detection requires resonant light which causes 
	fluorescence which is resonant as well.  Furthermore, 
	initialization produces resonant fluorescence. In the case of 
	resonant photons, even a single photon can affect the information 
	in the qubits, and one should try to avoid photons resonant with 
	a state used for storing quantum information.

\begin{figure}[!htb]
		\begin{center}
			\includegraphics[width=12cm]{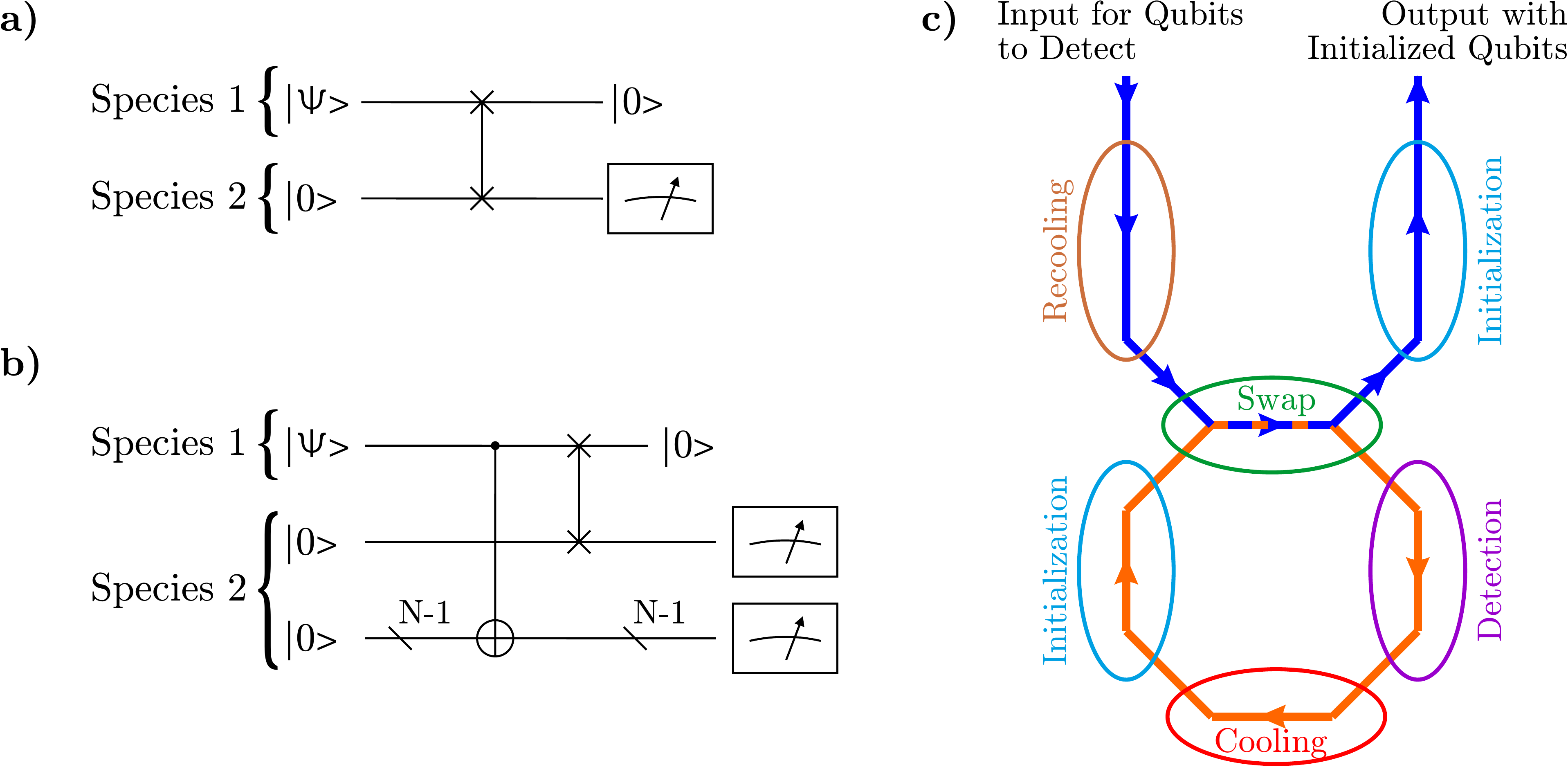}
			\caption{Circuit representation of a swap gate between two 
				ion species for detection and initialization (a) 
				and with an additional GHZ state generation for efficient 
				detection (b).  Panel (c) depicts a possible
				pipelining architecture in the detection zone.}
			\label{fig: detection pipeline}
		\end{center}
	\end{figure}

	This problem can be circumvented by using a second ion species 
	for detection.  For this, the state of the ion to be detected 
	has to be transferred onto the detection ion of a different species.  
	Entanglement between two ions of different species has been 
	demonstrated \cite{MultiSpeciesGates}.  The swapping operation, 
	illustrated in Fig.~\ref{fig: detection pipeline}~a in the 
	circuit model representation \cite{MikeIke}, requires only 
	near-resonant driving fields but no resonant fields.
	For detection, it transfers the quantum information to another 
	ion species, while initializing the main qubit for further processing. 
	Therefore, both state detection and initialization can be performed 
	with ions of the second ion species and one does not have to worry 
	about stray light resonant to the qubits in the quantum memory.
	This implies that initialization fidelity will depend on the 
	fidelity of entangling operations.  In order to increase the 
	initialization fidelity for qubits which have a quadrupole transition, 
	one can reinitialize the qubits additionally by optical pumping 
	via the quadrupole transition.

	High fidelity state detection needs to be performed fast which 
	requires high photon collection efficiency.  For example, in 
	Ca$^\mathrm{+}$, the qubit information is usually stored in the 
	ground state of the ion and, thus, an electron shelving pulse is 
	required to transfer the population of one qubit state into the 
	D$_\mathrm{5/2}$ state.  The D$_\mathrm{5/2}$ state has a limited
	lifetime and thus spontaneous emission causes errors in the 
	detection.  The life-time of the D$_\mathrm{5/2}$ state in 
	Ca$^\mathrm{+}$ is about 1~s \cite{Ca40-spec} which means that 
	the detection has to be performed in 10~$\mathrm{\mu}$s\footnote{In 
	this case, detection with the maximum likelihood method is not 
	considered \cite{HighFidelityReadout}.} to achieve it with a
	detection infidelity of 10$^\mathrm{-5}$.
	With a scatter rate of about 10~MHz in Ca$^\mathrm{+}$, an ion 
	emits about 100 photons during 10~$\mathrm{\mu}$s.  If one 
	requires 5 clicks on the detector for reliable detection, one has 
	to collect about 10~\% of the photons with a typical detector 
	efficiency of about 50~\%.  A collection efficiency of 10~\% 
	requires the detection optics to have NA~$>$~0.6.
	QEC requires ancilla qubits which have to be detected to extract 
	information on the occurred errors, and thus the detection region 
	has to be close to the QALU on the trap.  This might limit
	the optical access to the QALU.

	To increase the number of scattered photons in a certain period 
	of time, one can use the fluorescence of multiple ions by employing 
	Greenberger-Horne-Zeilinger (GHZ) states \cite{GHZ}, as 
	demonstrated in reference \cite{GHZdetection}.  The circuit 
	representation of this detection scheme can be seen in 
	Fig.~\ref{fig: detection pipeline}~b.  The input state 
	$\left|\psi\right\rangle=\alpha\left|0\right\rangle+\beta\left|1\right\rangle$
	is transferred onto $N$ ancilla qubits of a second ion species 
	to generate the GHZ state
	$\alpha~\left|00\cdots0\right\rangle~+~\beta~\left|11\cdots1\right\rangle$. %which is then detected.  
	With $N$ ancilla qubits, the count rate increases by a factor 
	of $N$ compared to the detection with just one ancilla qubit.  
	For detection in the same time interval, the collection 
	efficiency can be lower by a factor of $N$ compared to the 
	case with one ancilla qubit.  
	It is also possible to increase the detection fidelity for 
	longer detection times by performing a majority vote.
	As an example, if one chooses a detection time of 100~$\mathrm{\mu}$s
	for detection of Ca$^\mathrm{+}$, this will result in an 
	infidelity of $\approx$10$^\mathrm{-4}$ due to spontaneous 
	decay.  If one chooses 5 ancilla qubits and one can detect 
	how many ions are bright, 3 qubits will need to decay from 
	the D-state to the S-state for a wrong state detection.
	The probability for this to happen is 10$^\mathrm{-12}$.  
	Hence, the overall detection process will more likely be 
	limited by how efficiently one can generate the GHZ state 
	than by detection itself.

	Pipelining can also be incorporated in the detection/initialization 
	zone, as illustrated Fig.~\ref{fig: detection pipeline}~c.
	The incoming qubits are cooled to the ground state of motion
	before they are moved to the swapping zone. There, the
	CNOT-gates for GHZ state generation and the swap gate are 
	performed with initialized ancilla qubits of a second 
	ion species.  After the swap operation, the initialized 
	qubits are moved to another initialization zone where
	one can compensate the initialization error due to
	imperfect gates or leakage into other states during QIP.
	After the initialization, the ions are shuttled back to 
	other parts of the trapped ion quantum computer.  
	During the compensation of the initialization error, the 
	ancilla qubits of the second ion species are moved to a 
	detection zone where the (GHZ) state is detected.  
	After detection, the ions are 
	cooled and initialized in separate zones before they
	can be reused in the swap zone.  Experimentally, the challenge 
	will lie in protecting the quantum information between the 
	swap and the detection zone from (resonant) stray light of 
	the cooling and initialization zones.

	\subsection{Choice of ion species}
	\label{sec: 3 species TIQC}

	At first, one has to decide how many species one needs for this 
	architecture. One ion species is required for the qubit ions.  
	Another ion species is required for sympathetic cooling.  The 
	ions for detection can either be from the same ion species as 
	the ions for sympathetic cooling, or one can use a third ion species.  
	If one only uses two ion species, cooling in the memory region 
	and detection have to be pulsed so that resonant stray fields from 
	the memory region does not affect the quantum state in the detection 
	zone.  Whereas, using three ion species makes the detection zone
	independent from the memory region.  In the following, the three ion 
	species architecture will be discussed.

	\begin{table}[!htb]
		\caption{List of constants of ion species for QIP - part 1 \cite{SpontaneousPhotonScattering}.
			I is the nuclear spin,  {$\mathrm{\omega_0}$} is the zero field hyperfine splitting, 
			{$\mathrm{\lambda_{1/2}}$} and {$\mathrm{\lambda_{3/2}}$} are the wavelengths of the
			S$_\mathrm{1/2}$ to P$_\mathrm{1/2}$ and S$_\mathrm{1/2}$ to P$_\mathrm{3/2}$ transitions, and
			{$\mathrm{\Gamma_{1/2}}$} is the linewidth of the S$_\mathrm{1/2}$ to P$_\mathrm{1/2}$ transition.
			$^\mathrm{40}$Ca$^\mathrm{+}$, $^\mathrm{88}$Sr$^\mathrm{+}$, and $^\mathrm{138}$Ba$^\mathrm{+}$ are not
			in the list as their spin is 0, they do not have hyperfine splitting, and the transitions are only shifted
			by a frequency in the GHz regime compared to their respective isotopes in this list.}
		\label{tab: atomic constants}
		\centering
		\begin{tabular}{| c | c | c | c | c | c |}
			\hline
			\textbf{Ion} & \textbf{I} & {$\mathrm{\omega_0}$} \textbf{(GHz)} & {\boldmath$\mathrm{\lambda_{1/2}}$} 
				\textbf{(nm)} & {\boldmath$\mathrm{\Gamma_{1/2}}$} \textbf{(MHz)} & {\boldmath$\mathrm{\lambda_{3/2}}$} \textbf{(nm)} \\ \hline
			$^\mathrm{9}$Be$^\mathrm{+}$ & 3/2 & 1.25 & 313.1 & 19.6 & 313.0 \\ \hline
			$^\mathrm{25}$Mg$^\mathrm{+}$ & 5/2 & 1.79 & 280.3 & 41.3 & 279.6 \\ \hline
			$^\mathrm{43}$Ca$^\mathrm{+}$ & 7/2 & 3.23 & 396.8 & 22.5 & 393.4 \\ \hline
			$^\mathrm{67}$Zn$^\mathrm{+}$ & 5/2 & 7.2 & 206.2 & 62.2 & 202.5 \\ \hline
			$^\mathrm{87}$Sr$^\mathrm{+}$ & 9/2 & 5.00 & 421.6 & 21.5 & 407.8 \\ \hline
			$^\mathrm{111}$Cd$^\mathrm{+}$ & 1/2 & 14.53 & 226.5 & 50.5 & 214.4 \\ \hline
			$^\mathrm{137}$Ba$^\mathrm{+}$ & 3/2 & 8.04 & 493.4 & 20.1 & 455.4 \\ \hline
			$^\mathrm{171}$Yb$^\mathrm{+}$ & 1/2 & 19.7 & 369.4 & 19.7 & 328.9 \\ \hline
			$^\mathrm{199}$Hg$^\mathrm{+}$ & 1/2 & 40.51 & 194.2 & 54.7 & 165.0 \\ \hline
		\end{tabular}
	\end{table}

	\begin{table}[!htb]
		\caption{List of constants of ion species for QIP - part 2 \cite{QubitTypes}.  
			{$\mathrm{\lambda_{5/2}}$} is 
			the wavelength of the S$_\mathrm{1/2}$ to D$_\mathrm{5/2}$ transitions, and
			{$\mathrm{\tau_{5/2}}$} is the life time of the D$_\mathrm{5/2}$ state.}
		\label{tab: atomic constants 2}
		\centering
		\begin{tabular}{| c | c | c |}
			\hline
			\textbf{Ion} & {\boldmath$\mathrm{\lambda_{5/2}}$ \textbf{(nm)}} & {\boldmath$\mathrm{\tau_{5/2}}$} \textbf{(s)} \\  \hline
			$^\mathrm{43}$Ca$^\mathrm{+}$ & 729.1 & 1.17 \\ \hline
			$^\mathrm{87}$Sr$^\mathrm{+}$ & 674.0 & 0.36 \\ \hline
			$^\mathrm{137}$Ba$^\mathrm{+}$ & 1761.7 & 30 \\ \hline
			$^\mathrm{171}$Yb$^\mathrm{+}$ & 411.0 & 0.007 \\ \hline
			$^\mathrm{199}$Hg$^\mathrm{+}$ & 281.6 & 0.1 \\ \hline
		\end{tabular}
	\end{table}

	Table~\ref{tab: atomic constants} and Table~\ref{tab: atomic constants 2}
	show the properties for the choice of ion species in the quantum 
	von Neumann architecture for trapped ions.
	The criteria for the choice of the ion species of the qubit ions are:
	\begin{itemize}
		\item \textbf{Ground state qubit:} required for long storage time
		\item \textbf{Long coherence time:} low magnetic field dependence, 
			e.g. clock transitions in hyperfine qubits. The field strength
			should be large enough that motional sidebands do not overlap with 
			neighboring transitions, thus B~$>$~$\sim$10~Gauss.
		\item \textbf{Wavelength:} the lasers incorporated for the operation 
			of the quantum computer can hit the trap surface.  To avoid electron 
			emission, the work function of the surface material has to be higher 
			than the energy of a photon. Typical surface materials for segmented 
			Paul traps are gold and aluminum.  Gold has a work function of 5.3~eV 
			\cite{WorkfunctionGold}, which corresponds to 234~nm.  Aluminum 
			has a work function of 4.08~eV \cite{ModernPhysics}, which 
			corresponds to 304~nm.
		\item \textbf{Mass ratio:} for sympathetic cooling the mass ratio of 
			the ion species in the ion string should be close to 1 
			\cite{SympatheticCooling2,IonMassRatio1,IonMassRatio2}, so that all 
			modes of a mixed ion crystal can be efficiently cooled. Experimentally, 
			sympathetic cooling of two-ion crystals with a mass ratio of 3 has been 
			demonstrated \cite{SympatheticCooling3}.
		\item \textbf{Mass:} with the same electric field, lighter ions
			are accelerated faster, which is advantageous for ion movement.
			In the same trapping potential, lighter ions have higher trap 
			frequencies, which allows faster gate operations and less power is required
			for the entangling gate operations \cite{SpontaneousPhotonScattering}.
	\end{itemize}

	The criteria for the choice of the ion species of the cooling ions are:
	\begin{itemize}
		\item \textbf{Mass ratio:} as described for the qubit ion species.
		\item \textbf{Wavelength:} as described for the qubit ion species.
		\item \textbf{No nuclear spin:} this simplifies the level structure 
			and the laser system, if one does not require two beams with GHz
			detuning from one another.
	\end{itemize}

	The criteria for the choice of the ion species of the detection/initialization ions are:
	\begin{itemize}
		\item \textbf{Mass ratio:} as described for the qubit ion species.
		\item \textbf{Wavelength:} as described for the qubit ion species.
		\item \textbf{No nuclear spin:} as described for the cooling ion species.
		\item \textbf{Long lived D-state:} there are several different
			detection schemes like electron shelving 
			\cite{ElectronShelving1,ElectronShelving2,ElectronShelving3},
			or using sigma polarized light to cyclically drive a single transition
			\cite{Wineland1998}.  Both schemes allow high-fidelity state detection.  
			But in the same setup, electron shelving with an ion species which has 
			a long lived D$_\mathrm{5/2}$ state usually yields higher fidelity than 
			a detection scheme which is limited by off-resonant excitations 
			\cite{DetectionMonroe,HighFidelityReadout}.
	\end{itemize}

	\section{The Quantum 4004}
	\label{sec: quantum 4004}

	After the invention of integrated circuits, the first CPUs were developed
	\cite{Tanenbaum}. The Intel~4004 was one of the world's first 
	microprocessors and was the first microprocessor which customers could 
	program themselves.  It had the following technical specification\footnote{
	http://www.intel.com/Assets/PDF/DataSheet/4004\_datasheet.pdf - 
	website last visited April 30, 2017}:
	\begin{itemize}
		\item 4-bit CPU
		\item Instruction cycle time: 10.8~$\mu$s
		\item Instruction execution time: 1 or 2 instruction cycles
		\item Able to directly address 32768~bits (4096~bytes) of memory
	\end{itemize} 
	This was one of the starting points of the exponential increase in clock 
	speed and number of transistors of a CPU (Moore's law 
	\cite{MooresLaw,MooresLaw2}), leading to the computers that we have now.  
	Hence, it makes sense to assume that the speed and capabilities of first 
	generation quantum computers may be the equivalent of the Intel~4004 
	but in a quantum world.  Because of its simplicity, the Intel~4004 will 
	act as a blueprint for a quantum computer in the following.
	If Moore's law is applicable to the development of quantum computers, 
	the capabilities of quantum computers will increase exponentially over 
	time as well.
	
	Following the ideas of the previous section, this section only discusses
	the hardware of the trapped ion quantum computer.  The main focus is
	on simplicity of the hardware and, if possible, optimizations for certain QEC schemes
	or quantum algorithms are avoided\footnote{This choice is completely
	arbitrary but serves to highlight the simplicity of the hardware rather
	than showing how well the architecture can be used for certain QEC
	schemes or quantum algorithms.}.  This section only serves to exemplify
	the scalability and capabilities of such a quantum von Neumann architecture.

	This Quantum~4004 is based on the presented quantum von Neumann 
	architecture for trapped ions with design parameters corresponding to 
	the technical specification of the Intel~4004.
	\begin{itemize}
		\item The Intel~4004 had only one CPU. Thus, the Quantum~4004 will
			also work with only one QALU.
		\item The information is structured in ion chains of 4 qubits. - For 
			a 7-qubit Steane code \cite{SteaneCode}, it may be advantageous 
			to structure this architecture with 7 qubits in an ion chain.  
			But since this first design should not (by choice) have any 
			restrictions imposed by higher abstraction layers. Hence, the 
			following design will stick to the Intel~4004 model of 4 (qu)bits.
		\item Each qubit in the memory zone is DFS encoded. Thus, one ion 
			chain in the quantum memory contains 8 physical qubit ions. - The 
			memory of the Intel~4004 was structured in bytes (8 bits) as well.
		\item Since the proposed hardware should result in coherence times 
			of days or longer, a design with fast gate speed is not mandatory. 
			In analogy to the instruction execution time of 1 or 2 instruction 
			cycles (10.8~$\mu$s) in the Intel~4004, a single qubit operation 
			should be executed in 10~$\mu$s and an entangling gate operation 
			should be executed in 20~$\mu$s in the Quantum~4004.  
		\item The Intel~4004 was able to access 4~kBytes of RAM (32768~bits)
			and thus the architecture of the Quantum~4004 should be able to 
			store 32768 qubit ions. Due to DFS encoding, this would correspond 
			to 16384 physical (non-encoded) qubits.
	\end{itemize} 

	The QALU can be structured following the ideas illustrated in 
	Figure~\ref{fig: pipelining processing zone}.  In the region for QIP, 
	there will only be one string of 16 qubit ions (and cooling ions).  
	After DFS decoding, only 8 of these 16 qubit ions contain quantum
	information. Hence, the optics for QIP has to be optimized for an ion 
	string of 8 ions\footnote{Even if one only needs to perform 1 single qubit 
	operation on 1 ion, there will always be 8 qubit ions containing quantum 
	information in the QIP region.}, following
	the ideas presented in Appendix~\ref{sec: optical setup}.  
	It will be possible to perform 8 single qubit operations simultaneously.  
	Therefore, in an ideal case, it will be possible to perform 
	800 thousand single qubit operations or 50 thousand entangling gates 
	per second on 16384 physical qubits.
	For a syndrome measurement in a 7-qubit Steane code, one will need 
	4 entangling gates per syndrome, 3 syndrome measurements for bit flips 
	and 3 for phase flips.  This sums up to 24 entangling gates for 7 qubits.  
	If all qubits were encoded in a 7-qubit Steane code and one would need 
	1.12~s to perform a syndrome measurement on all 16384 physical qubits.  
	Due to coherence times of days, the 1.12~s for syndrome measurements 
	should allow for fault-tolerant QC.  
	These syndrome measurements would require about 14000 quantum state 
	detections to read out the syndromes.  Thus, serialized detection with 
	a single detection zone must detect a quantum state in 80~$\mu$s or less, 
	which is not possible with a single detection zone as detection involves 
	swap gates (consisting of 3 entangling gates, which each take 20~$\mu$s) 
	and CNOT-gates.  For detection in 80~$\mu$s with 7-qubit GHZ 
	states\footnote{A 7-qubit GHZ state would require 7 ions of the detection
	ion species and 1 ion of the qubit ion species in the QIP part of the 
	detection zone.  Thus, like in the QALU, one can use a setup optimized 
	for 8 ions.}, whose generation takes 180~$\mu$s\footnote{Due to pipelining
	in the detection zone, the detection time can be as long as 180~$\mu$s as well.}, 
	one will require at least 3 detection zones.  Ion shuttling must only 
	be fast enough to not (considerably) slow down pipelining in the QALU.
	Thus, the already demonstrated speeds of several 100~$\mu$m in a 
	couple of microseconds \cite{FastShuttling1,FastShuttling2} should be 
	sufficient.

	\begin{figure}[!htb]
		\begin{center}
			\includegraphics[width=13cm]{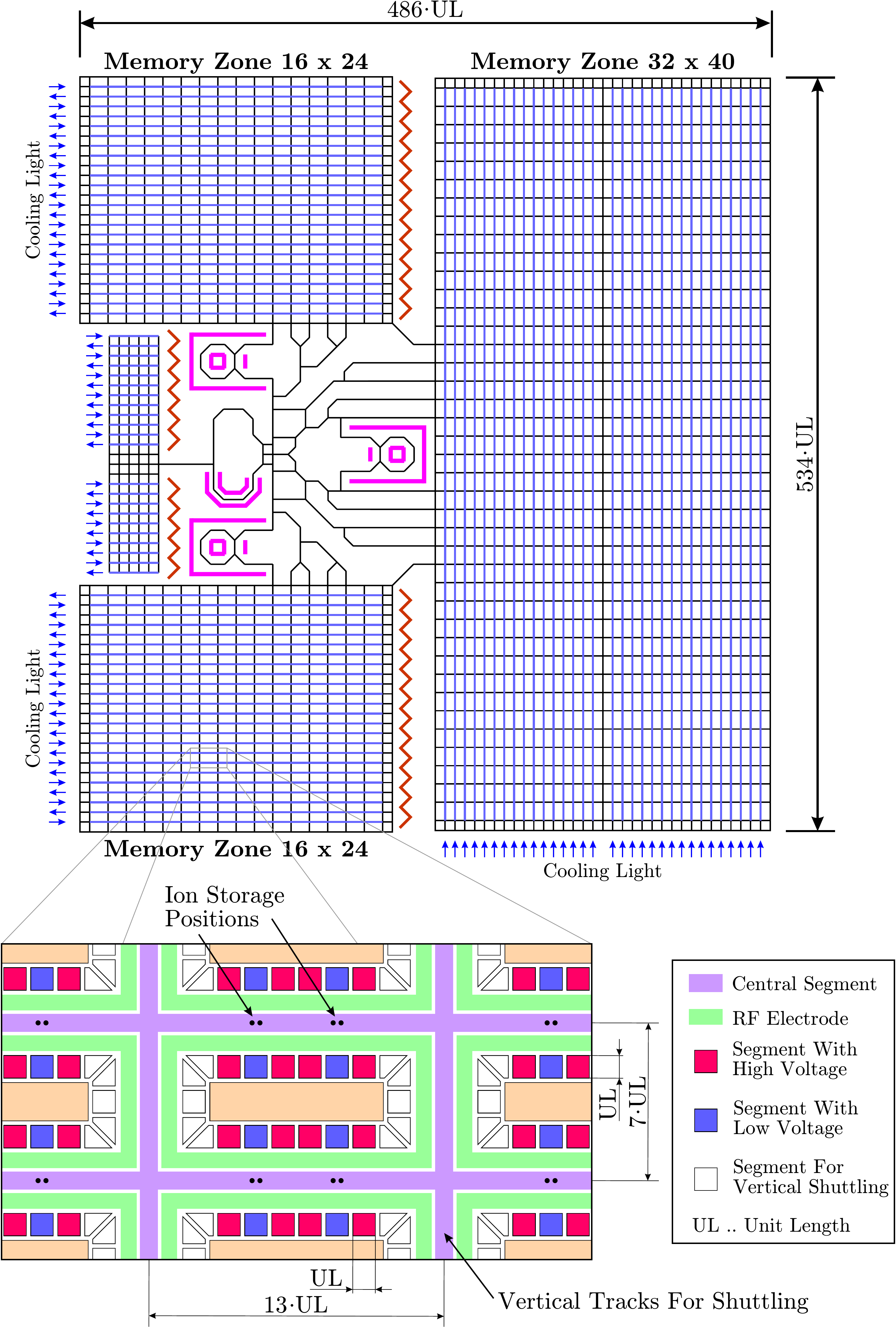}
			\caption{Layout of the Quantum 4004 with a cutout view of 2~$\times$~2 
				unit cells in the quantum memory.  For a detailed zoom on the
				central region, refer to Fig.~\ref{fig: layout 4004 zoom}.}
			\label{fig: layout 4004}
		\end{center}
	\end{figure}

	\begin{figure}[!htb]
		\begin{center}
			\includegraphics[width=12cm]{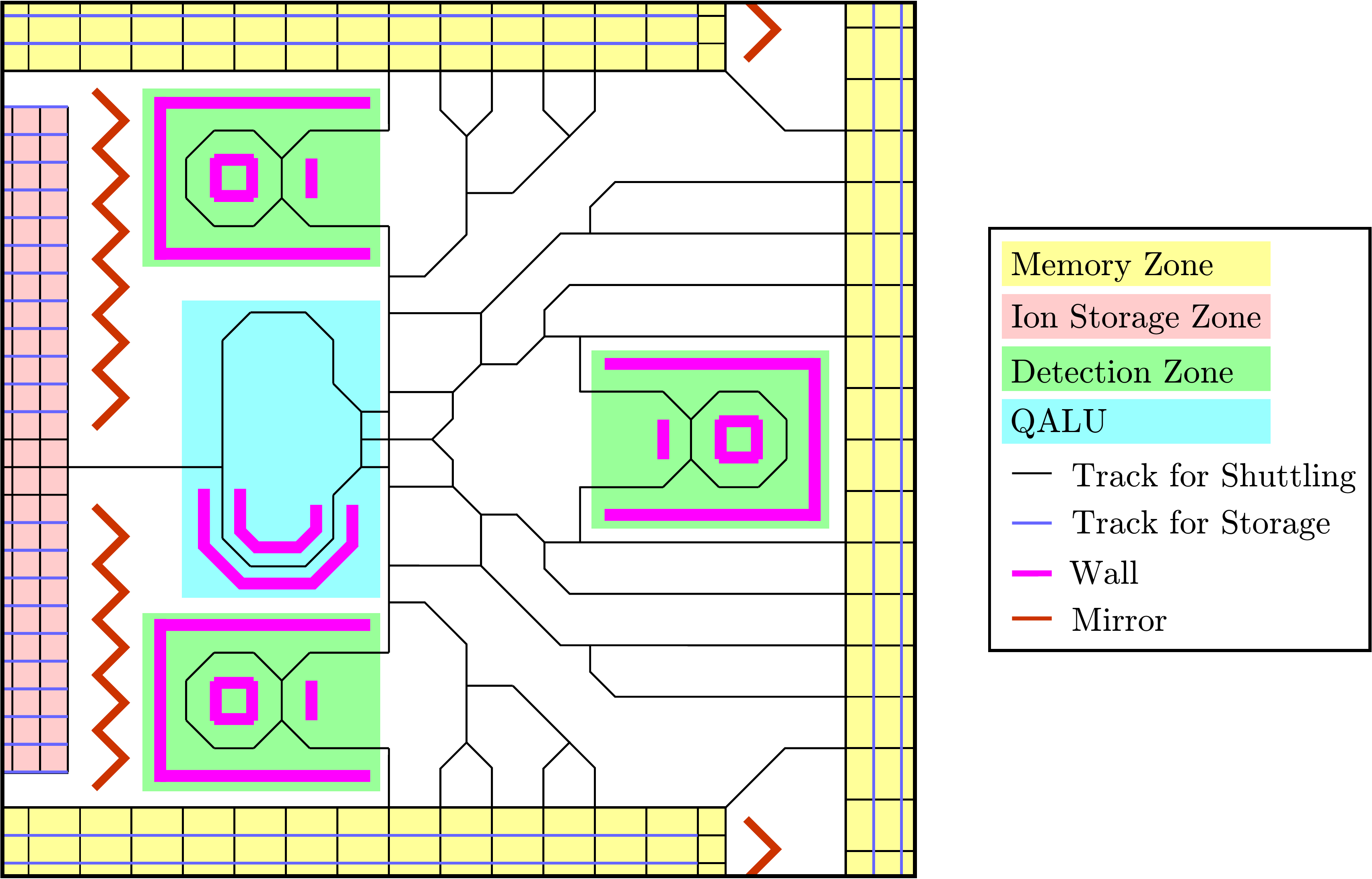}
			\caption{Zoom view of the layout of the Quantum 4004 from Fig.~\ref{fig: layout 4004}
				in the central region around the QALU.}
			\label{fig: layout 4004 zoom}
		\end{center}
	\end{figure}

	The layout of the Quantum~4004 with a subfigure of 2~$\times$~2 unit cells in
	the quantum memory is shown in Fig.~\ref{fig: layout 4004} and a zoom view
	of this layout near the QALU is depicted in Fig.~\ref{fig: layout 4004 zoom}.
	The QALU with pipelining is structured in a round shape, following the ideas
	illustrated in Fig.~\ref{fig: pipelining processing zone}.  The QIP
	region in the QALU is surrounded by walls to protect the quantum memory zones 
	from near resonant stray light.
	Left of the QALU in Fig.~\ref{fig: layout 4004 zoom} is the ion
	storage, or loading zone, which contains a grid for backup storage of ions 
	which can be used to replace lost ions.  
	The Quantum~4004 contains 3 detection zones and they are located in the 
	center of the trap, as shown in Fig.~\ref{fig: layout 4004 zoom}.
	The main surface area of the segmented Paul trap of the Quantum~4004 is 
	occupied by the quantum memory, which is divided in 3 parts in Fig.~\ref{fig: layout 4004}
	and structured following the ideas depicted Fig.~\ref{fig: memory zone}~b. 
	In the two $16\times24$ memory regions, the cooling beams travel from left 
	to right and are reflected back to reuse the same light a second time for cooling. 
	Whereas in the $32\times40$ memory region, the cooling beams travel from bottom to top.
	Efficient ion movement in the grid of a memory region is only possible parallel to
	the cooling beams, depicted by black lines.  On the tracks perpendicular to the 
	cooling beams, depicted by light blue lines, the ions are stored.
	When moving ions from storage to the QALU, it requires no additional ion shuttling 
	to move ion strings perpendicular to the cooling beams.  But the paths along the cooling beams
	are typically occupied by stored ions.  To simplify ion movement along the
	cooling beams, the three memory zones are surrounded by tracks which are not used
	for ion storage.

	\begin{table}[!htb]
		\caption{The estimated number of segments, number of DACs, and the size of the individual parts 
			of the Quantum~4004 architecture shown in Fig.~\ref{fig: layout 4004}.}
		\label{tab: segments}
		\centering
		\begin{tabular}{| l | c | c | c |}
			\hline
			\textbf{Region} & \textbf{Segments} & \textbf{Independent DACs} & \textbf{Size / UL} \\ \hline \hline
			Ion Storage Zone & $\approx$2734 & $\approx$10 & 35~$\times$~168 \\ \hline
			QALU & $\approx$298 & $\approx$150 & 47~$\times$~70 \\ \hline % 174: 2*14+1 zones @ 6 segments
			Detection Zones & $\approx 3\cdot200 $ & $\approx 3\cdot12$ & 3$\cdot$(57$\times$~41) \\ \hline %36: 4 zones * 6 + 2 zones * 6 for shuttling
			Memory Zones $16\times24$ & $\approx 2\cdot9820$ & $\approx2\cdot10$ & 2$\cdot$(238~$\times$~534) \\ \hline % segments: 31 * 24 * 16 + 5 * 2 * (31+24) = 12454
			Memory Zone $32\times40$ & $\approx$31486 & $\approx$30 & 222~$\times$~175 \\ \hline % segments: 63 * 40 * 16 + 5 * 2 * (63+40) = 41350
			Connecting Tracks & $\approx$1972 & $\approx$30 & - \\ \hline \hline
			\textbf{Total} & \textbf{\boldmath$\approx$56730} & \textbf{\boldmath$\approx$276} & \textbf{\boldmath 486~$\times$~534} \\ \hline
		\end{tabular}
	\end{table}

	Table~\ref{tab: segments} sums up the required hardware resources of the 
	Quantum~4004 architecture shown in Fig.~\ref{fig: layout 4004} and 
	Fig.~\ref{fig: layout 4004 zoom}.
	In total, one would require about 57000 individual segments for this architecture, 
	and thus the same amount of analog switches.  Multiplexing reduces the number of 
	independently controlled DACs to about 280.  Given that one has control over 32768 
	qubit ions (plus ions for sympathetic cooling and detection), one would need less 
	than 2 segments (= 1 segment pair) per qubit ion.  And one would need only one 
	independently controlled DAC for about every 120 qubit ions.

	To estimate the physical dimension of the trap, a unit length (UL) is defined as
	the segment size of a DC segment in the memory, as depicted in the inlet of
	Fig.~\ref{fig: layout 4004}.
	The required surface area of the planar trap in containing the Quantum~4004 is 
	486~UL~$\times$~534~UL. With a typical unit length of 80~$\mu$m, the trap would be
	38.9~mm~$\times$~42.7~mm with a diagonal of 57.8~mm length. 
	To achieve a relative magnetic field inhomogeneity for the whole trap of 
	10$^\mathrm{-6}$ with Helmholtz coils, one would need Helmholtz coils with a 
	radius of about 1~m, see Appendix~\ref{sec: magnetic field homogeneity} for details.  
	The dimensions of the superconducting shield would have to be a factor of 2 or 3 
	bigger than the dimension of the Helmholtz coils.  Hence, the superconducting shield 
	could be a cylinder with 3~m radius and 5~m length\footnote{If one only requires
	a relative inhomogeneity of 10$^\mathrm{-5}$, the dimensions of the setup will only
	be half as big.}.

	To connect the 57000 individual segments to external signals, one would have to place
	57000 bond pads on the trap, which would occupy nearly as much space as the trap
	itself.  Hence, if it is possible, one will have to integrate the analog switches
	and the multiplexing logic circuits into the trap, which will reduce the number of
	DC interconnections to 280 plus the control signals for the multiplexing logic circuits.

	For traps coated with aluminum, one can only work with Be$^\mathrm{+}$, Ca$^\mathrm{+}$, 
	Sr$^\mathrm{+}$, Ba$^\mathrm{+}$, and Yb$^\mathrm{+}$ ions to avoid photoelectric effect.  
	A desired mass ratio\footnote{Experimentally, ground state cooling has been shown with a 
	mass ratio of 3.} between 1/3 and 3 eliminates Be$^\mathrm{+}$
	from the pool of usable ion species.  As Ba$^\mathrm{+}$ has a long life time of 
	the D$_\mathrm{5/2}$ state of more than 30~s, it is the prime candidate
	for the detection ion species.  This leaves the choice between the following
	triples of ion species:
	\begin{itemize}
		\item Qubit ions: $^\mathrm{87}$Sr$^\mathrm{+}$, detection ions: $^\mathrm{138}$Ba$^\mathrm{+}$,
			cooling ions: $^\mathrm{40}$Ca$^\mathrm{+}$
		\item Qubit ions: $^\mathrm{87}$Sr$^\mathrm{+}$, detection ions: $^\mathrm{138}$Ba$^\mathrm{+}$,
			cooling ions: $^\mathrm{172}$Yb$^\mathrm{+}$
		\item Qubit ions: $^\mathrm{171}$Yb$^\mathrm{+}$, detection ions: $^\mathrm{138}$Ba$^\mathrm{+}$,
			cooling ions: $^\mathrm{88}$Sr$^\mathrm{+}$
	\end{itemize}
	The first choice has the advantage that the lowest required wavelength is 
	397~nm (for $^\mathrm{40}$Ca$^\mathrm{+}$) and only little optical power 
	of this wavelength is required for cooling.
	Thus, bleaching in fibers is least critical for this triple of ion species.
	But $^\mathrm{87}$Sr$^\mathrm{+}$ has the disadvantage that its nuclear
	spin is 9/2.  This makes state initialization more challenging than the use
	of $^\mathrm{171}$Yb$^\mathrm{+}$. 
	Whether to work with the first or the last triple of ion species is thus
	a decision whether it is easier to faithfully initialize $^\mathrm{87}$Sr$^\mathrm{+}$
	or to prevent bleaching in the fibers due to high-power state manipulation of 
	$^\mathrm{171}$Yb$^\mathrm{+}$ during constant operation for months
	or years\footnote{The highest optical powers will be required for high-fidelity
	entangling gates (up to several Watts \cite{SpontaneousPhotonScattering}).
	The transition wavelength between the S$_\mathrm{1/2}$ and P$_\mathrm{1/2}$ 
	in $^\mathrm{171}$Yb$^\mathrm{+}$ is 369~nm, whereas in $^\mathrm{87}$Sr$^\mathrm{+}$
	it is 422~nm.}.
	Since bleaching in fibers is a fundamental problem and state initialization of
	$^\mathrm{87}$Sr$^\mathrm{+}$ is a technical problem, the triple consisting
	of $^\mathrm{87}$Sr$^\mathrm{+}$ as the qubit ion species, $^\mathrm{138}$Ba$^\mathrm{+}$
	as the detection ion species, and $^\mathrm{40}$Ca$^\mathrm{+}$ and the cooling
	ion species is most favorable.

	For completeness, one should add that when using gold as the surface material 
	of the trap, the following triples of ion species are usable besides the already listed ones:
	\begin{itemize}
		\item Qubit ions: $^\mathrm{43}$Ca$^\mathrm{+}$, detection ions: $^\mathrm{88}$Sr$^\mathrm{+}$,
			cooling ions: $^\mathrm{24}$Mg$^\mathrm{+}$
		\item Qubit ions: $^\mathrm{25}$Mg$^\mathrm{+}$, detection ions: $^\mathrm{40}$Ca$^\mathrm{+}$,
			cooling ions: $^\mathrm{9}$Be$^\mathrm{+}$
	\end{itemize}
	And when extending the mass ratio to 1/3.5 to 3.5, the following triple of ion 
	species is usable besides the already listed ones:
	\begin{itemize}
		\item Qubit ions: $^\mathrm{43}$Ca$^\mathrm{+}$, detection ions: $^\mathrm{138}$Ba$^\mathrm{+}$,
			cooling ions: $^\mathrm{88}$Sr$^\mathrm{+}$
	\end{itemize}

	With the design of this chapter, it will be possible to construct a trapped ion
	quantum computer, which is capable of handling 32768 qubit ions on which one 
	can apply 800000 single qubit operations or 50000 entangling gate operations per second.

	To get a feeling for how difficult the fabrication the Quantum~4004 trap would be, 
	one can compare the layouts of Quantum~4004 (Fig.~\ref{fig: layout 4004}) 
	and the Intel~4004\footnote{Available at 
	http://www.4004.com/assets/redrawn-4004-schematics-2006-11-12.pdf and
	http://www.4004.com/assets/4004-masks-showing-fets-j3-.jpg - Website last 
	visited on April 30, 2017.}. While the Quantum~4004 requires more individual electrical 
	lines, the Intel~4004 has a higher complexity because of the interconnections 
	between the electrical lines.

	\subsection{Possible future quantum computers}
	\label{sec: possible future quantum computers}

	The Quantum~4004 is an example to illustrate a prototype, which is optimized
	for a simple hardware, and that quantum 
	computers with tens of thousands of qubits are technologically feasible.
	Of course, one will have to start by demonstrating that the individual
	parts of such a quantum computer work. But once that has been demonstrated,
	there is no reason why QC with more than 10000 qubits should not be
	possible in the (near) future.

	The optimization for a simple hardware in Section~\ref{sec: quantum 4004}
	limits the capabilities of the quantum computer.  For example, 	
	the choice of a single QALU slows the Quantum~4004 down.  
	Furthermore, in future publications, one should discuss architectures
	which are optimized for certain QEC schemes or even quantum algorithms.
	For example, when working with a Steane code, ion chains of length 7
	make more sense than length 4.
	If coherence times of days are achieved, one will not be restricted
	to only 32768 physical qubits but one will be able to work with millions
	of physical qubits (per processing zone).  The only disadvantage is that 
	even for 32768 qubits, most of the trap is already occupied by the quantum 
	memory. Hence, such quantum computers will require much bigger traps.
	In the example of the Quantum~4004 presented in this section, the vacuum
	chamber of the quantum computer has the size of about a room, which
	seems a feasible size.  When working with millions of qubits, one will have
	to find different ways how to generate a magnetic field with a big
	volume of high homogeneity or a way to allow QC with millions of qubits
	in a magnetic field with low homogeneity.

	\section{Summary}
	\label{sec: conclusion}

	As there exists no functioning quantum computer yet and it is not fully clear
	how to build a large-scale quantum computer, quantum computer architectural
	design becomes increasingly important.
	In this document, I presented knowledge of classical computer architecture,
	like von Neumann architecture or pipelining, which is applicable in
	quantum computer architectures.  Quantum computers are different from
	computers because decoherence in qubits destroys quantum information
	over time, which has to be compensated by executing QEC. 
	Whereas, classical information can, in principle, be stored infinitely long.
	Thus, quantum computers with a large memory will require parallelism
	for fault-tolerant computation.

	Currently, most quantum computer architectures favor a massively-parallel
	approach for which every qubit site has storage capability and computation
	capability.  This is the most sensible approach for small- and medium-scale
	QC, as QEC can be performed on all qubits simultaneously and it allows
	higher gate thresholds for fault-tolerant QC.  However, for large-scale
	systems, engineering challenges like obeying Rent's rule have to be overcome.
	This will most likely imply some kind of serialization of the computation.

	The presented Quantum von Neumann architecture is based on classical
	von Neumann architecture and has a specialized part, or section, for each task to 
	perform in the quantum computer, such as a QALU for QIP or a quantum
	memory for quantum information storage.  This architecture is applicable
	on systems with a long coherence time and the capability to move
	quantum information from one region of the quantum computer to another.

	After the general introduction, trapped ions are chosen to illustrate 
	how a quantum computer based on Quantum von Neumann architecture could 
	be built by incorporating multiplexing and pipelining technology.
	Furthermore, requirements for the quantum gate operations and the choice 
	of ion species are given.  At last, this theoretical knowledge is applied
	on a specific trapped ion quantum computer, the Quantum~4004, which has 
	a simple hardware and is the 
	quantum equivalent of the Intel~4004.  The Quantum~4004 has just one QALU
	and the computation speed is 10~$\mu$s for single qubit operations
	and 20~$\mu$s for two-qubit operations.  The quantum computer is
	structured in 4 qubits per ion string, which are DFS encoded. Thus, 
	there are 8 qubit ions per ion string.  In total, the Quantum~4004
	can work with up to 32768 qubit ions in a fault-tolerant way on a
	38.9~mm~$\times$~42.7~mm big ion trap with $\approx$57000 segments. 

	\section*{Acknowledgment}
	\label{sec: acknowlegment}
	\addcontentsline{toc}{section}{Acknowledgment}

	I would like to thank Philipp Schindler for helpful comments on this manuscript.

	\begin{appendix}
		\section{Appendix}
		\label{sec: appendix}

		\subsection{Magnetic field homogeneity}
		\label{sec: magnetic field homogeneity}

		In Section~\ref{sec: magnetic shielding}, a way to shield cryogenic experiments
		with a superconductor by exploiting Meissner effect was presented.
		Trapped ion qubits using a clock transition require a specific magnetic field,
		which is on the order of 10~mT.  The generation of a constant field inside the 
		magnetic shield requires a persistent current in superconducting coils.
		Typical trap dimensions are 40 to 60~mm, see Section~\ref{sec: quantum 4004} 
		for details, and most of the surface is occupied by the quantum memory.
		Ideally, one wishes to have the same magnetic field strength over the
		whole trap.  But as coils cannot become arbitrarily big and as the ion trap's
		permeability is different from vacuum's permeability,  one cannot have
		a perfectly homogeneous magnetic field. Of course, one can measure the
		local magnetic field at all (relevant) positions on the trap and 
		calculate the phase evolution of each qubit in the trap.  This gets
		more and more laborious with increasing size of the trap and increasing
		number of qubits. Thus, it is advantages to have a highly homogeneous field
		at the position of the ion trap.

		The sensitivity to magnetic fields at clock transitions is typically on the
		order of 100~kHz/(mT)$^2$ at a bias field of about 10~mT \cite{YbClockQubit,BeClockQubit,HighFidGateHarty}.  
		In order to achieve coherence times of a day, one requires the magnetic 
		deviation to be smaller than 10~nT, which is a relative deviation of 
		about 10$^{-6}$ of the typical bias field. 10~nT is also the offset that the
		superconducting shield might produce, see Section~\ref{sec: magnetic shielding} 
		for details.

		The highest magnetic homogeneity inside a coil can be achieved with a solenoid
		coil or a cylindrical coil.  Such structures limit the optical access
		to the trap and as all electrical signals have to be sent to the
		trap along the coil axis, it is not clear if that has an effect on the
		bias field.  Therefore, a solenoid coil is an option for future designs 
		but in the following, it will not be considered.

		\begin{figure}[!htb]
			\begin{center}
				\includegraphics[width=14cm]{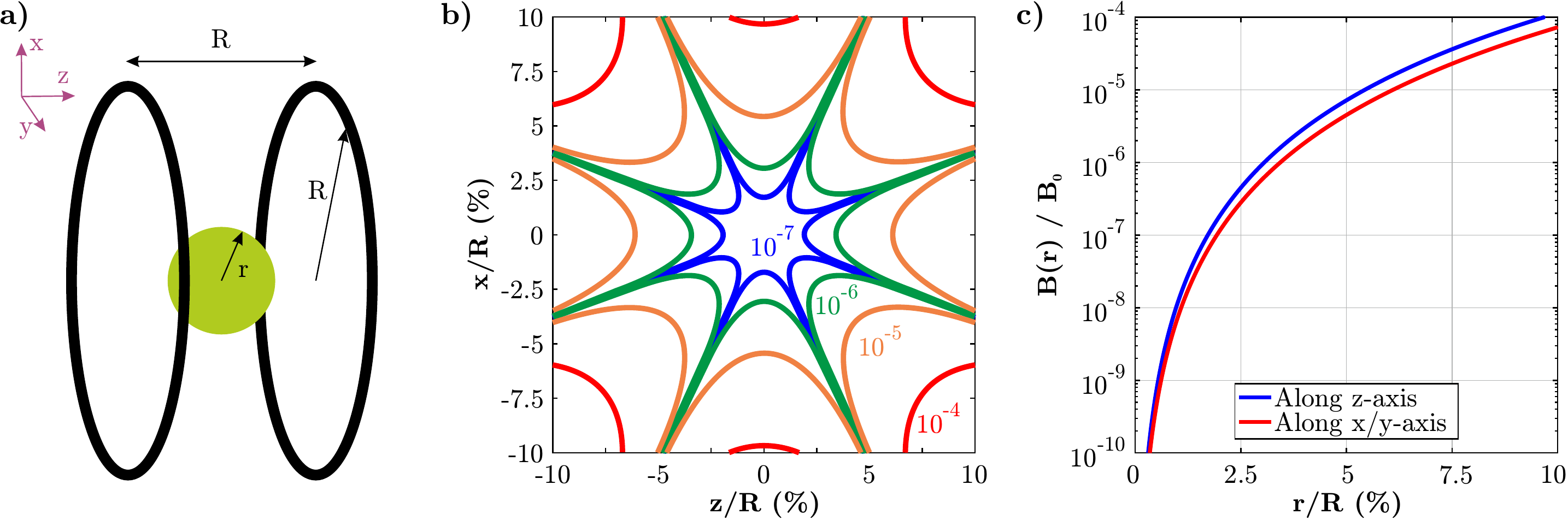}
				\caption{Panel (a) shows a Helmholtz coil.
					The contour plot in panel (b) illustrates the homogeneity of
					the magnetic field produced by Helmholtz coils, and panel (c)
					shows the relative deviation of the magnetic field along the axes of
					the coordinate system with respect to the magnetic field in the center.}
				\label{fig: magnetic homogeneity Helmholtz}
			\end{center}
		\end{figure}

		Helmholtz coils consist of two coils with radius R on the same
		coil axis, and the distance between the coils is R, as depicted in 
		Fig.~\ref{fig: magnetic homogeneity Helmholtz}~a.  They combine high
		magnetic field homogeneity with high optical access.  The volume with
		same relative deviations from the field strength in the center is 
		approximated by a sphere with radius r.
		Fig.~\ref{fig: magnetic homogeneity Helmholtz}~b shows a contour plot
		of a simulation of the magnetic field distribution of the Helmholtz 
		coils in the XZ-plane.  The coordinates in this simulation are 
		normalized to the coil radius\footnote{The simulations assume that the 
		coils are infinitely narrow.}, and the results along the axes of the
		coordinate system are plotted in Fig.~\ref{fig: magnetic homogeneity Helmholtz}~c.
		The results show that if one tries to get a spherical volume with a 
		magnetic homogeneity with relative deviations of less than 10$^\mathrm{-6}$, 
		the radius of the sphere will only be 3~\% of the coil radius.  
		As an example, the trap described in Section~\ref{sec: quantum 4004}
		has a diagonal length of about 60~mm.  Therefore, Helmholtz coils that 
		produce magnetic homogeneity with relative deviations of less than 
		10$^\mathrm{-6}$ have to be at least 1~m in radius.
		To avoid influencing the magnetic homogeneity in the center, the 
		dimensions of the superconducting shield have to be chosen a factor 
		of 2 or 3 bigger than the dimensions of the Helmholtz coils to influence
		the magnetic homogeneity as little as possible.  
		The resulting diameter of several meters of the superconducting
		shield requires a vacuum chamber of the size of a room to provide
		the required spatial homogeneity.  These dimensions enable optical 
		alignment inside the shield.

		\begin{figure}[!htb]
			\begin{center}
				\includegraphics[width=10cm]{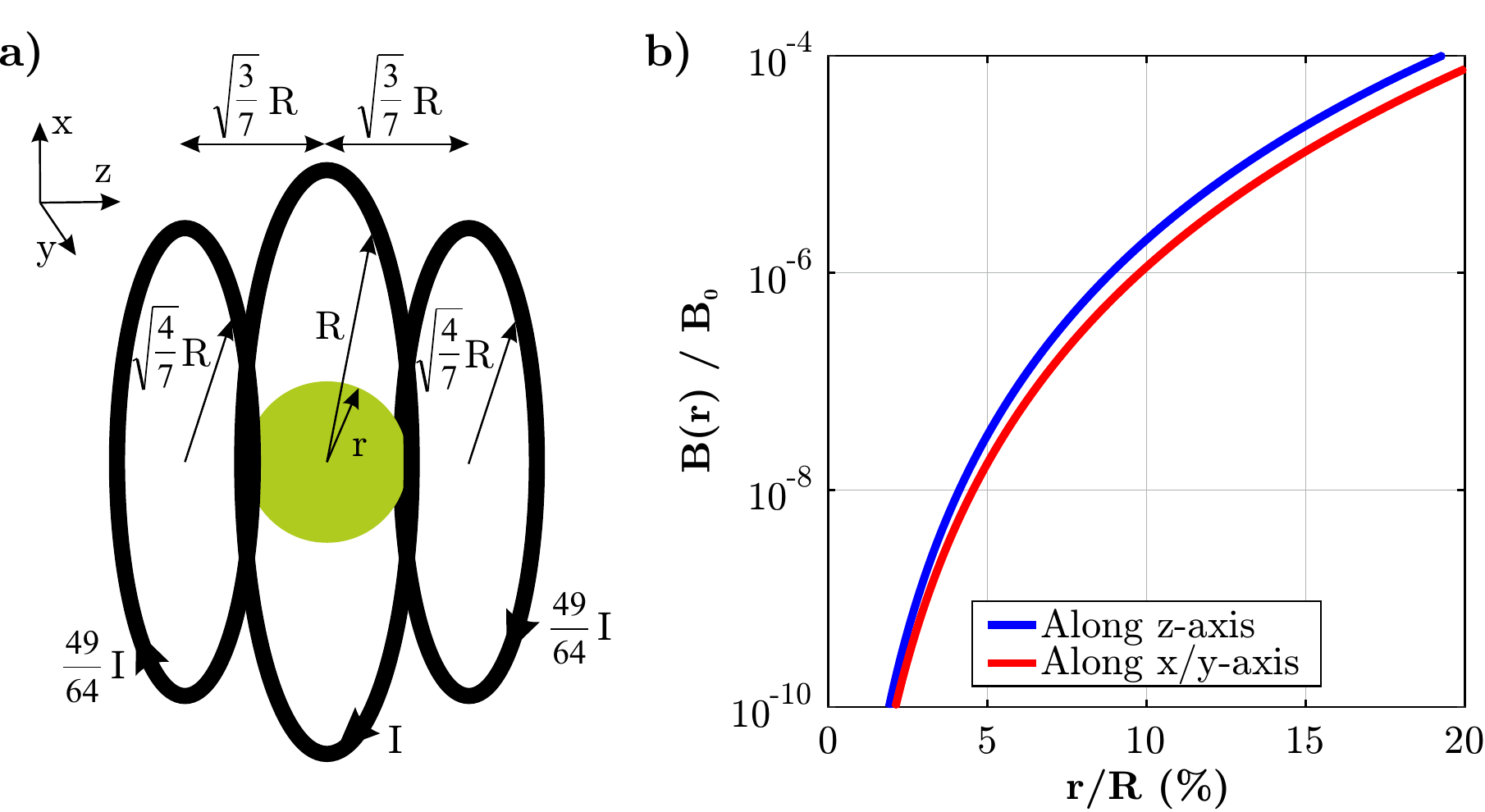}
				\caption{Panel (a) shows a Maxwell coil and panel (b) 
					depicts the relative deviation of the magnetic field along the axes of
					the coordinate system with respect to the magnetic field in the center.}
				\label{fig: magnetic homogeneity Maxwell}
			\end{center}
		\end{figure}

		In order to decrease the size of the coils while maintaining the spatial 
		homogeneity and high optical access, one can use Maxwell coils as illustrated 
		in Fig.~\ref{fig: magnetic homogeneity Maxwell}~a.  The simulations of the
		magnetic homogeneity along the axes, depicted in Fig.~\ref{fig: magnetic homogeneity Maxwell}~b,
		show that a sphere with radius of about 9~\% of the main radius R of the 
		Maxwell coils yields deviations of the magnetic field of less than 
		10$^\mathrm{-6}$ with respect to the field in the center.  Hence, using 
		Maxwell coils reduces the dimension of the setup by a factor of about 3 
		compared to Helmholtz coils.

		So far, the medium in the center of the coils was assumed vacuum.  But no 
		material is perfectly non-magnetic.  Therefore, placing a trap in the center 
		of the coils will cause additional inhomogeneity.  By the choosing 
		low-magnetic materials in trap fabrication, one can try to minimize the 
		effect of the trap on the magnetic field inhomogeneity.

		\subsection{Cryogenic system}
		\label{sec: cryogenic system}

		\begin{figure}[!htb]
			\begin{center}
				\includegraphics[width=16cm]{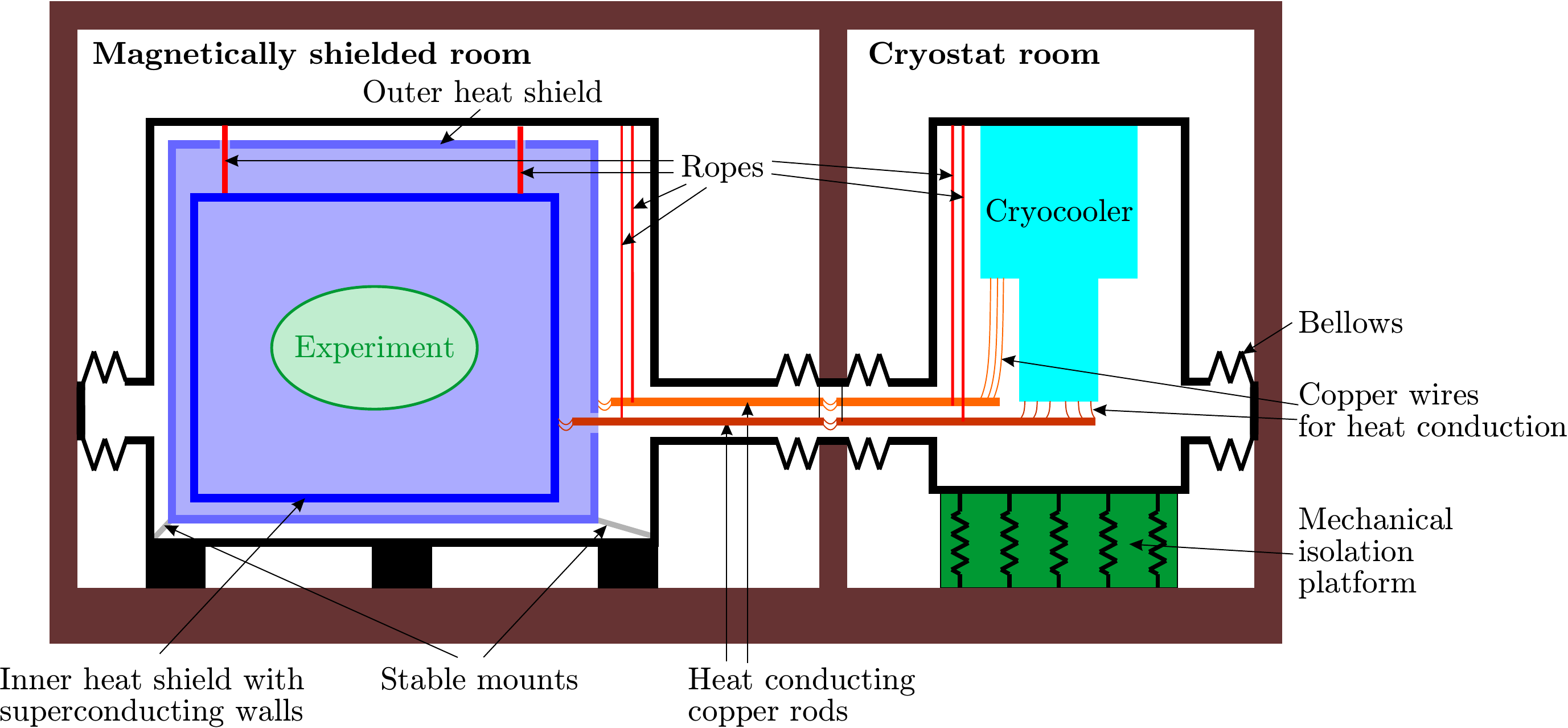}
				\caption{A setup with a closed cycle cryostat in separate room.  The heat transfer from the heat
					shield with the ion trap to the cold finger happens via suspended transfer rods.
					For vibration isolation, the inner heat shield is suspended from the vacuum chamber via ropes.}
				\label{fig: two room cryo}
			\end{center}
		\end{figure}

		The superconducting shield, which was introduced in 
		Section~\ref{sec: magnetic shielding}, has to become superconducting in a 
		zero-field environment.  Hence, if one wants to work with a closed-cycle 
		cryostat, like a Pulse-Tube cryocooler, which produces a magnetic field, 
		the cryocooler will have to be placed in a separate room so that the experiment 
		can be shielded against magnetic field from the cryocooler, as depicted in 
		Figure~\ref{fig: two room cryo}.  The heat transfer can be maintained via 
		copper rods that connect the coldfinger of the cryocooler with the heat 
		shields which contain the experiment.  For the stability of the optical
		alignment inside the superconducting shield, vibration isolation is
		required \cite{VibIsolation}.  Therefore, the copper rods are suspended 
		with ropes and the heat connection to the coldfinger is maintained via 
		thin copper wires.  To avoid vibration transport through the vacuum chamber, 
		bellows on both sides of the magnetically shielding wall divide the vacuum 
		chamber in a part with vibrations and a part without vibrations.  
		Furthermore, the cryocooler should be mounted on a vibration isolation 
		platform to not mechanically couple the different parts of the vacuum chamber 
		via the floor.

		If the light is transmitted via fibers to the experiment, the outer heat 
		shield can be mounted rigidly to the vacuum chamber.  However for good 
		vibration isolation, the inner shield with the experiment should 
		be suspended  with ropes from the vacuum chamber, or the outer shield.
		As mentioned in Section~\ref{sec: magnetic field homogeneity}, the 
		dimensions of the inner heat shield might be several meters in all directions 
		so that one can generate a  homogeneous magnetic field for the experiments.  
		Such a heat shield might weight a ton or more. When suspended with ropes, 
		this heavy mass will reduce the vibrations inside the shield due to its inertia.
		Furthermore, the dimensions the size of a room reduce the surface to volume 
		ratio in the vacuum chamber which will allow lower vacuum pressures.  

		If the magnetic shielding with one layer is not sufficient, one can add a 
		second layer of superconducting shielding.  This second layer has to be on 
		a separate heat shield, so that one can make sure to first enter the 
		superconducting phase with the inner magnetic shield and afterward with
		the outer magnetic shield.  Otherwise magnetic fields, which get locally
		pinned in the magnetic shields during transition to the superconducting 
		phase, will increase magnetic gradients at the position of the ion trap.
		The disadvantage is that multiple separate heat shields might complicate 
		the cryogenic setup.  If the vacuum pressure is too high for storage times of days 
		with a negligible amount of collision, one will have to add a dilution 
		refrigerator to the cryogenic system to freeze out even more background gas. 

		\subsection{Lasers and optical setup}
		\label{sec: optical setup}

		Locked lasers have servo bumps at the edge of the locking bandwidth, 
		which is typically on the order of 1~MHz \cite{GersterMSc}.  This frequency 
		is similar to the frequencies of motion in the trap.  Hence, if one drives 
		a sideband transition, the servo bump might be on resonance with the carrier
		and cause undesired quantum state evolution.  To avoid this, one can use 
		frequency-doubled laser systems.  As the second harmonic generation is a 
		non-linear process, the servo bumps are less efficiently converted than the 
		main laser line and thus attenuated compared to the main laser line.

		To avoid high frequency amplitude and phase noise in the laser light,
		one can use clean-up cavities, as they are used in measurements for 
		gravitational wave detection \cite{GWDIntensityStab}.  The remaining 
		(low frequency) amplitude noise can be compensated via AOMs or EOMs, 
		as typically done in trapped ion experiments. Furthermore, the clean 
		up cavities also reduce servo bumps.  

		The optical gate operations are performed off-resonantly on an S-to-P 
		transition with blue light, for exact wavelengths see Table~\ref{tab: atomic constants}.
		Blue light can cause bleaching in the fibers, which is a problem for 
		the continuous operation of the quantum computer, as such fibers have 
		to be replaced.  But in recent years, fibers which show low loss for 
		ultra-violet light have been developed \cite{YvesFiber}. 

		\begin{figure}[!htb]
			\begin{center}
				\includegraphics[width=13cm]{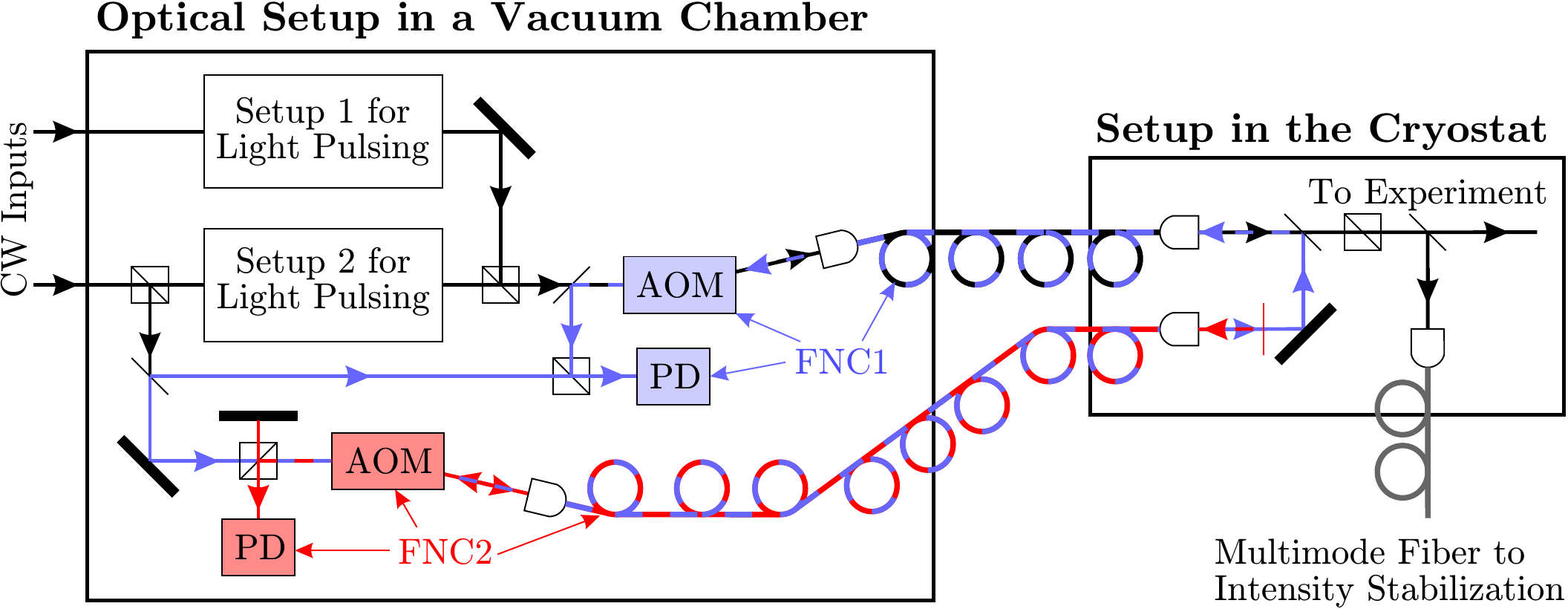}
				\caption{An optical setup for two Raman beams from CW sources with FNC and  
					pulsed intensity stabilization.}
				\label{fig: optical setup cryo}
			\end{center}
		\end{figure}

		For high fidelity gate operations, both the amplitude and the phase 
		of the light have to be stabilized.  Since the light is transmitted 
		into the cryogenic setup with a fiber, the light polarization at the 
		output of the fiber might fluctuate. Figure~\ref{fig: optical setup cryo} 
		shows an optical setup, in which a polarizer cleans the polarization 
		of the beam before being sent to the experiment.  After the polarizer, 
		a beam sampler reflects light partially into a multimode fiber whose 
		output is used for intensity stabilization, resulting in a stabilization 
		of both the intensity and the polarization. 

		As the corresponding wavelength of a clock transition in a hyperfine
		trapped ion qubit \cite{QubitTypes} is in the range of millimeters,
		phase stabilization is not required for single qubit operation.
		However, entangling operations may require phase stabilization
		during their execution \cite{SpontaneousPhotonScattering,MultiSpeciesGates}.
		Narrow linewidth light in a fiber experiences fiber noise \cite{FiberNoise}, 
		which might broaden the light to a linewidth of as high as 10~kHz.
		As entangling gate operations take about 100~$\mu$s, fiber noise
		has to be canceled for entangling gates. Fiber noise cancelation
		(FNC) setups modulate the phase or frequency of an AOM to
		cancel the effect of the fiber \cite{FiberNoiseCancelation}.
		FNC only works efficiently in continuous wave (CW) operation.
		In the setup of Figure~\ref{fig: optical setup cryo}, there
		are two fibers used to send light into the cryostat.
		A fraction of the main light in the optical setup is 
		coupled out and used for FNC of the second single mode fiber (FNC2).  
		Inside the cryostat, the output of the second fiber is reflected 
		back into the main fiber, and this CW light can be used for FNC of 
		the main fiber (FNC1).
		The light used for the gate operations can be pulsed and is sent
		via the FNC1 setup into the cryostat.  The remaining phase drifts 
		are from thermal drifts and acoustics coupling into the optical setup.
		Of these drifts, optical path length fluctuations between the 
		output of the fiber in the cryostat and the ions in the trap cannot 
		be stabilized actively.  However, inside the cryostat, there is no 
		acoustic noise and the temperature is stabilized.  Therefore, one 
		can neglect acoustic vibrations in the vibration isolated cryostat.  
		The remaining drifts are thermal, and thus slow.  In the firmware,
		one has to choose gate operations which can cope with these
		slow drifts, for example such as demonstrated in reference \cite{MultiSpeciesGates}.
		It is helpful to place the optical setup for pulsing the light and the 
		FNCs inside a small vacuum chamber and thermally stabilize it as well.
		That way, one can minimize the phase drifts during the experiment.

		In order to save space inside the inner heat shield, one should
		miniaturize the optical setup, shown on the right hand side of 
		Figure~\ref{fig: optical setup cryo}, consisting
		of 3 fibers and a polarizer, as such a setup will be required
		for each beam used for QIP.

		\begin{figure}[!htb]
			\begin{center}
				\includegraphics[width=11cm]{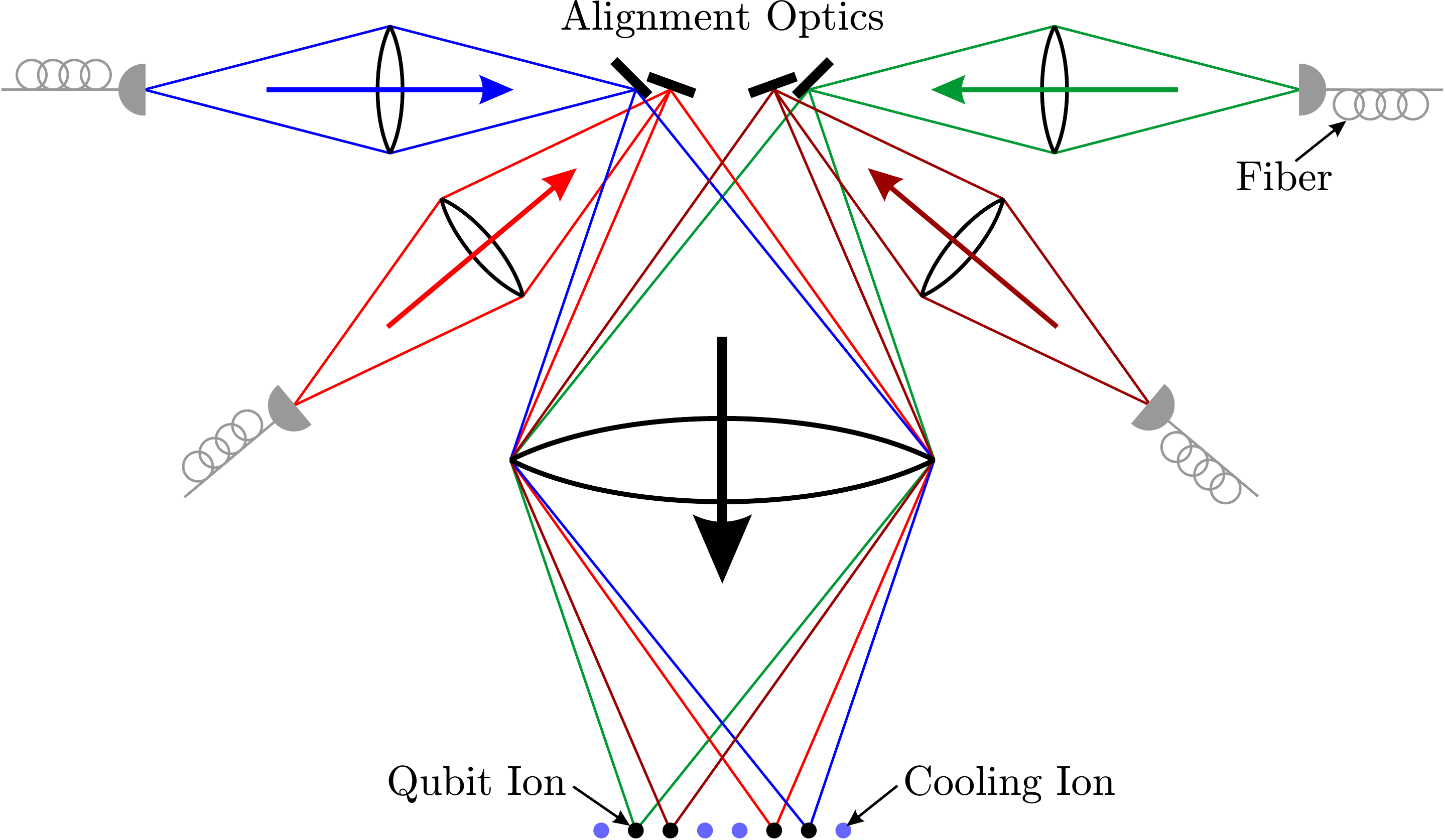}
				\caption{Possible addressing setup in the cryostat.}
				\label{fig: addressing in cryo}
			\end{center}
		\end{figure}

		The optical setup described in Section~\ref{sec: quantum gates - optical vs RF}
		requires single ion addressing capability for all ions
		in the QIP zone of the QALU.  A possible alignment strategy
		is illustrated in Fig.~\ref{fig: addressing in cryo}, where
		each ion that should be illuminated has its own optical setup.  
		The light from a fiber is imaged onto a mirror of a mirror
		array from where the light for the different ions is sent
		towards the ion string in the Paul trap. Therefore, an 
		optical setup images the mirror array onto the ion string.
		The mirror array can either be an array of fixed mirrors
		or can be a microelectromechanical systems (MEMS) array.

		\subsection{Detectors}
		\label{sec: detectors}

		It is advantageous to include detectors for photon counting in the 
		cryogenic setup.  There, transition edge sensors (TESs) may be
		the prime choice, as they have a quantum efficiency of up to 95~\% 
		\cite{TESefficiency}. These sensors exploit the strongly temperature-dependent
		resistance at a superconducting phase transition.  When a photon
		is absorbed in the superconductor, the material locally enters
		the normally conducting phase which can be detected electronically.

		During operation, electric current flows through the TES so that it 
		is close the phase transition.  This current is temperature dependent
		and changes when a photon is detected.  As the presented quantum von 
		Neumann architecture (for trapped ions) requires long coherence times 
		and thus high spatial homogeneity and temporal stability of the magnetic 
		field, TESs should not be operated on or close to the trap\footnote{
		Such sensor have been integrated into traps \cite{SupercondDetectors}.}
		in this architecture.  Ideally, the TESs are located near the superconducting
		shield or even outside of it so that the magnetic field generated 
		by their operation does not influence the coherence time in the system.  
		For such detection, one can use a detection lens close to the trap 
		and a second lens to focus the light onto the detector.  The idea
		is the same as for addressing, depicted in Fig.~\ref{fig: addressing in cryo} 
		only that the direction of the light is reversed. Another idea is to integrate
		a fiber into the trap for light collection.  The output of the fiber
		can then be directed onto the detector \cite{FiberCouplingDetection}.

		\subsection{Trap Design}
		\label{sec: trap design}

		A major design challenge for large planar traps with thousands of
		segments pairs is shunting the RF on the DC segments.  The traces on the
		trap might be too long for efficient shunting near the trap with shunt 
		capacitors or on the trap with trench capacitors.  Furthermore,
		trench capacitors on the trap would drastically increase the size of the trap.
		Hence, the RF will have to be shunted near the trapping zone of the trap.

		\begin{figure}[!htb]
			\begin{center}
				\includegraphics[width=11cm]{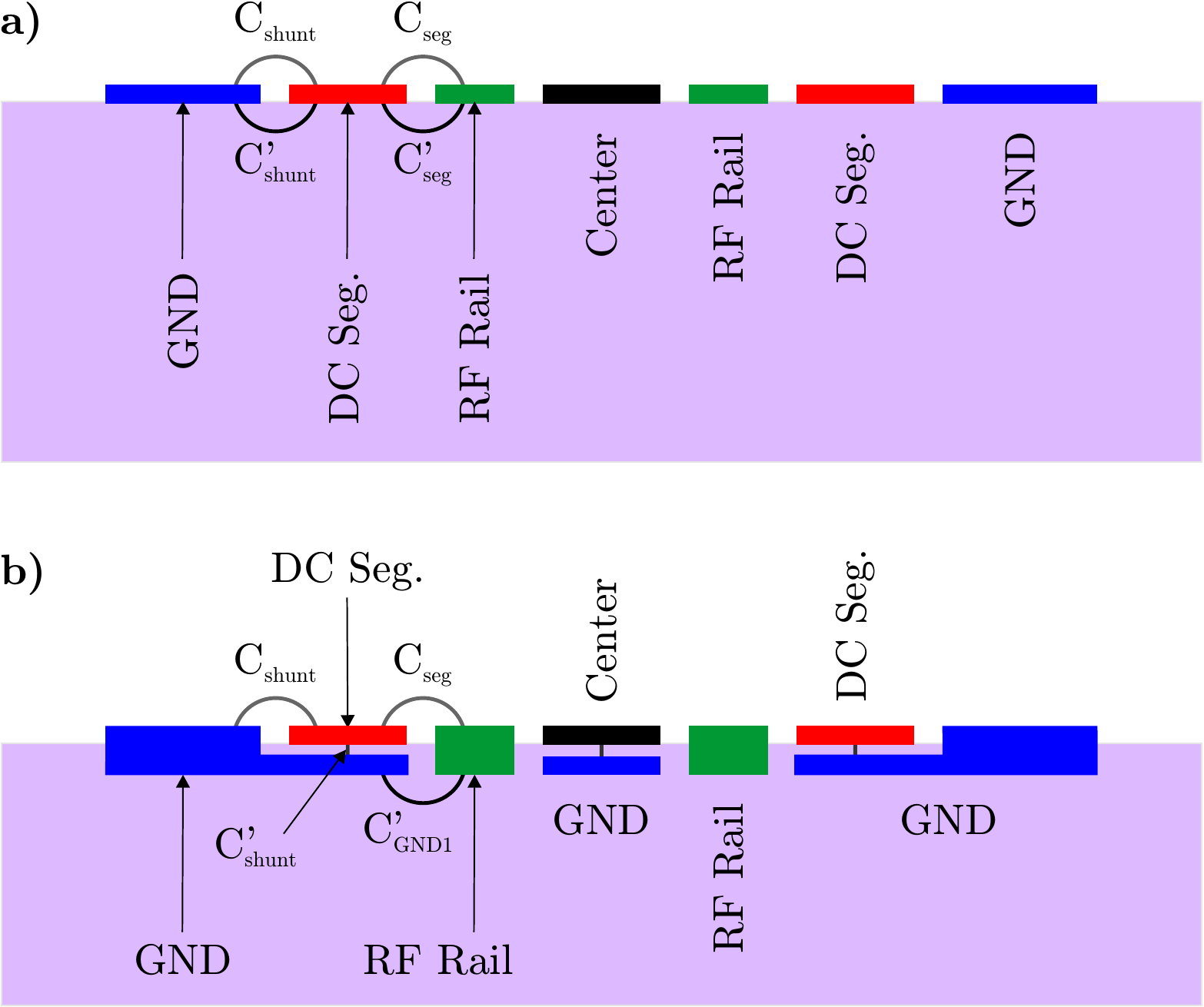}
				\caption{Cutaway views of a planar, segmented Paul trap. Panel~(a)
					displays the standard traps, and in panel~(b), ground planes
					below the DC segments are added.}
				\label{fig: trap suggestions}
			\end{center}
		\end{figure}

		The occurring capacitances of a planar Paul trap are illustrated in the 
		cutaway view of Figure~\ref{fig: trap suggestions}~a.  A DC segment has the 
		capacitances C$_\mathrm{shunt}$ and C$_\mathrm{shunt}^\prime$ to ground, 
		where C$_\mathrm{shunt}$ is the capacitance through vacuum
		and C$_\mathrm{shunt}^\prime$ is the capacitance through the trap material.
		C$_\mathrm{shunt}^\prime=\epsilon_\mathrm{r}$C$_\mathrm{shunt}$ with
		$\epsilon_\mathrm{r}$ the relative dielectric constant of the trap material,
		e.g. SiO$_\mathrm{2}$: $\epsilon_\mathrm{r}=3.8$.
		A DC segment has the capacitances C$_\mathrm{seg}$ and C$_\mathrm{seg}^\prime$
		to the next RF rail, where C$_\mathrm{seg}$ is the capacitance through vacuum
		and C$_\mathrm{seg}^\prime$ is the capacitance through the trap material.
		The ratio R=(C$_\mathrm{seg}$+C$_\mathrm{seg}^\prime$)/(C$_\mathrm{shunt}$+C$_\mathrm{shunt}^\prime$)
		defines the RF voltage on the DC segment without an additional shunting network
		and is typically around 1.  For efficient shunting, this should be 1/100 or 1/1000.

		In order to decrease the RF voltage on the trap, one can decrease the
		capacitance to the RF rail through the trap material and increase the 
		capacitance to ground by placing a ground electrode underneath each the 
		DC segment, as depicted in Figure~\ref{fig: trap suggestions}~b.  
		In typical planar traps, the gaps between segments are between 10 and
		30~$\mathrm{\mu}$m.  The skin depth in gold or aluminum for typical
		trap drive frequencies between 10 and 50~MHz is about 1~$\mathrm{\mu}$m
		at cryogenic temperatures.  Hence, the metal layers on the trap only
		need to be 2-5~$\mathrm{\mu}$m thick.  These dimensions cause C$_\mathrm{seg}^\prime$ 
		to vanish which reduces the total capacitance of a DC segment to the RF 
		rail by a factor $\epsilon_\mathrm{r}+1$.  Instead of C$_\mathrm{seg}^\prime$, 
		the capacitance C$_\mathrm{GND}^\prime$ to ground appears, where 
		C$_\mathrm{GND}^\prime$ is roughly equal to C$_\mathrm{seg}^\prime$ of 
		the old geometry.  Therefore, the capacitance for the RF rail does not 
		change, and the RF resonator will experience the same capacitive load. 

		In this new geometry, C$_\mathrm{shunt}^\prime$ is mainly dominated 
		by the parallel plate capacitor underneath the segment which results 
		in a higher capacitance than in the previous geometry.  For example, 
		SiO$_\mathrm{2}$ has a dielectric strength of about 30~V/$\mathrm{\mu}$m.  
		Thus, 1 or 2~$\mathrm{\mu}$m should be enough to safely operate the trap
		with DC-voltages of about $\pm$10~V.  As an example, a parallel plate 
		capacitor with an area 100~$\mathrm{\mu}$m~$\times$~100~$\mathrm{\mu}$m 
		and a 1~$\mathrm{\mu}$m thick SiO$_\mathrm{2}$ isolation layer results 
		in a capacitance of 0.33~pF. The capacitance C$_\mathrm{seg}$ of such
		a DC segment to the RF rail will be orders of magnitude smaller.
		In order to increase the capacitance C$_\mathrm{shunt}^\prime$ further, 
		one can place the ground plane underneath the DC segment all the way 
		between the bond pad and the trapping zone. 
		To sum up, in the new geometry, C$_\mathrm{seg}^\prime$ can be neglected 
		and the higher capacitance C$_\mathrm{shunt}^\prime$ reduces the ratio R 
		without changing the capacitances C$_\mathrm{seg}$ and C$_\mathrm{shunt}$ 
		which are important for the behavior of the trap.

		Another aspect that one has to take into account with planar
		traps, which contain hundreds of junctions, is that one has to 
		make sure that the RF potential in both rails is always in phase 
		with each other.  Axial micromotion arises due to the wiring, when
		the electrical path length in one rail is longer than in the other 
		rail by a fraction of the wavelength of the RF trap drive, 
		typically 3 to 5$^\circ$.  To circumvent this problem, one can short
		all connecting RF rails of a junction underneath it.  But this comes
		at the expense of a higher capacitance between the RF rails and
		ground (or DC segments). 

		\subsection{DAC design}
		\label{sec: dac design}

		With multiplexing architectures as described in Section~\ref{sec: multiplexing ion storage and movement},
		large scale planar traps with thousands of segment pairs may still require
		hundreds of controllable voltages.  An architecture in which a single 
		field-programmable gate array (FPGA) controls all DACs will limit the 
		number of controllable DACs to the number of pins on the FPGA and serial 
		communication will be required to reduce the number of interconnections 
		per DAC. 

		To circumvent the possibly slow serial communication, one can use a single 
		FPGA with a RAM chip for one or two DACs only.  
		From the central experiment control, the FPGA receives a command what segment 
		to control and what ion movement to execute. For example, "control segment 
		number 491 and move an ion string from left to right".  Standard segments will 
		have only few commands, like "move from left to right", and "move from right 
		to left".  Segments at junctions may require a couple more commands and segments 
		in the QALU will have the most complicated instruction set, as it has to include 
		things like splitting and recombining of ion chains.  All these different voltage 
		ramps can be stored in the RAM.

		Such a DAC-architecture allows incorporation of fast 16-bit DACs with more 
		than 100~MSamples/s.  The RAM only needs to contain one DAC output value 
		every 100~ns or 1~$\mu$s, and the FPGA can interpolate the values in between 
		to minimize quantization noise. The quantization noise of these DACs will have
		a frequency of more than 100~MHz and can easily be filtered so that it cannot 
		perturb the QIP. 

		Ideally, one wants to integrate a system consisting of the FPGA, RAM, and
		DAC on one single chip, like a special-purpose direct digital synthesis 
		(DDS)\footnote{DDS are devices with fast DACs that output a series of voltage 
		values with the help of look-up tables. This is the same operation as requires 
		in the system described above.}.  But even on a PCB, the DAC-unit for one 
		segment pair can be miniaturized.  To minimize the capacitive load on a DAC-unit, 
		the DAC-units should be located close to the trap, which would imply that they 
		need to be cryo-compatible.  In general, the DAC-units can be placed
		outside the vacuum chamber as well.  However, the multiplexing logic circuits 
		controlling the analog switches will have to be located in the cryogenic 
		environment to reduce the number of interconnects in the cryostat.  
		FPGAs, which can be used for the multiplexing logic circuits, 
		and analog switches are successfully operated in cryogenic environments 
		\cite{FPGAinCryo,AnalogSwitchinCryo}.
	\end{appendix}

	\clearpage
	\bibliography{SerialQC}

\end{document}